\documentclass[11pt]{article}
\pdfoutput=1

\usepackage[usenames,dvipsnames,table]{xcolor}
\usepackage[utf8]{inputenc}

\usepackage{jheppub} 

\usepackage{tikz-cd} 
\usepackage{circuitikz}
\usepgfmodule{shapes}
\usetikzlibrary{decorations.pathmorphing}
\usetikzlibrary{decorations.pathreplacing,decorations.markings,snakes}
\usetikzlibrary{backgrounds, arrows,calc,shapes,decorations.pathreplacing, automata,positioning}

\usepackage{cancel} 
\usepackage{xcolor}

\usetikzlibrary{shapes.misc}

\tikzset{cross/.style={cross out, draw=black, minimum size=5*(#1-\pgflinewidth), inner sep=0pt, outer sep=0pt},
cross/.default={2pt}}

\tikzset{snake it/.style={decorate, decoration=snake}}

\tikzset{
  hyper/.style    = { thick, double, 
  double distance = 3pt }}

\pgfdeclarepatternformonly{coarsedots}{\pgfqpoint{-1pt}{-1pt}}{\pgfqpoint{5pt}{5pt}}{\pgfqpoint{12pt}{12pt}}{
    \pgfpathcircle{\pgfqpoint{.5pt}{.5pt}}{.5pt}
    \pgfusepath{fill}}
    
\tikzset{gauge/.style={rounded rectangle, draw=black!100, thick, minimum size=5mm},  gaugeD/.style={rounded rectangle, draw=black!100,double,thick,minimum size=5mm},  empty/.style={rounded rectangle, draw=white!100, thick, minimum size=5mm}, flavor/.style={rectangle, draw=black!100, thick, minimum size=5mm},flavorD/.style={rectangle, draw=black!100, double,thick, minimum size=5mm}}

\usepackage{graphicx}
\usepackage{amsmath, amssymb}
\usepackage{youngtab}
\usepackage{changepage}

\newcommand{\be}{\begin{eqnarray}}
\newcommand{\ee}{\end{eqnarray}}
\newcommand{\ba}{\begin{array}}
\newcommand{\ea}{\end{array}}
\newcommand{\bea}{\begin{eqnarray}}
\newcommand{\eea}{\end{eqnarray}}
\newcommand{\bpic}{\begin{tikzpicture}}
\newcommand{\epic}{\end{tikzpicture}}

\newcommand{\nn}{\nonumber}
\newcommand{\bn}{\begin{enumerate}}
\newcommand{\en}{\end{enumerate}}

\def\CC{{\cal C}}

\def\CF{{\cal F}}

\def\cN{{\cal N}}
\def\CN{{\cal N}}

\def\CW{{\cal W}}
\def\cW{{\cal W}}

\def\Tr{\mathop{\text{Tr}}\nolimits}

\def\b{\beta}

\def\d{\delta}

\def\r{\rho}

\def\s{\sigma}

\def\u{\upsilon}

\def\M{\mathfrak{M}}

\title{Deconfinements, Kutasov--Schwimmer dualities and $D_p[SU(N)]$ theories}

\author[a]{Sergio Benvenuti,}
\author[b,c]{Riccardo Comi,}
\author[b,c]{Sara Pasquetti,}
\author[d]{Matteo Sacchi}

\affiliation[a]{INFN, Sezione di Trieste, SISSA, via Bonomea 265, 34136 Trieste, Italy}
\affiliation[b]{Dipartimento di Fisica, Università di Milano-Bicocca, Piazza della Scienza 3, I-20126 Milano, Italy}
\affiliation[c]{INFN, Sezione di Milano-Bicocca, Piazza della Scienza 3, I-20126 Milano, Italy}
\affiliation[d]{Mathematical Institute, University of Oxford, Andrew-Wiles Building, Woodstock Road, Oxford, OX2 6GG, United Kingdom}

\emailAdd{benve79@gmail.com, r.comi2@campus.unimib.it, sara.pasquetti@gmail.com, matteo.sacchi@maths.ox.ac.uk}

\abstract{Kutasov--Schwimmer (KS) dualities involve a rank-$2$ field with a polynomial superpotential. We derive KS-like dualities via deconfinement, that is assuming only Seiberg-like dualities, which instead just involve fundamental matter. Our derivation is split into two main steps. The first step is the 
construction of two families of  linear quivers with $p\!-\!1$ nodes that confine into a rank-$2$ chiral field with degree-$(p\!+\!1)$ superpotential. Such chiral field is an $U(N)$ adjoint in 3d  and an $USp(2N)$ antisymmetric in 4d. In the second step 
we use these linear quivers  to derive, via deconfinement, in a relatively straightforward fashion, two classes of KS-like dualities: the Kim--Park duality for $U(N)$ with adjoint in 3d and the Intriligator duality for $USp(2N)$ with antisymmetric in 4d.
We also discuss the close relation of our 3d family of confining unitary quivers  to the 4d $\mathcal{N}\!=\!2$ $D_p[SU(N)]$ SCFTs by circle compactification and various deformations. 
}

\begin{document}

\maketitle

\flushbottom

\section{Introduction and summary}

Infrared (IR) dualities are an interesting phenomenon that can characterize the low energy dynamics of quantum field theories (QFT). They involve a pair of different microscopic QFTs that describe the same low energy physics. Several instances of IR dualities are known, with the most important example being the Seiberg duality \cite{Seiberg:1994rs} for 4d $\CN=1$ supersymmetric quantum chromodynamics (SQCD) with $SU(N)$ gauge group  and fundamental matter. Soon after its discovery, Kutasov and Schwimmer (KS) \cite{1Kutasov_1995,Kutasov_1995} proposed a duality for $SU(N)$ with $F$ fundamental matter fields and an adjoint $\Phi$, with polynomial superpotential $\CW=\Tr(\Phi^{p+1})$. The dual theory theory has $SU(pF-N)$ gauge group, $F$ flavors and one adjoint $\phi$ with the same $\CW=\Tr(\phi^{p+1})$ superpotential, plus $p$ sets of singlets flipping the mesons dressed with the adjoint $\phi$
. The KS duality can be understood as a generalization of the Seiberg duality to which it reduces for $p=1$, since the superpotential becomes a mass term for the adjoint field which is then integrated out in the IR.

There exist other variants of this duality, to which we will sometime refer as \emph{KS-like dualities}. In \cite{Intriligator:1995ff} Intriligator proposed a version of the duality for symplectic gauge group (IKS). In this case we have a  $USp(2N)$ theory with $2F$ fundamental chirals 
and one antisymmetric $\Phi$ with $\mathcal{W}=\Tr(\Phi^{p+1})$ superpotential.\footnote{Throughout this paper, traces over $USp(2N)$ groups are taken with the insertion of the $2N\times 2N$ totally antisymmetric tensor $J_N=i\sigma_2\otimes \mathbb{I}_N$.} Moreover, there exist also three-dimensional $\mathcal{N}=2$ analogues of the KS duality: the Kim--Park (KP)  duality for a $U(N)$ theory with $F$ fundamental flavors and an adjoint chiral \cite{Kim_2013}, and its $SU(N)$ variant by Park--Park  \cite{Park_2013}. 

These KS-like dualities and their tests are subtle in many ways. The most straightforward test is anomaly matching. The mapping of chiral gauge invariant operators can also be worked out, however it is more intricate than in the ordinary Seiberg duality because of the truncation in the chiral ring of dressed operators due to the polynomial superpotential. The most powerful test of supersymmetric dualities is usually the matching of supersymmetric partition functions such as the supersymmetric index \cite{Kinney:2005ej,Romelsberger:2005eg,Dolan:2008qi,Bhattacharya:2008zy,Bhattacharya:2008bja,Kim:2009wb,Imamura:2011su,Kapustin:2011jm,Dimofte:2011py}, but this presents various difficulties for the KS-like dualities. Contrary to other dualities no index identities are known in the math literature and perturbative tests are hard to push to high order since the dual gauge group grows quickly (however see \cite{Dolan:2008qi} for the matching of the index in the KS duality in the Venziano limit).

During the last few years, various works managed to prove many dualities involving gauge theories with rank-$2$ matter fields assuming only the validity of \emph{Seiberg-like dualities}, that is dualities for simple classical gauge groups and only fundamental matter. This technique goes under the name of \emph{deconfinement} and it schematically works as follows. We trade the rank-2 field for an additional gauge node using some known s-confining duality in the sense of \cite{Csaki:1996zb} for which the dual theory is just a Wess--Zumino (WZ) model of chiral fields with no gauge group, such as the Seiberg duality for $F=N+1$. This gives an auxiliary quiver gauge theory that can be further dualized using Seiberg-like dualities until we reach the desired dual frame.

The idea of the deconfinement first appeared in \cite{Berkooz:1995km,Pouliot:1995me,Luty:1996cg}, but it has been recently revamped in many ways and in various dimensions (see e.g.~\cite{Pasquetti:2019uop,Sacchi:2020pet,Benvenuti:2020gvy,Etxebarria:2021lmq, Benvenuti:2021nwt,  Bajeot:2022kwt, Bottini:2022vpy, Bajeot:2022lah, Bajeot:2022wmu, Bajeot:2023gyl,Amariti:2023wts,Amariti:2024sde,Amariti:2024gco}). Examples of successes of the deconfinement program are the derivation \cite{Bajeot:2022kwt} of most of the s-confining dualities with classical gauge group of \cite{Csaki:1996zb}, and the derivation of a self-duality for $USp(2N)$ with one antisymmetric and $8$ fundamental chirals \cite{Benvenuti:2020gvy,Bajeot:2022lah}.

The deconfinement technique fits into the larger program of trying to derive all IR dualities from a small set of fundamental Seiberg-like dualities. Along this line of research, a related approach is the \emph{local dualization algorithm}, which
allows us to prove mirror dualities with eight \cite{Hwang:2021ulb,Comi:2022aqo,Giacomelli:2023zkk,Giacomelli:2024laq} and four supercharges \cite{Benvenuti:2023qtv},
assuming only Seiberg-like dualities that are either the Intriligator--Pouliot (IP) duality \cite{Intriligator:1995ne} in 4d or the Aharony duality \cite{Aharony:1997gp} in 3d.

In this paper we tackle the problem of proving KS-like dualities 
using the deconfinement. In particular, we focus on the 3d KP duality for $U(N)$ gauge group with adjoint matter and the 4d IKS duality for $USp(2N)$ gauge group with antisymmetric matter, and show how it is possible to derive them assuming only Seiberg-like dualities. These are again the IP duality in 4d and the Aharony duality in 3d. In fact in three dimensions we will also make use of dualities with monopole superpotentials \cite{Benini:2017dud}, which however can still be obtained as a deformation of the Aharony duality. Let us point out that the derivation of the 3d KP duality provides also strong evidence for the validity of the 4d KS $SU(N)$ duality. Indeed, as shown in \cite{Amariti:2014iza,Nii_2015}, the 3d KP duality follows from the 4d $SU(N)$ KS duality via dimensional reduction plus deformations.\footnote{The 3d KP duality can also be obtained by a different dimensional reduction from the 4d IKS duality
along the lines of
\cite{Benini:2017dud,Amariti:2018wht,Pasquetti:2019hxf,Bajeot:2022lah,Bottini:2021vms,Comi:2022aqo,Bajeot:2023gyl}. 
From this perspective, our 4d derivation already contains the 3d one since we can take such dimensional reduction limit at any of the steps. However, for clarity of exposition we will discuss the 3d and the 4d cases separately.

}

It is important to emphasize again that in the mathematical literature there is no proof of the equality of supersymmetric partition functions and indices associated to KS dual theories. Since the integral identities corresponding to Seiberg-like dualities have instead been proven \cite{2003math......9252R,Fokko,Spiridonov:2009za,Spiridonov:2011hf}, our arguments amount to a proof of the integral identities corresponding to KS-like dualities.

Our derivation is somewhat of a tour-de-force, the strategy involves two main steps. The first one consists of constructing families of linear quivers with $p-1$ nodes, which are infrared dual to a chiral superfield $\Phi$ in a rank-$2$ representation of the flavor symmetry with polynomial superpotential $\CW=\Tr(\Phi^{p+1})$. We build two such classes of theories:
the 3d $\cN=2$ $\mathcal{C}_p[SU(N)]$ family deconfines an $SU(N)$ adjoint, while the 4d $\cN=1$ $\mathsf{C}_p[USp(2N)]$ family deconfines a $USp(2N)$ antisymmetric. As an example, the $\mathcal{C}_p[SU(pr+l)]$ ($0 < l < p$) theories are given by the following quiver:
\begin{align}\nn\label{fig:Cp_quiver_ln00}
\includegraphics[scale=1.05]{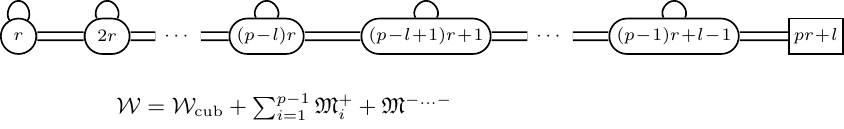}
\end{align}
where $\mathcal{W}_{\text{cub}}$
consists of the cubic couplings between all the adjoint fields and the adjacent bifundamentals,  while the non-perturbative monopole terms
break supersymmetry to $\mathcal{N}=2$
and are responsible for the confinement of the theory.

The proof that these linear quivers are IR dual to the WZ model with a single tensor field and degree-$(p+1)$ superpotential is highly non-trivial and works by induction. Each inductive step involves taking the mirror quiver, and using sequentially the appropriate Seiberg-like dualities.

The second step is relatively straightforward and consists in using our new confining dualities to deconfine the rank-$2$ field in the KS-like dualities. We start from the electric side of the KS-like duality and deconfine the rank-$2$ field into the $p-1$ nodes quiver, obtaining a linear quiver of lenght $p$. Now we apply iteratively the Seiberg-like duality from one side to the other of the quiver. This process involves $p$ dualizations, hence it produces precisely $p$ sets of flipping singlets. Moreover, the final quiver is precisely the magnetic gauge group predicted by the KS-like duality attached to the $p-1$ nodes quiver that deconfines the rank-$2$ field $\phi$ in the appropriate representaiton and with $\CW=\Tr(\phi^{p+1})$ interaction. Hence, we can reconfine the whole quiver into the magnetic theory. In this way, we manage to derive the 3d $\CN=2$ KP duality for $U(N)$ with adjoint and $F$ fundamental flavors and the 4d $\CN=1$ IKS duality for $USp(2N)$ with antisymmetric and $2F$ fundamentals.

One important aspect of our story is the relation between the 3d $\CN=2$ $\mathcal{C}_p[SU(N)]$ family and the 4d $\CN=2$ $D_p[SU(N)]$ superconformal field theories (SCFTs), which are non-Lagrangian theories that can be realized either via geometric engineering or in class $\mathcal{S}$ \cite{Cecotti:2012jx,Cecotti:2013lda,Wang:2015mra}. Recently \cite{Maruyoshi:2023mnv}, building on some earlier results \cite{Xie:2021omd,Kang:2023dsa,Bajeot:2023gyl}, it was argued that a $D_p[SU(N)]$ theory with $\gcd(p,N)=1$ deformed by a Coulomb Branch (CB) operator of dimension $p/(p+1)$ flows in the IR to an $SU(N)$ adjoint chiral with $\mathcal{W}=\Tr(\Phi^{p+1})$ superpotential. This result is closely related to our new confining dualities, in particular our 3d $\mathcal{N}=2$ $\CC_p[SU(N)]$ quiver theories are very similar to the   3d $\mathcal{N}=4$ quivers  describing the dimensional reduction of $D_p[SU(N)]$, as already pointed out in \cite{Bajeot:2023gyl} for $p=2$ and $N=2r+1$. Indeed, despite being non-Lagrangian in 4d, the circle compactification of $D_p[SU(N)]$ admits a Lagrangian description in terms of a linear quiver \cite{Closset:2020afy,Giacomelli:2020ryy}. When $\gcd(p,N)=1$ this has exactly the same matter content and perturbative superpotential, i.e.~excluding the monopole terms, of our $\mathcal{C}_p[SU(N)]$ confining theory.  
Interestingly for other values of $p$ and $N$ our 3d $\mathcal{C}_p[SU(N)]$ confining theories are still very closely related (by simple deformations) to the circle reduction of $D_p[SU(N)]$. Specifically, compactifying on $S^1$ the non-Lagrangian 4d $\CN=2$ $D_p[SU(N)]$ theory for $\gcd(p,N)\neq1$ and $N\neq pr$, one still gets a linear 3d $\CN=4$ quiver of length $p-1$, but with mixed unitary and special unitary gauge groups. 
One then has to perform a combination of a deformation breaking the special unitary groups to unitary, dualizations and deformations by monopole superpotential to land on $\mathcal{C}_p[SU(N)]$. 

The rest of the paper is organized as follows. In Section \ref{sec3dandDp} we introduce the 3d $\mathcal{N}=2$ $\mathcal{C}_p[SU(N)]$ theory, discuss various of its properties and show that it confines to an $SU(N)$ adjoint with $\CW=\Tr(\Phi^{p+1})$. In Section \ref{sec:KPproof} we use this new confining duality to provide a derivation of the 3d KP duality via deconfinement and sequential Aharony dualizations. Section \ref{sec:DpSUN} deals with the relation between our 3d $\mathcal{N}=2$ $\mathcal{C}_p[SU(N)]$ theories and the 4d $\CN=2$ $D_p[SU(N)]$ SCFTs. In Section \ref{sec:4dCp} we discuss the 4d uplift of the $\mathcal{C}_p[SU(N)]$ theory, which we call $\mathsf{C}_p[USp(2N)]$, and its confining duality into a $USp(2N)$ antisymmetric with $\CW=\Tr(\Phi^{p+1})$. In Section \ref{sec:KSI} we use this to provide a derivation of the 4d IKS duality via deconfinement and sequential IP dualizations. We conclude in Section \ref{conclusions} with some comments on various open questions and possible future directions. The main text is supplemented by Appendix \ref{appendix}, in which we review some important facts about the Aharony duality and the variants with monopole superpotentials.

\paragraph{Note added:} While this manuscript was being finalized we learned about the work of \cite{Hwang} which has a substantial overlap with our manuscript.
We thank the authors of \cite{Hwang} for sharing their draft with us and for coordinating the submission.

\paragraph{Notation for flippers.} Flipping fields $f_O$ are gauge singlet chiral fields that enter the superpotential only once, linearly and multiplying some operator $O$. In other words the superpotential term is $\mathcal{W}_{\text{flip}}=f_O \, O$. As a consequence, the F-term equation of $f_O$ set to zero $O$ in the chiral ring. We denote such singlet fields simply as $F[O]$. Moreover, we write the superpotential term as $\CW_{\text{flip}} = \text{Flip}[O]$.

\section{The 3d $\cN=2$ $\mathcal{C}_p[SU(N)]$ family}\label{sec3dandDp}

In this section we introduce a class of 3d $\mathcal{N}=2$ theories named $\mathcal{C}_p[SU(N)]$, where $p$ and $N$ are integers such that $N \geq p+1$. As we shall see, these theories are constructed by turning on an appropriate monopole superpotential in certain 3d $\mathcal{N}=4$ theories which explicitly breaks supersymmetry down to $\mathcal{N}=2$. We will show that the $\mathcal{C}_p[SU(N)]$ theories confine to a Wess--Zumino (WZ) model of a chiral $\Phi$ in the adjoint representation of a $SU(N)$ flavor symmetry with $\cW \sim \Phi^{p+1}$. Along the way, we also construct the mirror description of the $\mathcal{C}_p[SU(N)]$ theories.

\subsection{The  $\mathcal{C}_p[SU(N)]$ family}\label{sec:supfam}

The $\mathcal{C}_p[SU(N)]$ theories are 3d $\cN=2$ theories whose global symmetry,on top of the $U(1)_R$ R-symmetry, is\footnote{In the following we will only focus on the $SU(N)$ Lie algebra of the non-abelian flavor symmetry and ignore the global structure. We will however pay attention to the finite $\mathbb{Z}_{p+1}$ symmetry.}
\begin{equation}
    PSU(N)\times \mathbb{Z}_{p+1}\,.
\end{equation}
They are defined as a quiver theory with $p-1$ unitary gauge nodes of the form\footnote{In the drawing each circle node is a gauge symmetry, the square node is the flavor symmetry, the double lines between nodes represent pairs of bifundamental chirals in conjugate representations, and arcs are adjoint chirals.}
\begin{align}\label{fig:Cp_quiver}
\includegraphics[scale=1.2]{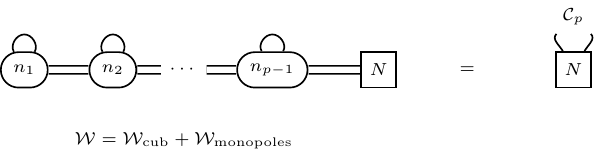}
\end{align}
The $\CC_p[SU(N)]$ theory is also depicted in short as given on the right. This notation is motivated by the fact that, as we shall see, the basic operator charged under the $SU(N)$ symmetry transforms in the adjoint representation and more generally the only representations that appear in the spectrum of gauge invariant operators are uncharged under the $\mathbb{Z}_N$ center, so that the faithful symmetry is actually $SU(N)/\mathbb{Z}_N$.

It is convenient to define
\begin{equation}
    N = p r + l \,,\qquad l=0,\cdots,p-1\,,
\end{equation}
so that the ranks of all the gauge nodes can be obtained via the formula
\begin{align}\label{eq:ranks_Cp}
    n_j=j r+ [l-p+j]^+\,,\qquad j=1,\cdots, p-1 \,,
\end{align}
where $[k]^+=k$ if $k>0$ and zero otherwise.
From \eqref{eq:ranks_Cp}, it is possible to compute that all the gauge nodes are balanced, i.e.~the number of flavors is twice their ranks, except for the $j=p-l$ node which is overbalanced with excess number $1$, i.e.~the number of flavors is twice the gauge rank plus one. 

The superpotential contains two types of terms. The first one $\CW_{\text{cub}}$ consists of the standard cubic couplings between all the adjoint fields and the adjacent bifundamentals that characterizes $\mathcal{N}=4$ theories when presented in terms of $\mathcal{N}=2$ fields. The second one $\CW_{\text{monopoles}}$, which we shall specify momentarily, involves monopole operators and breaks supersymmetry to $\mathcal{N}=2$. In particular, it only preserves one combination of the two abelian factors in the Cartan subgroup of the $\mathcal{N}=4$ $SU(2)_H\times SU(2)_C$ R-symmetry, which is identified with the $\mathcal{N}=2$ R-symmetry. For any generic $l$, the R-charges of the chiral fields under such symmetry are
\begin{align}
    & R[\text{bifundamental}] = \frac{1}{p+1} \,, \nn \\
    & R[\text{adjoint}] = 2 - \frac{2}{p+1} \,.
\label{ras}
\end{align}
The remaining combination is instead broken to a finite subgroup. As we will see more in details later, this is a $\mathbb{Z}_{p+1}$ symmetry under which the charges of the chirals are\footnote{Despite the fact that the bifundamental chirals have fractional charge, as we shall see below the minimal charge appearing in the spectrum of gauge invariant operators is the unit one in this parametrization.}
\begin{align}\label{eq:finitechir}
    & q[\text{bifundamental}] = \frac{1}{2}\text{ mod } (p+1)\,, \nn \\
    & q[\text{adjoint}] =-1\text{ mod } (p+1)= p\text{ mod } (p+1) \,.
\end{align}
This finite symmetry is also a residue of the $p-1$ topological symmetries, so that the charge of the monopoles under it also has a bare contribution on top of the 1-loop one due to the fermions in the chirals \cite{Borokhov:2002ib,Borokhov:2002cg,Gaiotto:2008ak,Benna:2009xd,Bashkirov:2010kz}. Specifically, the remnant of the topological symmetry is the one that mixes with coefficient $+1$ for the balanced nodes and with coefficient $\frac{3}{2}$ for the unbalanced $(p-l)$-th node in the $l\neq 0$ case. In this way, a monopole with magnetic fluxes $m^{(j)}_{a_j}$ under the $j$-th node, with $j=1,\cdots,p-1$ and $a_j=1,\cdots,n_j$, has bare charge (in this convention $n_p=0$ when $l=0$)
\begin{align}
    q_0\left[m^{(j)}_{a_j}\right]=
        \sum_{\substack{{j=1} \\ j\neq p-l}}^{p-1}\sum_{a_j=1}^{n_j}m^{(j)}_{a_j}+\frac{3}{2}\sum_{a_{p-l}=1}^{n_{p-l}}m^{(p-l)}_{a_{p-l}}\,,
\end{align}
to which one should add the contribution of the fermions in the chirals with charges \eqref{eq:finitechir} with the usual formula.

Let us now discuss in more details the two possible cases $l \neq 0 $ and $l=0$.

\paragraph{\boldmath$l \neq 0$.} In this case $\CW_{\text{monopoles}}$ consists of a linear monopole superpotential, involving the monopoles $\M_i^+$ ($i=1,\cdots,p-1$) with magnetic fluxes $+1$ for each gauge node separately and a monopole $\mathfrak{M}^{- \ldots -}$ with magnetic flux $-1$ for all gauge nodes simultaneously. The theory can be depicted as
\begin{align}\label{fig:Cp_quiver_ln0}
\includegraphics[scale=1]{Figures/tikz/Cp_quiver_ln0}
\end{align}
This theory starts with a balanced tail of $p-l-1$ nodes, then we have an overbalanced node and then a sequence of $l-1$ balanced nodes.


\paragraph{\boldmath$l=0$.} In this case $\CW_{\text{monopoles}}$ is similar to that of the $l \neq 0$ case in \eqref{fig:Cp_quiver_ln0}, with the difference being that the $\M^{- \ldots -}$ monopole is now dressed with an adjoint chiral,\footnote{This can be taken to be any of the adjoint chirals, since monopoles with the same level of dressing are identified due to quantum relations.} denoted by $\M_A^{- \ldots -}$, and there is also a superpotential flipping the $\M^{- \ldots -}$ monopole. The theory can be depicted as
\begin{align}\label{fig:Cp_quiver_l0}
\includegraphics[scale=1]{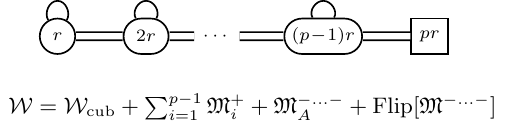}
\end{align}
Notice that now all the nodes are balanced.

\paragraph{Effects of the monopole superpotential (\boldmath$l\neq 0$).}
If we turn off all the monopole superpotentials in $\CC_p[SU(N)]$, then this coincides with the 3d $\CN=4$ $T^\sigma_\rho[SU(N)]$ theory \cite{Gaiotto:2008ak} with
\begin{equation}\label{eq:Cppartitions}
    \sigma=[1^N] \,,\qquad \rho=[ (r+1)^l, r^{p-l} ]\,,
\end{equation}
in particular for $l=0$ we just have $\rho=[r^p]$. The theory is

\begin{align}
    \includegraphics{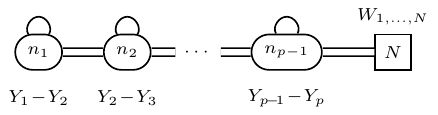}
    \label{trho}
\end{align}
where $Y_i-Y_{i+1}$ is the FI parameter for the topological symmetry of the $i$-th node,
$W_1\ldots W_N$ the mass parameters for the flavor symmetry and we  pick the following convention for the R-charges and charges under the $U(1)_A$ axial symmetry (that is the commutant of the $\mathcal{N}=2$ R-symmetry inside the $\mathcal{N}=4$ R-symmetry):
\begin{align}\label{eq:mir_charges}
    & R[\text{hypers}] = \frac{1}{2} \quad, \quad A[\text{hypers}] = 1 \,, \nn \\
    & R[\text{adjoint}] = 1 \quad , \quad A[\text{adjoint}] = -2 \ \,.
\end{align}

From the partition $\rho$ we see an $S[U(l)\times U(p-l)]$ symmetry acting on the Coulomb branch operators, which is enhanced from the topological symmetries of the two tails of balanced nodes.
In this theory the monopoles are charged under the various abelian symmetries, as it can be computed using the usual monopole formula. For convenience, we encode these charges as the coeffients in the linear combination of the parameters in the $S^3$ partition function for the corresponding symmetries. Focusing on the monopoles appearing in the superpotential for $l\neq 0$, we have 
\begin{align}\label{eq:monopcharge_def}
& \M_i^+:\quad  2\left(i \frac{Q}{4}-m_A\right)+Y_i-Y_{i+1}       \,, \qquad i=1,\cdots,p-1\,,\,\, i\neq p-l \,, \nn\\
 &\M_{p-l}^+:\quad   3\left(i \frac{Q}{4}  -m_A\right)+Y_{p-l}-Y_{p-l+1}     \nn  \,,\\
&\mathfrak{M}^{- \ldots -}:\quad 3\left(i \frac{Q}{4}  -m_A\right)+Y_p-Y_1\,,
\end{align}
where $m_A$ is the real mass parameter for $U(1)_A$ and $i\tfrac{Q}{2}$ is related to the background field for the $\CN=2$ R-symmetry, meaning that the R-charge can be read as the coefficient of $i\tfrac{Q}{2}$. 
Turning on $\CW_{\text{monopoles}}$ requires all of these monopoles to have charge two, which corresponds to setting the combinations of parameters in \eqref{yc} equal to $iQ$.
From the positive monopoles we see that all the $Y_i$ parameters are then specialized in terms of the axial mass $m_A$ as
\begin{align}
& Y_i-Y_{i+1}= i\frac{Q}{2}+2m_A
\,, \qquad i=1,\cdots,p-1\,,\,\, i\neq p-l \,, \nn\\
 &Y_{p-l}-Y_{p-l+1}= i\frac{Q}{4}+3m_A\,.
 \label{yc}
\end{align}
Furthermore, from the negative monopole we then get the further constraint
\begin{align}\
 2(p+1)m_A  = i\frac{Q}{2} (1-p)  \,,
 \label{mc}
\end{align}
which consistently yields the R-charge assignment in \eqref{ras}. From this perspective one can also show that there is a finite $\mathbb{Z}_{p+1}$ subgroup of the axial and the topological symmetries that is preserved. We will come back to this point around equation \eqref{imco}.
The $l=0$ case is analogous.

\paragraph{Background BF couplings.} If we treat the flavor symmetry as $U(N)$ rather than $SU(N)$, as we will do in the later section where such symmetry will be gauged, then we should also include in the definition of the theories a BF coupling between the R-symmetry and the $U(1)$ part of $U(N)$. This can be encoded in its contribution to the $S^3$ partition function as follows:
\begin{align}\label{eq:BFCpSUN}
\begin{cases}
    \mathrm{e}^{2 \pi i \frac{iQ}{2}\frac{1-l}{p+1} \sum_{i=1}^{pr+l} W_i} & l\neq 0 \\
    \mathrm{e}^{2 \pi i \frac{iQ}{2}\frac{1-p}{p+1} \sum_{i=1}^{pr} W_i} & l=0
\end{cases}
\end{align}
Notice that these only appear in the combination $\sum_{i=1}^NW_i$ corresponding to the $U(1)$ part of $U(N)$, so the BF couplings trivialize if we treat the symmetry as $SU(N)$ setting $\sum_{i=1}^NW_i=0$.

\paragraph{Spectrum of gauge invariant operators.} 

Below we discuss some interesting gauge invariant operators of the $\CC_p[SU(N)]$ theory which we will later map to the singlets in the confined dual frame \eqref{fig:conf_genk}.  First of all we have the operator $\mu$ in the adjoint of $SU(N)$. From the quiver theory in \eqref{fig:Cp_quiver} this operator is constructed as 
\begin{equation}
    \mu=Q \tilde{Q}\,,
\end{equation}
where $Q$ and $\tilde{Q}$ are the last bifundamental chirals, charged under $SU(N)$. Notice that this coincides with the moment map of the flavor $SU(N)$ in the $\mathcal{N}=4$ theory obtained by turning on only $\CW_{\text{cub}}$ and not $\CW_{\text{monopoles}}$, and by some abuse of notation we will sometimes refer to it as ``moment map" even in the $\mathcal{N}=2$ $\CC_p[SU(N)]$ theory. This operator has R-charge
\begin{align}
    R[\mu]=\frac{2}{p+1} 
\end{align}
and it has charge under the finite $\mathbb{Z}_{p+1}$ symmetry
\begin{align}
    q[\mu]=1\text{ mod } (p+1)\,.
\end{align}

In addition to this operator we also have a set of gauge invariant chiral operators that are singlets under the $SU(N)$ flavor symmetry. In \eqref{fig:Cp_quiver} we can construct monopoles that have $-1$ magnetic flux under a string of $j$ consequent balanced gauge nodes, which we denote by $\M^{\ldots -^j \ldots}$. We recall that for $l \neq 0$ the theory has two sequences of balanced nodes of length $p-l-1$ and $l-1$ respectively, as explained below \eqref{fig:Cp_quiver_ln0}. For $l = 0$ all the nodes are balanced, however the monopole with magnetic flux $-1$ under all the gauge nodes is flipped, so that in total we have $p-2$ such monopoles. The R-charge of a monopole charged under a string of $j$ balanced gauge nodes is\footnote{This can be easily calculated starting from the $\CN=4$ theory with the same strategy used around \eqref{eq:monopcharge_def}. A monopole with $-1$ magnetic flux under a string of consequent balanced nodes starting from the node $k$ and extending to the node $k+j$ has the following charges:
   $$ \M^{\ldots -^j \ldots}:\quad 2\left(i \frac{Q}{4}- m_A\right)-Y_{k}+Y_{k+j}   =\left(2 - \frac{2(j+1)}{p+1}\right)i\frac{Q}{2} \,,$$
when using the constraints in eqs.~\eqref{yc} and \eqref{mc}. Notice that the result does not depend on the choice of $k$ but only on the length $j$.
} 
\begin{align}\label{eq:monop_rch}
    R\left[\M^{\ldots -^j \ldots}\right] = 2 - \frac{2(j+1)}{p+1} \,.
\end{align}
Instead, the charge under the $\mathbb{Z}_{p+1}$ symmetry is
\begin{align}
    q\left[\M^{\ldots -^j \ldots}\right]=-(j+1)\text{ mod } (p+1)=p-j\text{ mod } (p+1) \,.
\end{align}
Indeed, in the $\CC_p[SU(N)]$ theory there are many such monopoles. We claim that monopoles with the same length constructed from the same balanced tail are degenerate. This leaves us with two sets
\begin{align}\label{eq:Cp_monop}
    l \neq 0 \quad : \quad & \begin{cases}
    & \M_L^{\ldots -^j \ldots} \qquad j=1,\ldots,p-l-1 \qquad \text{from the left tail} \nn \\
    & \M_R^{\ldots -^j \ldots} \qquad j=1,\ldots,l-1 \qquad \text{from the right tail}
    \end{cases} \,, \nn \\
    l = 0 \quad : \quad & \M^{\ldots -^j \ldots} \qquad j=1,\ldots,p-2 \qquad \text{from the only tail} \,. 
\end{align}
Finally, for $l=0$ we also have the flipper $F[\mathfrak{M}^{- \ldots -}]$, whose R-charge is $\tfrac{2p}{p+1}$ and whose $\mathbb{Z}_{p+1}$ charge is $p$.

The operators that we have just discussed are the most elementary ones, however the theory also possesses composite operators, some of which in particular parametrize a conformal manifold. Indeed the operator obtained as the product of $\M_{L/R}^{\ldots -^j \ldots}$ and $\M_{L/R}^{\ldots -^{p-j-1} \ldots}$ has R-charge equal to 2 and is not charged under any global symmetry for every $j$, therefore it represents an exactly marginal deformation. One may also consider other exactly marginal deformations given by the product of more than two $\M_{L/R}^{\ldots -^j \ldots}$ operators, such that the R-charge of the result is 2.

\subsubsection{Statement of the confining duality}\label{sec:deconfdual}

We claim that these theories confine into an adjoint $SU(N)$ chiral with $\cW \sim \Tr(\Phi^{p+1})$ plus extra singlets
\begin{align}\label{fig:conf_genk}
\includegraphics[scale=1.2]{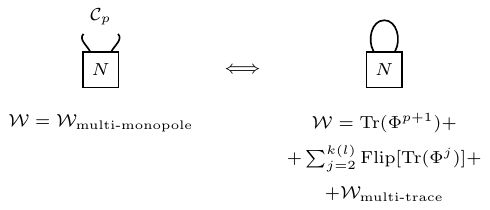}
\end{align}
where in the figure we are depicting the $\CC_p[SU(N)]$ theory in short as the label $\CC_p$ embedded in an arc to recall that its spectrum contains only one operator charged under the $SU(N)$ global symmetry, which is in the adjoint representation.

On the r.h.s.~of \eqref{fig:conf_genk} we have flipping fields for traces of $\Phi$ from power 2 up to $k(l)$, which is given by (recall $N=pr+l$)
\begin{align}
	k(l) = \text{min}(l,p-l) \,.
\end{align}
Notice that the above formula enjoys the property $k(l) = k(p-l)$. This will turn out to be important in the next section when we will use these theories to provide a derivation of the Kim--Park duality, which maps the gauge rank $N=p r +l$ to $N'=pr' +(p-l)$.

In the duality \eqref{fig:conf_genk}, the superpotential $\CW_{\text{multi-monopole}}$ on the l.h.s.~contains all the possible exactly marginal deformations given by the product of many $\M_{L/R}^{\ldots -^j \ldots}$ operators, as described in the end of the previous subsection. Correspondingly on the r.h.s.~we are turning on all the exactly marginal deformations given as multi trace operators, encoded in $\CW_{\text{multi-trace}}$. The reason why the duality holds at such point of the conformal manifold is motivated by the $S^3$ partition function perspective. The derivation of the duality that we will discuss in field theory in the next section can indeed be rigorously applied at the level of the partition function. However, the downside of this approach is that the partition function is not sensitive to the precise structure of the superpotential at each step of the derivation, but only to the charges of the fields induced by it. In particular, in the partition function computation one finds contributions of gauge singlet fields with charges that are compatible with exactly marginal massive deformations. In order to find the duality \eqref{fig:conf_genk} one has to simplify the contribution of such fields, which should be interpreted as being on a point of the conformal manifold where the mass deformations are activated and the singlets are integrated out. Additionally, some of the steps of our derivation will involve confining gauge nodes and in this process all possible superpotential terms compatible with the charge assignments can be generated. 
A similar situation has been discussed in \cite{Amariti:2021lhk,Amariti:2022dyi}.

For these reasons, we only claim that the duality \eqref{fig:conf_genk} holds on a generic point of the conformal manifold. 
See Section \ref{sec:lemma2} for more comments on this point in the case $N=p+1$.

The superpotential $\Tr(\Phi^{p+1})$ in the WZ theory uniquely determines the R-charge of $\Phi$ to be
\begin{align}
    R[\Phi]=\frac{2}{p+1}
\end{align}
and it also preserves a finite $\mathbb{Z}_{p+1}$ symmetry such that
\begin{align}
    q[\Phi]=1\text{ mod } (p+1)\,.
\end{align}
Such charge assignment is compatible with the following operator map across the duality. The operator $\mu$ in the adjoint of $SU(N)$ on the electric side gets mapped to the adjoint chiral in the dual WZ model
\begin{align}
    \mu\quad &\longleftrightarrow \quad \Phi\,.
\end{align}
We can also map the two set of monopoles $\M_{L/R}^{\ldots -^j \ldots}$ of the $\CC_p[SU(N)]$ theory to flavor singlets in the WZ model. Suppose that $\M_L^{\ldots -^j \ldots}$ is the biggest of the two sets, that is $p-l \geq l$. Then the map is
\begin{equation}\label{eq:confmapmono}
\left\{
\begin{tabular}{c}
 $\M_L^{\ldots,-^j \ldots}$  \\
 $\M_R^{\ldots,-^j \ldots}$
\end{tabular} 
\right\}
\qquad \longleftrightarrow \qquad
\left\{
\begin{tabular}{c}
 $\Tr(\Phi^{p-j})$  \\
 $\CF[\Tr(\Phi^{j+1})] $
\end{tabular}
\right\}
\end{equation}
If $p-l < l$  instead the map is
\begin{equation}\label{eq:confmapmono}
\left\{
\begin{tabular}{c}
 $\M_L^{\ldots,-^j \ldots}$  \\
 $\M_R^{\ldots,-^j \ldots}$
\end{tabular}
\right\}
\qquad \longleftrightarrow \qquad
\left\{
\begin{tabular}{c}
 $\CF[\Tr(\Phi^{j+1})]$  \\
 $\Tr(\Phi^{p-j})$
\end{tabular}
\right\}
\end{equation}
The operator map also tells that some of the monopoles are actually composite because they are mapped to traces of powers of $\Phi$, so there are quantum relations in the electric theory that identify such monopoles with traces of powers of $\mu$. For $l=0$ the electric theory also has the flipper $F[\mathfrak{M}^{- \ldots -}]$, however looking at its charges we claim that this is also involved in a relation with $\Tr(\mu^p)$ so that the independent combination is mapped to $\Tr(\Phi^p)$.

\subsubsection{The mirror theories $\check{\CC}_p[SU(N)]$}
To prove the confinement we need the mirror dual theories $\check{\CC}_p[SU(pr+l)]$ which are given for $l\neq 0$ and $l=0$ by
\begin{align}\label{fig:mirCp_ln0}
\includegraphics[scale=0.8]{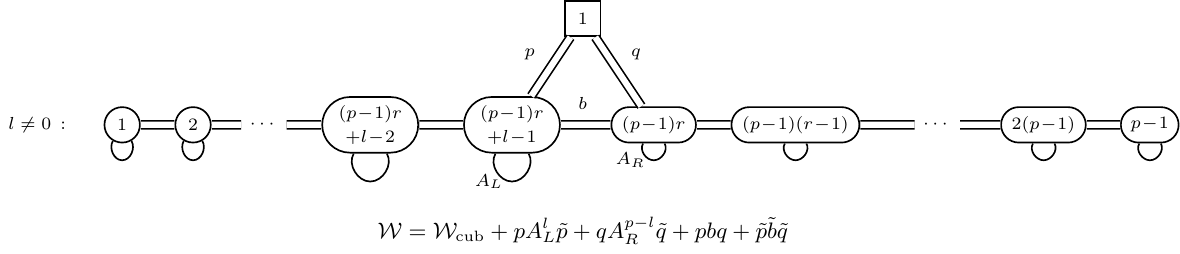}
\end{align}
\begin{align}\label{fig:mirCp_l0}
\includegraphics[scale=0.9]{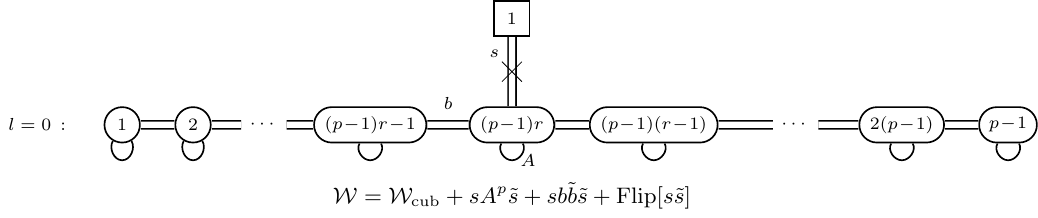}
\end{align}
where we are indicating with $\CW_{\text{cub}}$ all the cubic superpotential terms coupling the bifundamentals with the adjacent adjoint chirals. In the second theory, notice that we can equivalently use the bifundamental on the right of the flavor to construct the $s b \tilde{b} \tilde{s}$ operator due to F-term relations. 

Both for $l \neq 0$ and $l = 0$, the R-charges of the bifundamentals and adjoint chirals are fixed by the superpotential to be
\begin{align}
    & R[\text{bifundamental}] = 1 - \frac{1}{p+1} \,, \nn \\
    & R[\text{adjoint}] = \frac{2}{p+1} \,. 
\end{align}
The superpotential also constraints the R-charge of the flavors to be
\begin{align}
    & l \neq 0 \quad : \quad \begin{cases}
        R[p,\tilde{p}] = 1 - \frac{l}{p + 1} \\
        R[q,\tilde{q}] = 1 - \frac{p-l}{p + 1}
    \end{cases} \, , \nn \\
    & l = 0 \quad : \quad R[s,\tilde{s}] = 1 - \frac{p}{p+1} \,.
\end{align}

For any generic $l$, the map between chiral ring generators relates the $\mu$ operator of the $\CC_p[SU(N)]$ theory with a matrix where on the diagonal there are the traces of the adjoint fields, while the other entries are filled by extended mopoles (as is customary for mirror dualities). In the mirror theory we can also construct dressed mesonic operators. These are different depending on whether $l \neq 0$ or $l = 0$\footnote{For $l \neq 0$ the operators $qA^{l-1}\tilde{q}$ and $p A^{p-l-1} \tilde{p}$ are related to $b\tilde{b}$ due to the F-terms of the adjoint chiral fields. Therefore there is only one independent operator that we map to the trace of any adjoint chiral in the $\CC_p[SU(N)]$ theory, that are all identified in the chiral ring. Similarly, for $l=0$ the operator $sA^{p-1}\tilde{s}$ is set to zero in the chiral ring due to F-term relations.}
\begin{align}
    & l \neq 0 \quad : \quad \begin{cases}
    R[p A_L^j \tilde{p}] = 2 - 2 \frac{l-j}{p+1} \quad \text{for} \quad  j = 0 , \ldots , l-2 \\
    R[q A_R^j \tilde{q}] = 2 - 2 \frac{p-l-j}{p+1} \quad \text{for} \quad j = 0 , \ldots , p-l-2
    \end{cases} \,, \nn \\
    & l = 0 \quad : \quad R[s A^j \tilde{s}] = 2\frac{j+1}{p+1} \quad \text{for} \quad j = 1, \ldots, p-2 \,.
\end{align}
These operators are mapped to the set of monopoles in $\CC^p[SU(N)]$ as
\begin{align}\label{eq:Cpmir_opmap}
    pA_L^j\tilde{p} \quad &\leftrightarrow \quad \M_R^{\ldots -^{l-1-j} \ldots} \qquad j=0,\ldots,l-2 \, \,, \nn \\
    qA_R^j\tilde{q} \quad &\leftrightarrow \quad \M_L^{\ldots -^{p-l-1-j} \ldots} \qquad j=0,\ldots,p-l-2 \,.
\end{align}
For $l=0$ the map is given by
\begin{align}
    sA^j\tilde{s} \quad &\leftrightarrow \quad \M^{\ldots -^{p-1-j} \ldots} \qquad j=0,\ldots,p-2 \,.
\end{align}

If we consider the $\CC_p[U(N)]$ theories, that are defined as the $\CC_p[SU(N)]$ theory with the extra BF coupling defined in \eqref{eq:BFCpSUN}, the mirror $\check{\CC}_p[U(N)]$ coincides exactly with $\check{\CC}_p[SU(N)]$, without extra BF couplings. We shall show this momentarily.

\subsubsection*{Deriving the mirror theories}

The $\check{\CC}_[SU(N)]$ mirror theories are derived as a deformation of some known 3d $\mathcal{N}=4$ mirror pair. Indeed, as we have already mentioned, if turn off all the monopole superpotentials in $\CC_p[SU(N)]$, then this coincides with some $T^\sigma_\rho[SU(N)]$ theory.  We can then proceed by taking the mirrors of these $\mathcal{N}=4$ theories (which are simply obtained by swapping the partitions $\sigma\leftrightarrow \rho$) and studying the effect of the monopole superpotential across the duality.

\paragraph{\boldmath$l \neq 0$.} In this case the procedure to obtain the mirror theory is schematically depicted below.
\begin{align}\label{fig:mirr_proof_ln0}
    \includegraphics[scale=0.8]{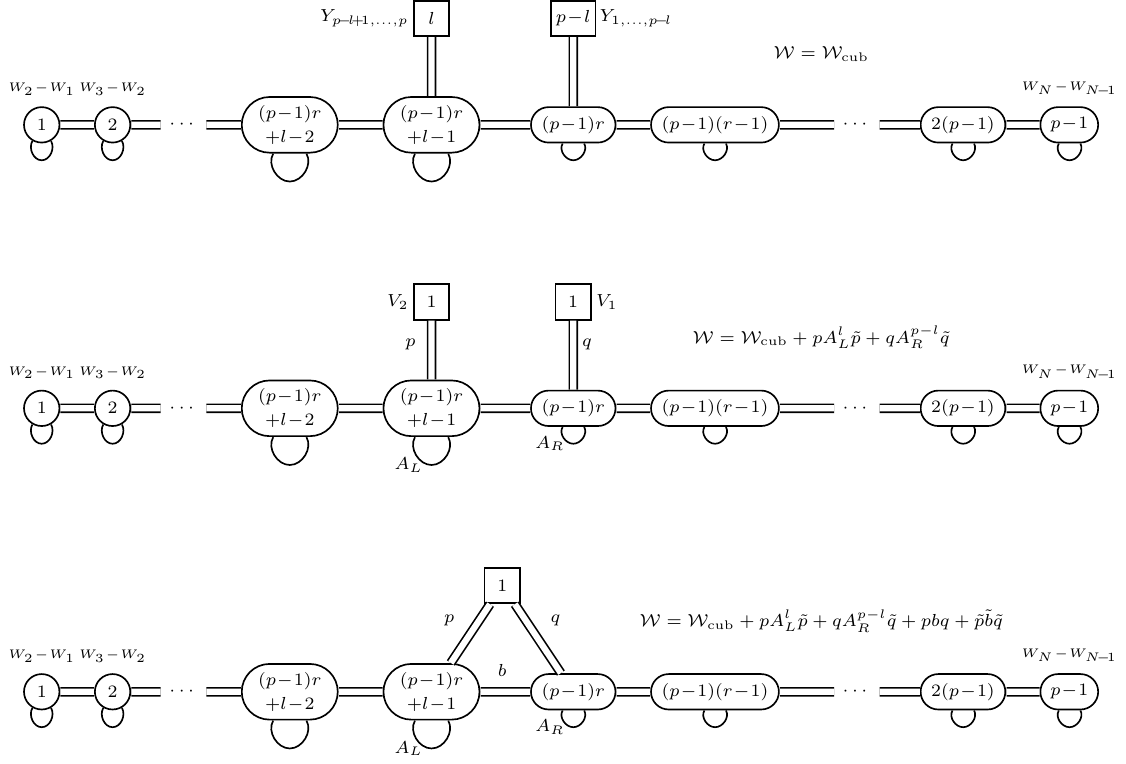}
\end{align}
We start from the theory $T^\r_\s[SU(N)]$ theory with $\sigma=[1^N]$ and $\rho=[(r+1)^l,r^{p-l}]$ in the first line of the figure above. In the figure we have also specified the parameterization of the global symmetries of the theory. The FI of the $j$-th gauge node is $W_{j+1}-W_j$ while the labels beside the flavor symmetries are the corresponding mass parameters.
We pick the following assignation of charges:
\begin{align}\label{eq:mir_charges}
    & R[\text{hypers}] = \frac{1}{2} \quad, \quad A[\text{hypers}] = -1 \quad, \quad \eta[\text{hypers}] = -\frac{1}{2} \,, \nn \\
    & R[\text{adjoint}] = 1 \quad , \quad A[\text{adjoint}] = 2 \quad, \quad \eta[\text{adjoint}] = 1 \,.
\end{align}
We have also introduced a redundant symmetry $U(1)_\eta$ which can be reabsorbed by $U(1)_A$ transformations. This symmetry will be useful to observe the discrete $\mathbb{Z}_{p+1}$ symmetry in the resulting theory after the $\mathcal{N}=2$ deformation. Notice that the charge assignment is such that the moment map on the Coulomb branch has charge $+1$ under $U(1)_\eta$.

In the mirror we  have two sets of respectively $l$ and $p-l$ flavors rotated by a flavor $S[U(l)\times U(p-l)]$ symmetry. We now want to turn on a superpotential deformation which is the mirror of the monopole superpotential in \eqref{fig:Cp_quiver_ln0}. It is convenient to first study the effect of turning on the monopoles with magnetic fluxes $+1$ under all the balanced nodes, that is all nodes except the $(p-l)$-th one.
In the mirror, this corresponds to the next-to-maximal nilpotent mass for both the mesons in the adjoint of $U(l)$ and $U(p-l)$, with the effect of breaking them down to two $U(1)$ symmetries. The result of the deformation is depicted in the second line of \eqref{fig:mirr_proof_ln0}. 

The breaking of symmetries due to this deformation is encoded in the following specialization of the parameters:
\begin{align}\label{eq:specFIlneq0}
	& Y_j = \frac{p-l+1-2j}{2}\left(i\frac{Q}{2}+2m_A+\eta \right) + V_1 \qquad \text{for} \quad j=1,\ldots,p-l \,, \nn \\
	& Y_{p-l+j} = \frac{l+1-2j}{2}\left(i\frac{Q}{2}+2m_A+\eta \right) + V_2 \qquad \text{for} \quad j=1,\ldots,l\,,
\end{align}
where we recall that the set of $Y_j$ for $j=1,\ldots,p-l$ are the real mass parameters for the $U(l)$ symmetry and the $Y_j$ for $j=p-l+1, \ldots, p$ are those of the $U(p-l)$ symmetry. 
Moreover, $\eta$ is the real mass parameters for the $U(1)_\eta$ symmetry, where notice that this can be completely reabsorbed by a shift of $m_A$, the parameter for the axial $U(1)_A$ symmetry. 
Finally, $V_{1,2}$ are the parameters for the two $U(1)$'s left after the nilpotent mass, which breaks $U(l) \to U(1)_{V_2}$ and $U(p-l) \to U(1)_{V_1}$. However, we can always reabsorb an overall flavor $U(1)$ symmetry by a compensating gauge transformation, which in this case allows us to get rid of the diagonal combination $V_1+V_2$, so that only the off-diagonal $V_1-V_2$ is physical.

In the theory in the second line of \eqref{fig:mirr_proof_ln0} all the bifundamentals and adjoints have the same R-charges as stated in \eqref{eq:mir_charges}, while the two flavors surviving the nilpotent mass deformation have 
\begin{align}\label{eq:mir_charges2}
    & R[p,\tilde{p}] = 1 - \frac{l}{2} \quad , \quad A[p,\tilde{p}] = -l \quad, \quad \eta[p,\tilde{p}] = -\frac{l}{2} \,, \nn \\
    & R[q,\tilde{q}] = 1-\frac{p-l}{2} \quad , \quad A[q,\tilde{q}] = -(p-l) \quad, \quad \eta[q,\tilde{q}] = -\frac{p-l}{2} \,.
\end{align}
Turning on the monopole with magnetic flux $+1$ for the $(p-l)$-th node
and the one with magnetic flux $-1$ under all the gauge node simultaneously in the electric theory corresponds to adding two triangle superpotentials in the mirror, which break the aforementioned $U(1)_{V_1-V_2}$ and one combination of the $\mathcal{N}=2$ R-symmetry and of the $U(1)_A$ axial symmetry. We can compute how these two symmetries are broken starting from the charges given in \eqref{eq:mir_charges} and \eqref{eq:mir_charges2}. The breaking condition written in term of $S^3$ partition function variables is\footnote{Recall that the variable $\eta$ is redundant, as in can be reabsorbed into $m_A$. However, we do not do this to explicitly isolate the residual $\mathbb{Z}_{p+1}$ symmetry.}
\begin{align}\
    & V_1 = V_2 = 0 \,, \nn \\
    & (p+1)(2m_A + \eta) = i\frac{Q}{2} (1-p)  \,.
    \label{imco}
\end{align}
From the above condition we see that $m_A$ is fixed as
\begin{align}\label{eq:ma_spec}
    m_A = i \frac{Q}{2} \frac{1-p}{2(p+1)} \,,
\end{align}
with $\eta$ a $\mathbb{Z}_{p+1}$ variable satisfying
\begin{align}
    (p+1)\eta = 0 \,.
\end{align}
We thus end up with the $\mathcal{N}=2$ theory depicted in the last line of \eqref{fig:mirr_proof_ln0}, with enhanced $SU(N)$ topological symmetry and a $\mathbb{Z}_{p+1}$ finite symmetry, that is the mirror $\check{\CC}_p[SU(N)]$ theory.

\paragraph{\boldmath$l = 0$.} In this case the procedure to obtain the mirror theory is schematically depicted below.
\begin{align}\label{fig:mir_proof_l0}
    \includegraphics[scale=0.8]{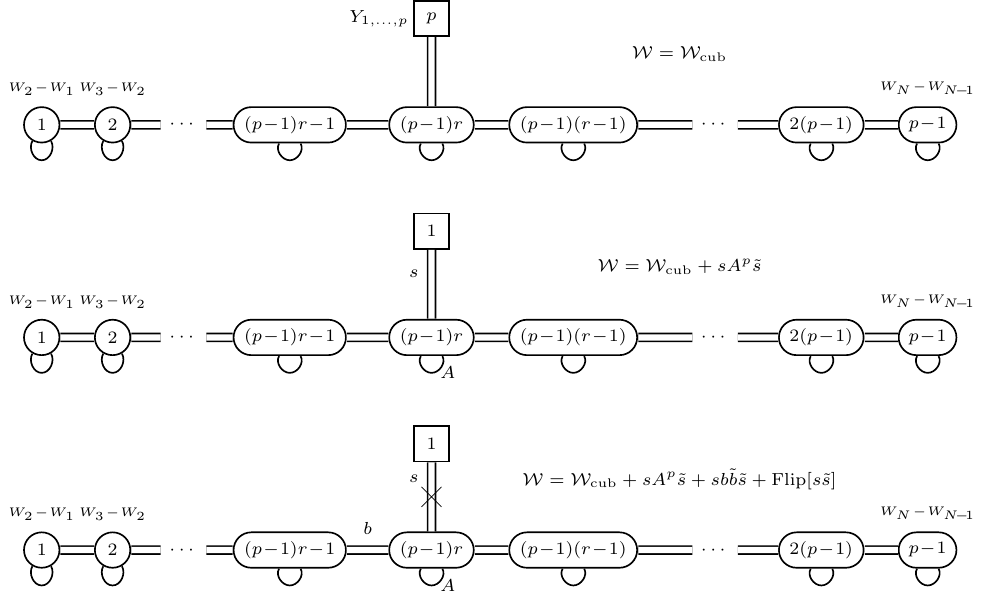}
\end{align}
Here we start from the $T^\r_\s[SU(N)]$ theory with $\sigma=[1^N]$ and $\rho=[r^p]$ depicted in the first line of \eqref{fig:mir_proof_l0}. From the partition $\rho$ we see an $SU(p)$ symmetry which in the electric theory acts on the Coulomb branch operators, which is enhanced from the topological symmetries of the nodes which are all balanced. In the mirror we therefore have $p$ flavors rotated by a flavor $SU(p)$ symmetry. Turning on the positive monopoles of all the nodes in the electric theory maps to a next-to-maximal nilpotent mass for the meson in the adjoint of such $SU(p)$, therefore breaking it completely. The resulting theory is the one in the second line of \eqref{fig:mir_proof_l0}, where the naive $U(1)$ flavor symmetry can be as usual reabsorbed with a gauge transformation.
Now the dressed monopole with magnetic flux $-1$ under all nodes maps to the quartic coupling between the flavor and the bifundamental $\d \CW = q \tilde{q} b \tilde{b}$, which breaks a combination of the $\mathcal{N}=2$ R-symmetry and of the axial symmetry. Finally, the superpotential $\text{Flip}[\M^{- \ldots -}]$ is mapped to $\text{Flip}[q\tilde{q}]$ in the mirror theory. We thus reach the last line of \eqref{fig:mir_proof_l0}. 

As a final comment, one can observe that the actual global symmetry of the resulting theory has a $\mathbb{Z}_{p+1}$ factor whose presence can be observed using the same strategy implemented in the previous paragraph for the $l \neq 0$ case.

\paragraph{Comments on the BF couplings.}
We now comment on the origin of the BF couplings \eqref{eq:BFCpSUN} in the $\mathcal{C}_p[U(N)]$ theories. In particular, we show that such a choice of BF couplings implies that there are none in their mirror duals $\check{\CC}_p[U(N)]$. As we will see in the next section, this is the only step in the derivation of the confining duality \eqref{fig:conf_genk} of $\mathcal{C}_p[U(N)]$ in which the BF couplings can change, so that there will be none on the WZ side as well.

Let us start from the case $l\neq0$. We follow the steps of \eqref{fig:mirr_proof_ln0} and trace how the BF couplings change after these steps. We start from the $\CN=4$ mirror pair
between the theory in figure \eqref{trho} and the top quiver in figure \eqref{fig:mirr_proof_ln0}.
More precisely for this mirror duality to hold, one should include the following BF couplings on the two sides:
\begin{align}
	&\text{electric:} \qquad \mathrm{e}^{2\pi i Y_p \sum_{j=1}^{pr+l} W_j} \,, \nn \\
	&\text{magnetic:} \qquad \mathrm{e}^{2\pi i \left( (\sum_{a=1}^l Y_{p-l+a}) (\sum_{j=1}^{(p-1)r+l-1} W_j) + (\sum_{a=1}^{p-l} Y_a ) (\sum_{j=1}^{(p-1)r+l} W_j)\right)} \,.
\end{align}
These can be derived using the mirror dualization algorithm \cite{Hwang:2021ulb,Comi:2022aqo}. The general rule is that in the electric theory the flavor mass parameters couple to the FI's for the nodes on their right, while in the mirror we have the opposite. Applying the specializations in \eqref{eq:specFIlneq0} and \eqref{eq:ma_spec}, the BF couplings become
\begin{align}
	&\text{electric:} \qquad \mathrm{e}^{2 \pi i \frac{iQ}{2}\frac{1-l}{p+1} \sum_{j=1}^{pr+l} W_j} \,, \nn \\
	&\text{magnetic:} \qquad \mathrm{e}^0 \,.
\end{align}

For the $l=0$ case the $\CN=4$ mirror pair involves the BF couplings
\begin{align}
	&\text{electric} \qquad \mathrm{e}^{2\pi i Y_p \sum_{j=1}^{pr} W_j} \,, \nn \\
	&\text{magnetic} \qquad \mathrm{e}^{2\pi i \left( (\sum_{a=1}^l Y_a) (\sum_{j=1}^{(p-1)r} W_j)\right)} \,.
\end{align}
Implementing the specializations we obtain
\begin{align}
	&\text{electric} \qquad \mathrm{e}^{2 \pi i \frac{iQ}{2}\frac{1-p}{p+1} \sum_{j=1}^{pr} W_j} \,, \nn \\
	&\text{magnetic} \qquad \mathrm{e}^{0} \,.
\end{align}
From this analysis we thus see that including the BF couplings \eqref{eq:BFCpSUN} in the definition of the $\mathcal{C}_p[U(N)]$ theories results in having no BF coupling in their mirrors $\check{\mathcal{C}}_p[U(N)]$.


\subsection{Derivation of the 3d confining duality for $\CC_p[SU(N)]$}\label{deriv3dconf}

In this section we provide a derivation of the confining duality \eqref{fig:conf_genk} for the $\CC_p[SU(N)]$ theory. 

\subsubsection{Inductive derivation}

The derivation is inductive and is based on the following two preliminary results, which we will in turn derive in the following sections.

\paragraph{Inductive step.} If $N = pr+l > p+1$, the $\CC_p[SU(N)]$ theory enjoys a duality in terms of a model composed of a $\CC_p[SU(N-1)]$ theory plus singlets
\begin{align}\label{lemma1}
\includegraphics[scale=1.1]{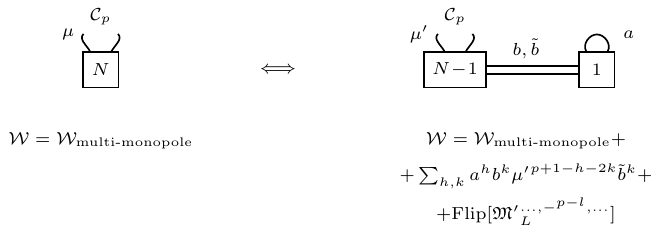}
\end{align}
where the $SU(N)$ moment map and the monopole operators on the r.h.s.~are primed in order to distinguish them from those on the l.h.s.~On the r.h.s.~of \eqref{lemma1} the superpotential contains all the possible couplings between $a$, $b$, $\tilde{b}$, $\mu'$ of degree $p+1$, such that $b$ and $\tilde{b}$ appear with the same powers. Moreover, we have a flipping term for the ${\M'}_L^{\ldots -^{p-l} \ldots}$ operator. Notice that for $l=0,1$ such operator does not exists in the $\CC_p[SU(N-1)]$ theory,\footnote{More precisely, for $l=0$ it does not exist since the length of the quiver is only $p-1$, while for $l=1$ it exists but it is not part of the chiral ring since it is already turned on in the superpotential.} therefore we do not have such flipping term and thus there is no extra singlet on the r.h.s.~of \eqref{lemma1}.
We also point out that the action of the duality on the gauge ranks $N=pr+l$ and $N'=N-1=pr'+l'$ implies that for $l \neq 0$ we have $r' = r$ and $l' = l-1$, while for $l = 0$ we have $r' = r - 1$ and $l' = p-1$. 

For $l\neq 0$ the map of the operators in the duality \eqref{lemma1} is
\begin{align}\label{eq:lemma1_opmap}
    \mu &\quad \longleftrightarrow \quad  
\renewcommand{\arraystretch}{1.2}
    \left( \begin{array}{c|c} 
        \mu' & b \\ 
        \hline 
        \tilde{b} & a
    \end{array} \right)  \,, \nn \\
    \M_L^{\ldots -^j \ldots} &\quad \longleftrightarrow \quad {\M'}_L^{\ldots -^j \ldots} \,, \nn \\
    \M_R^{\ldots -^j \ldots} &\quad \longleftrightarrow \quad \{ {\M'}_R^{\ldots -^j \ldots} , \CF[{\M'}_L^{\ldots -^{p-l} \ldots}] \} \,.
\end{align}
Let us comment on the last two lines of the map. The set of $\M_L^{\ldots -^j \ldots}$ monopoles in the second line always contains one less element with respect to the dual ${\M'}_L^{\ldots -^j \ldots}$ set, however the extra operator ${\M'}_L^{\ldots -^{p-l} \ldots}$ is flipped and so $j=1,\ldots,p-l-1$. Conversely the set $\M_R^{\ldots -^j \ldots}$ in the third line is always bigger by one unit than the dual set ${\M'}_R^{\ldots -^j \ldots}$, which is then completed with the flipper the flipper $\CF[{\M'}_L^{\ldots -^j \ldots}]$ and so $j=1,\ldots,l-1$ where $j=l-1$ corresponds to the flipper. Notice in particular that this third line is absent for $l=1$. Moreover, we stress that some of the monopole operators involved in this map are composite, as we have already commented below \eqref{eq:confmapmono}.

For $l= 0$ the situation is simpler, since on the electric side we have a single set of monopole operators $\M^{\ldots -^j \ldots}$. The operator map is then simply
\begin{align}\label{eq:lemma1_opmap}
    \mu &\quad \longleftrightarrow \quad  
\renewcommand{\arraystretch}{1.2}
    \left( \begin{array}{c|c} 
        \mu' & b \\ 
        \hline 
        \tilde{b} & a
    \end{array} \right)  \,, \nn \\
    \M^{\ldots -^j \ldots} &\quad \longleftrightarrow \quad  {\M'}_R^{\ldots -^j \ldots} \,,
\end{align}
where $j=1,\ldots,p-2$ since on the l.h.s.~the monopole $\M^{\ldots -^{p-1} \ldots}=\mathfrak{M}^{- \ldots -}$ is flipped and, as we have already commented below  \eqref{eq:confmapmono}, its flipper is not independent from $\Tr(\mu^p)$ and so it does not need to be mapped. Notice also that on the r.h.s.~the set $\M_L^{\ldots -^j \ldots}$ is empty and therefore does not need to be mapped.

\paragraph{Inductive basis.} The basis of the inductive derivation consists in the confining duality \eqref{fig:conf_genk} for the $\CC_p[SU(p+1)]$ theory, that is $N=p+1$ or equivalently $r=1$, $l=1$
\begin{align}\label{lemma2}
\includegraphics[scale=1.1]{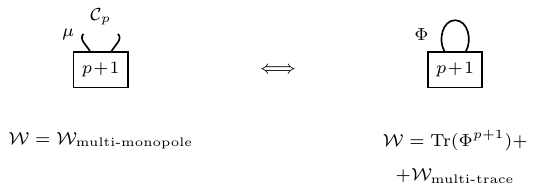}
\end{align}

We will provide a derivation of the two dualities \eqref{lemma1} and \eqref{lemma2} in Sections \ref{sec:lemma1} and \ref{sec:lemma2}, respectively, by iteration of the Aharony duality only, while now we will use them to derive inductively the general confining duality \eqref{fig:conf_genk}.

\paragraph{Inductive derivation.} Let us start from a $\CC_p[SU(N)]$ theory with $N=pr+l>p+1$. Applying \eqref{lemma1} we reach $\CC_p[SU(N-1)]$ coupled to $SU(N-1)$ fundamentals and flavor singlets.
Applying \eqref{lemma1} again to $\CC_p[SU(N-1)]$ and decomposing the other fields appropriately lowers the rank of the flavor symmetry by another unit and produces fields which can be recombined with those produced at the previous step into $SU(N-2)\times SU(2)$ bifundamentals and an $SU(2)$ adjoint
\begin{align}\label{iter2}
\includegraphics[scale=1.1]{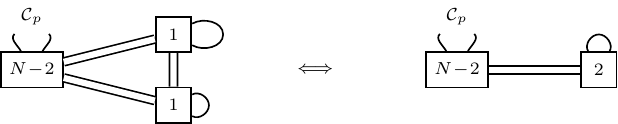}
\end{align}
From now on, to avoid cluttering, we do not give the exact superpotential at each step. One can reconstruct it from that given in \eqref{lemma1}, recalling that all the fields, including the adjoint operator of the $\CC_p$ theories, have R-charge $\tfrac{2}{p+1}$ by writing the most general superpotential involving all the operators with R-charge 2 that are singlets under the $SU(N)$ global symmetry.
After $k$ iterations of \eqref{lemma1} we reach
\begin{align}\label{iterk}
\includegraphics[scale=1.1]{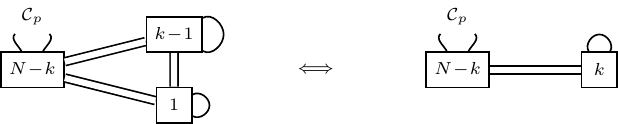}
\end{align}
For  $k=N-p-1$ we reach $\CC_p[SU(p+1)]$ coupled to an $SU(p+1)\times SU(N-p-1)$ bifundamental and an $SU(N-p-1)$ adjoint. The confined  frame is then obtained using the inductive basis \eqref{lemma2}: we are left with $SU(p+1)\times SU(p(r-1)+l-1)$ bifundamentals $b,\tilde b$ and  two adjoints $a, a'$ which recombine in the $SU(N)$ adjoint $\Phi$
 \begin{align}\label{finalframe}
\includegraphics[scale=1]{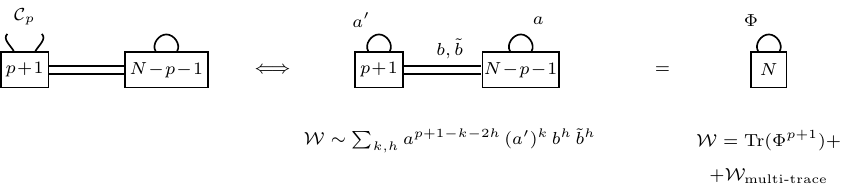}
\end{align}
All the superpotential terms of degree $p+1$ reconstruct the superpotential $\Tr(\Phi^{p+1})$ plus all exactly marginal multi-trace terms.
The adjoint $\Phi$ in the confined frame is a $N \times N$ matrix and it can be mapped across each step of the inductive derivation by collecting together the adjoint operator $\mu$ of the $\CC_p[SU(N-k)]$ theory, the $SU(N-k)\times U(k)$ bifundamentals $b,\tilde{b}$ and the $U(k)$ adjoint $a$ as:
\begin{align}\label{eq:phi_decomposition}
\renewcommand{\arraystretch}{1.3}
    \Phi_{N\times N}= \left( 
\begin{array}{c|c} 
 \mu_{(N-k) \times (N-k)} & b_{k \times (N-k)} \\ 
  \hline 
  \tilde{b}_{(N-k) \times k} & a_{k\times k}
\end{array} 
\right)\,.
\end{align}

\paragraph{Comments on the flipping fields.}
The derivation we have just presented was schematic and missed some important details, in particular the flipping fields $F[\Tr(\Phi^j)]$ that are produced at each step. These can be efficiently tracked (including their charges under the global symmetries) by performing the derivation at the level of the $S^3$ partition function, but here we will only make some comments on their pattern.

Each inductive step produces, in addition to the bifundamentals $b, \tilde b$ and the adjoint $a$ fields discussed above, an extra singlet $F[{\M'}_L^{\ldots -^{p-l} \ldots}]$ that has R-charge $2-\frac{2l}{p+1}$ for $l\neq 0,1$. After a generic number of iterations some of the singlets have R-charges that sum up to 2. Hence, there are quadratic deformations that are exactly marginal, so if we are on a generic point of the conformal manifold these will be activated with the effect of making the singlets massive. We claim that starting from $N=pr+l$ and reaching $N=p+1$ (the base of the induction) we produce exactly the flipping terms in the superpotential as in \eqref{fig:conf_genk}.

More precisely, we can see the following general pattern. The singlet produced for a given $l>1$ becomes massive together with the singlet produced for $l'=p+1-l$ which has R-charge $2-\frac{2(p+1-l)}{p+1} = \frac{2l}{p+1}$. We can then see that if we do $p$ iterations we produce the set of singlets
\begin{align}
	\{ 2 - \frac{2j}{p+1} \, | \, j=2,\ldots,p-1 \} = \{ 2 - \frac{4}{p+1} , 2 - \frac{6}{p+1} , \ldots , \frac{6}{p+1} , \frac{4}{p+1} \} \,.
\end{align}
The first singlet of the set acquires a mass together with the last one, same for the second singlet and the second-to-last, and so on. Therefore we conclude that $p$ iterations do not produce any singlet, after taking into account mass relations. Suppose now that we are trying to prove the confinement for $N=p r + l$. To reach the base of the induction $\CC_p[SU(p+1)]$ we first perform $p(r-1)$ iterations to reach $\CC_p[SU(p+l)]$ which do not produce any singlet as explained above, and then perform $l-1$ iterations to get to $\CC_p[SU(p+1)]$. The produced singlets will be of the form
\begin{align}
	\{ 2 - \frac{2j}{p+1} \, | \, j=2,\ldots,l \} \,.
\end{align}
If $l < \lfloor p/2 \rfloor$ one can notice that there are no massive singlets in the set. However if $l \geq \lfloor p/2 \rfloor $, then the $j = \lfloor p/2 \rfloor + k$ singlet will become massive together with $j' = \lfloor p/2 \rfloor - k$. Therefore for generic $l$ we are left with the following set of singlets:
\begin{align}
	\{ 2 - \frac{2j}{p+1} \, | \, j = 2, \ldots, \, \text{min}(l,p-l) \, \} \,,
\end{align}
which are exactly those presented in \eqref{fig:conf_genk}, both in number and in their charges which are compatible with them flipping the $j$-th power of the adjoint chiral  in the confined frame.

\subsubsection{Derivation of the inductive step}\label{sec:lemma1}

To prove the inductive step \eqref{lemma1} we start from the mirror of the $\CC^p[SU(N)]$ theory, as depicted in \eqref{fig:mirCp_ln0}. The leftmost node is $U(1)$ whose adjoint chiral is simply a singlet, hence we can dualize it with the Aharony duality. The main effects of the dualization are listed below, for more details see Appendix \ref{app:aharony}.
\begin{itemize}
    \item The dual gauge rank is $F-N_c$, where $F$ is the total number of flavors seen by the dualized node whose rank is $N_c$. Notice that when we dualize a balanced node $F=2N_c$ the rank does not change.
    \item $F^2$ singlets are produced, which flip the adjoint mesonic operator constructed from the flavors and maps to the meson $Q\tilde{Q}$ of the theory before the dualization. If the $SU(F)$\footnote{Notice that actually in Aharony duality the global symmetry of the $U(N)$ SQCD with $\CW=0$ is $SU(F)^2$ rotating independently the fundamentals and anti-fundamentals (see Appendix \ref{app:aharony}). However, in all the calculations that we will perform in the following sections the $SU(F)^2$ symmetries are always broken down to the vector-like diagonal $SU(F)$ subgroup due to gaugings or superpotentials.} symmetry is broken, for example down to $SU(F_1) \times SU(F_2)$ with $F_1+F_2=F$, one has to decompose the adjoint operator according to the relevant branching rule, for example the $SU(F)$ adjoint decomposes in $SU(F_1)$ and $SU(F_2)$ adjoints and two bifundamentals of $SU(F_1) \times SU(F_2)$.
    \item If the flavors are involved in a cubic superpotential, such as the $\mathcal{N}=4$ superpotential $\mathcal{W}_{\text{cub}}=\Phi Q\tilde{Q}$, after the dualization we get a quadratic superpotential for the singlet in the adjoint of $SU(F)$ and the third field $\Phi$ involved in the interaction. These fields are then massive and integrated out.
    \item Two extra singlets that do not transform under the non-abelian global symmetry are produced, these are flippers for the two monopoles of the dualized node.
\end{itemize}

After dualizing the first $U(1)$ node, this stays the same as it has two flavors. However, from the second and third bullet point we observe that the adjoint chiral of the $U(2)$ gauge node that is next to the one we dualized gets flipped. Therefore we can now apply the Aharony duality again to this second node and keep iterating this procedure. At each dualization, the adjoint chiral of the following node is removed, while the one at the previous node is restored. After we have dualized all the nodes in the balanced tail, starting from $U(1)$ up to $U((p-1)r+l-2)$, we reach the following theory:
\begin{align}\label{fig:inductive1}
\includegraphics[scale=0.85]{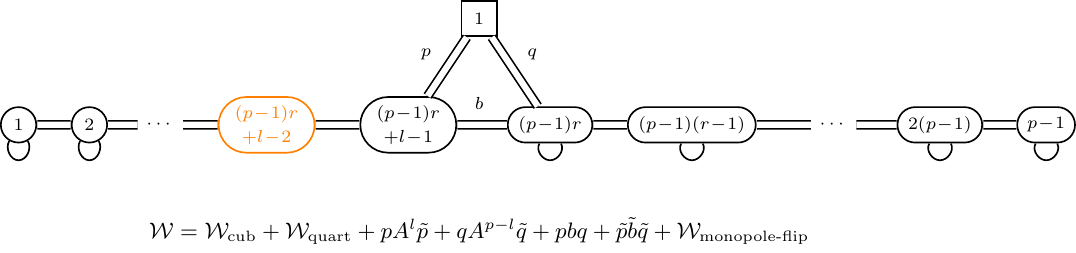}
\end{align}
where the last dualized node is highlighted in orange. 
As usual, $\CW_{\text{cub}}$ encodes the cubic couplings between adjoint chirals and bifundamentals on its sides. In addition we have $\CW_{\text{quart}}$ which contains quartic superpotentials between the bifundamentals on the sides of a gauge node without an adjoint chiral. These superpotential terms are generated by Aharony duality when the adjoint chiral is removed.
The mapping of the monopoles in the quiver after each dualization can be worked out using the results of \cite{Benvenuti:2020wpc} (see Appendix \ref{app:monopoles} for a review).
In particular  $\mathcal{W}_{\text{monopole-flip}}$ contains the contribution of all the flippers of the monopoles generated by the iterations of the Aharony duality.

We now dualize the following node with rank $(p-1)r+l-1$.
This is not balanced and its rank changes as
\begin{align}
    [(p-1)r+l-2+(p-1)r] - [(p-1)r+l-1]  = (p-1)r - 1 \,.
\end{align} 
The result of this dualization is
\begin{align}\label{fig:inductive2}
\includegraphics[scale=0.85]{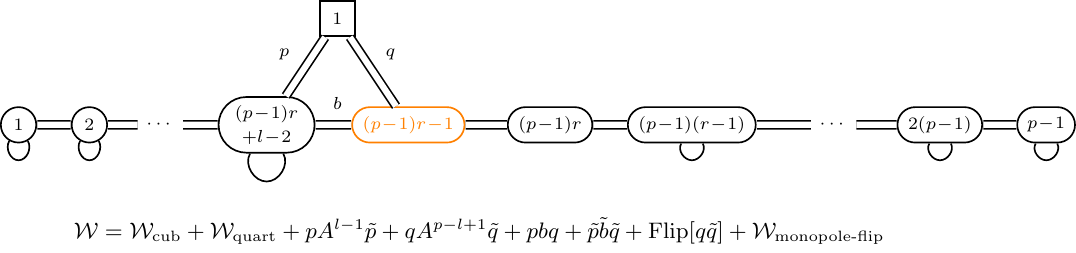}
\end{align}
The flippers produced by the Aharony duality have the effect, as usual, of flipping the adjoint of the following node and restoring the adjoint on the previous one. In addition the $q,\tilde{q}$ fields in \eqref{fig:inductive1} are flipped and the $p,\tilde{p}$ fields in \eqref{fig:inductive2} are generated. 

We then dualize the following node with rank $(p-1)r$. The rank of this node changes as
\begin{align}
    [(p-1)r-1+(p-1)(r-1)] - [(p-1)r] = (p-1)(r-1) \,,
\end{align}
meaning that its rank decreased by $p-1$ units.
We keep iterating the Aharony duality towards the right, each time decreasing the rank of the dualized node by $p-1$ units.
Therefore when we perform the Aharony duality on the last $U(p-1)$ gauge node it confines. We thus reach the final frame
\begin{align}\label{fig:inductive3}
\includegraphics[scale=0.9]{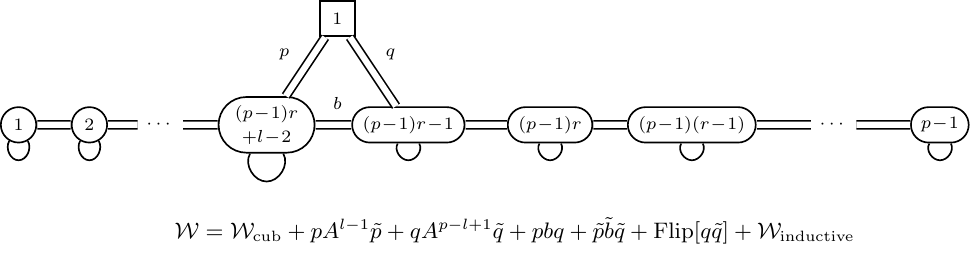}
\end{align}

All in all the effect of iterating the Aharony duality is to shorten the quiver by one node, resulting in the mirror $\check{\CC}_p[SU(N-1)]$ theory with the addition of an extra singlet $F[q\tilde{q}]$ which corresponds to the flipper $F[\M^{\ldots -^{p-l} \ldots}]$ on the r.h.s.~of figure \eqref{lemma1} obtained by mirroring back to $\CC_p[SU(N-1)]$, as can be read from the operator map in \eqref{eq:Cpmir_opmap}. 
In addition all the monopole flippers generated by the iterations of the 
Aharony duality  reconstruct the $SU(N-1) \times U(1)$ bifundamentals $b,\tilde{b}$. Finally, the adjoint of the first dualized node, which is a gauge singlet, is the $a$ field. 
Notice that when the last node is confined, all the possible superpotential couplings will be generated in $\mathcal{W}_{\text{inductive}}$, reconstucting the superpotential in  \eqref{lemma1}.

For $l=0,1$ there are subtleties related to the fact that for $l=0$ the $\check{\CC}^p[SU(N)]$ theory is different. However, a slight modification of the same strategy we just presented provides the expected result also for these cases.

\subsubsection{Derivation of the inductive basis}\label{sec:lemma2}

We are left to prove the base of the induction, which is the case $r=l=1$ of \eqref{fig:conf_genk}. 
The sketch of the derivation is given below in \eqref{fig:prooflemma2} and it involves the one- and two-monopole Aharony dualities introduced in \cite{Benini:2017dud}  (see also Appendix \ref{app:aharony}).
\begin{align}\label{fig:prooflemma2}
\includegraphics[scale=1.1]{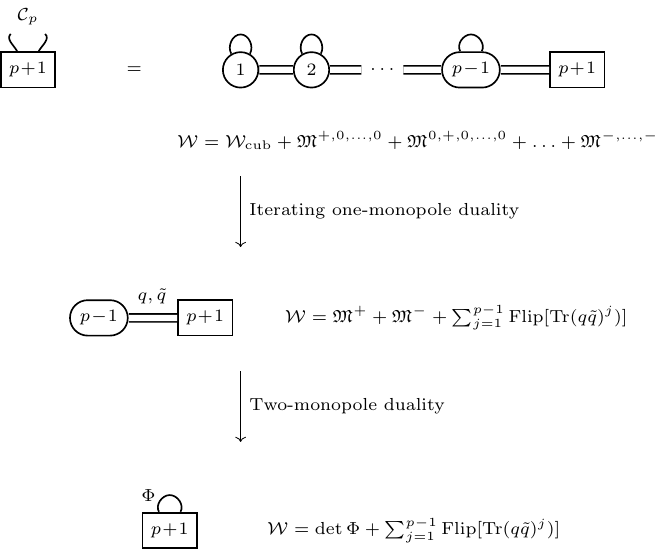}
\end{align}
We start from the $\CC_p[SU(p+1)]$ theory, whose quiver description is depicted in the first line of \eqref{fig:prooflemma2}.
Focusing on the first $p-2$ nodes we see that each of them has the fundamental positive monopole linearly turned on.
Starting from the leftmost $U(1)$ node we can use iteratively the one-monopole duality in the confining case relating
$U(N)$ with $N+1$ flavors and $\mathcal{W}=\M^+$ to a WZ model with $(N+1)^2$ singlets $M^i{}_{j}$ and a singlet $\gamma$ with $\mathcal{W}=\gamma \det M$, to sequentially confine the first $p-2$ nodes. This is because, similarly to the previous derivation of the inductive step, the $U(1)$ adjoint is a singlet and we can indeed use the one-monople duality which confines it in a WZ removing the adjoint of the next $U(2)$ node, so that the duality can be iterated.

The sequential confinement of the first $p-2$ nodes, due to the positive monopoles at each node, is explained in great details in Section $3.1$ of \cite{Aprile:2018oau} (see also \cite{Hwang:2020wpd}) and  yields a $U(p-1)$ theory with $p+1$ flavors
 and    $\mathcal{W}=\sum_{j=1}^{p-1}\gamma_j \Tr((q\tilde q)^j)+ \M^++\M^-$ .
The flippers of the powers of the meson
are produced by the $p-2$ applications of the one-monopole duality
and their effect is to break the non-abelian flavor symmetry from $SU(N+1)\times SU(N+1)$ to $SU(N+1)$,
while the monopole superpotential is the residual of the  $\mathcal{C}_p[SU(p +1)]$ theory original superpotential and breaks the abelian axial and topological symmetries.
Notice in particular that as explained in \cite{Benvenuti:2020wpc} (see Appendix \ref{app:monopoles}) the monopole with negative charge under all nodes is shortened to  $\M^-$.

The last step  is performed by applying the confining two-monople duality relating $U(N)$ with $N+2$ flavors and $\mathcal{W}=\M^++\M^-$  to a WZ model with $(N+2)^2$ singlets $M^i{}_{j}$ and $\mathcal{W}= \det M$. Hence, we obtain that the
$\mathcal{C}_p[SU(p +1)]$ theory  confines to an $SU(p+1)$ adjoint with  $\cW = \det \Phi + \sum_{j=1}^{p-1}\gamma_j \Tr(\Phi^j)$.

One can rewrite the superpotential in terms of the traces of $\Phi$ and use the F-term relations to simplify the superpotential expression.
Indeed, in \eqref{lemma2} we are considering the $\CC_p[SU(p+1)]$ theory at a generic point of the conformal manifold, where $\CW_{\text{multi-monopole}}$ is turned on. Correspondingly, on the r.h.s.~we turn on all the exactly marginal deformations. These terms include the product of two flipping fields $F[\Tr(\Phi^j)]F[\Tr(\Phi^{p-1-j})]$. Using the F-term equations all the flipping fields can be integrated out and in turn we obtain the $\CW_{\text{multi-trace}}$ superpotential, which contains all the possible exactly marginal deformations obtained from the product of two or more traces of the adjoint $\Phi$. Notice that still the flipping field $F[\Tr(\Phi)]$ is present, so that $\Phi$ is traceless even after solving the F-term equations. With all these considerations taken into account we obtain the duality in \eqref{lemma2} for $r=l=1$, which is the inductive basis \eqref{lemma2}.

\section{Derivation of the 3d $\CN=2$ $U(N)$ Kim--Park duality via deconfinement}\label{sec:KPproof}

In this section we use the confining duality of the 3d $\CN=2$ theories $\mathcal{C}_p[SU(N)]$ to provide a derivation the Kim--Park (KP) duality \cite{Kim_2013} for $U(N)$ gauge group. Since this was derived using only the Aharony duality and a mirror duality which can also be obtained from the Aharony duality via the mirror dualization algorithm \cite{Hwang:2021ulb,Comi:2022aqo}, it means that our derivation of the KP duality assumes only the validity of the Aharony duality for $U(N)$ gauge group.

The KP duality relates  3d $\CN=2$ unitary gauge theories with adjoint and fundamental matter as follows:\footnote{The theory has a continuous non-abelian $SU(F)\times SU(F)$ flavor symmetry where each factor acts on $Q$ and $\tilde{Q}$ independently. In order to simplify the notation in the quivers of this section we identify these two $SU(F)$ into a single one and represent it with a square box. However our discussion equally applies considering the full $SU(F)\times SU(F)$ symmetry. }
\begin{align}\label{KP}
\includegraphics[scale=1.1]{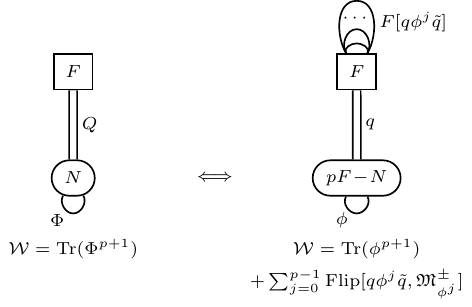}
\end{align}
Notice that on the r.h.s.~the flippers $F[q\phi^j\tilde{q}]$ are not singlet under the non-abelian global symmetries and, indeed, they flip all the components of the $q\phi^j\tilde{q}$ matrix. These singelts are also included in the picture as arcs. In the case $p=1$, the adjoint chirals are massive on both sides and we recover the Aharony duality.
The mapping of the chiral ring generators reads
\begin{equation}
\left\{
\begin{tabular}{c}
 $Q \Phi^j \tilde{Q}$  \\
 $\M^{\pm}_{\Phi^j}$  \\
 $\Tr(\Phi^j)$ 
\end{tabular}
\right\}
\qquad \longleftrightarrow \qquad
\left\{
\begin{tabular}{c}
 $F[q \phi^{p-j} \tilde{q}]$  \\
 $F[\M^{\mp}_{\phi^{p-j}}]$  \\
 $\Tr(\phi^j)$ 
\end{tabular}
\right\}
\qquad
\begin{tabular}{c}
$ j = 0, 1, \ldots, p-1 $ \,,\\
$ j = 0, 1, \ldots, p-1 $ \,,\\
$ j = 2, 3, \ldots, p-1 $ \,.
\end{tabular}
\end{equation}
where $\M^{\pm}_{\Phi^j}$ denotes the monopole with magnetic flux $\pm1$ dressed with the $j$-th power of the adjoint field of the residual $U(1)\times U(N-1)$ gauge group in the monopole backgroud, see e.g.~\cite{Cremonesi:2013lqa}.

Given our construction of the $\CC_p[SU(N)]$ theories, the proof of the duality \eqref{KP} assuming Aharony duality is rather straightforward. We first deconfine the adjoint of the electric theory into a $\CC_p[SU(N)]$, then we sequentially dualize all the $p$ gauge nodes, this builds a quiver which is a $U(pF-N)$ node attached to a $\CC_p[SU(pF-N)]$, so we can reconfine the $\CC_p[SU(pF-N)]$ theory into an adjoint, obtaining the magnetic side of the KP duality. We are going to discuss derivation in the simpler case of $p=2$, $l=1$ first and then in full generality.

\subsection{A simple example: $p=2$, $l=1$}
Let us first discuss the simplest case, namely cubic superpotential and odd rank of the gauge groups. Here only the S-confining duality for $U(r)$ with $2r+1$ flavors, found in \cite{Bajeot:2023gyl}, is needed. The 3d KP duality in this case reads
\begin{align}\label{fig:KPp=2_proof1}
    \includegraphics[]{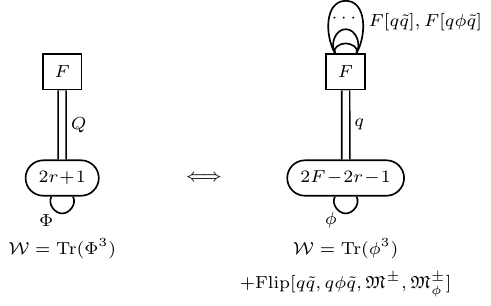}
\end{align}
Starting from the electric theory, we  can use the confining duality \eqref{fig:conf_genk} to replace the adjoint $\Phi$ plus the $\Tr(\Phi^3)$ interaction with a $\CC_2[SU(2r+1)]$, which in turn is given by a $U(r)$ theory with adjoint and $2r+1$ fundamentals. So the electric theory in \eqref{fig:KPp=2_proof1} is dual to the following quiver\footnote{Notice that the $p=2$ case has no conformal manifold. Therefore the deconfinement of the adjoint through the $\CC_2[SU(N)]$ theory does not involve subtleties about exactly marginal deformations or the presence of flipping fields. This will not be the case for $p > 2$, as we will discuss in the next subsection.}
\begin{align}\label{fig:KPp=2_proof2}
    \includegraphics[]{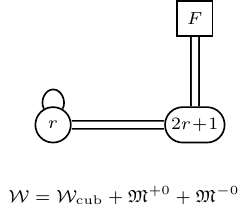}
\end{align}

Now we dualize the right node, killing the $U(r)$ adjoint and producing the first set of flippers
\begin{align}\label{fig:KPp=2_proof3}
    \includegraphics[]{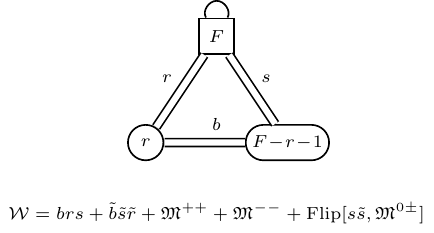}
\end{align}
Notice that the monopole terms in the superpotential of \eqref{fig:KPp=2_proof2} $\M^{+0} + \M^{-0}$ \emph{extended} to $\M^{++} + \M^{--}$, following the rules of \cite{Benvenuti:2020wpc} reviewed in Appendix \ref{app:monopoles}.

Dualizing the left node in \eqref{fig:KPp=2_proof3} produces the second set of flippers and \emph{shortens} the two monopoles in the superpotential
\begin{align}\label{fig:KPp=2_proof4}
    \includegraphics[]{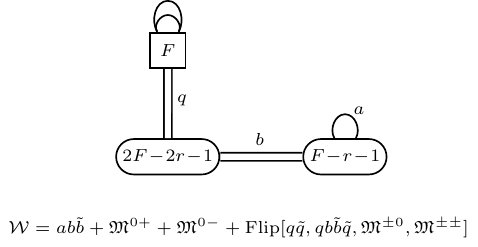}
\end{align}
We can at this point recognize the quiver of the theory $\CC_2[SU(2F-2r-1)]$ attached to a $U(2F-2r-1)$ node. Replacing the $\CC_2[SU(2F-2r-1)]$ theory with an adjoint $\phi$ with $\Tr(\phi^3)$ interaction by means of the confining duality \eqref{fig:conf_genk}, we arrive at
\begin{align}\label{fig:KPp=2_proof5}
    \includegraphics[]{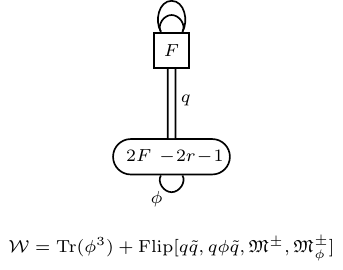}
\end{align}
which is the magnetic side of the duality \eqref{fig:KPp=2_proof1}.

In the last step, from \eqref{fig:KPp=2_proof4} to \eqref{fig:KPp=2_proof5}, we claim that the monopoles $\M^{\pm 0}$ and $\M^{\pm \pm}$ map to $\M^{\pm}$ and $\M^{\pm}_\phi$ respectively, leading to the superpotential in \eqref{fig:KPp=2_proof5}.
We have checked this claim by carefully keeping track of all the specialized FI parameters and BF couplings discussed in Section \ref{sec3dandDp}.

\subsection{General derivation}
In this subsection we go through the steps involved in the proof of \eqref{KP} for generic $p$ and $N$.

Let us start from the electric theory, the $U(N)$ theory with one adjoint $\Phi$, $F$ flavors $Q$, $\tilde{Q}$ and $\cW = \Tr(\Phi^{p+1})$. The first step is to use \eqref{fig:conf_genk} to deconfine the adjoint into a $\CC_p[SU(N)]$ theory with the last flavor node gauged as a $U(N)$ with $F$ flavors
\begin{align}\label{fig:KPproof1}
\includegraphics[scale=1.1]{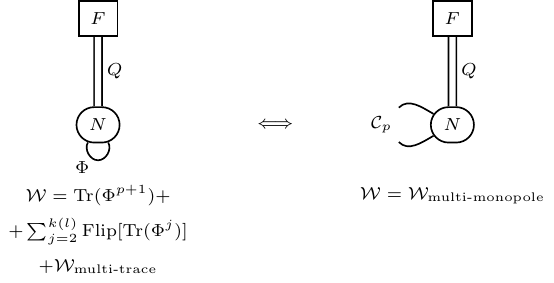}
\end{align}
In order to deconfine the adjoint chiral $\Phi$ with a $\CC_p[SU(N)]$ theory we have added all the exactly marginal deformations given by multi trace operators on the l.h.s.~and by multi monopole operators on the r.h.s., as discussed below the confining duality in \eqref{fig:conf_genk}. In the following steps we will not write the $\CW_{\text{multi-monopole}}$ terms anymore to avoid cluttering. In the last step, when we will reconfine the $\CC_p[SU(N)]$ theory we will reintroduce them. We also added the correct flipping terms required by the duality \eqref{fig:conf_genk}, recalling that given $N=pr+l$ we have
\begin{align}\label{eq:flipperkl}
    k(l) = \min(l,p-l) \,.
\end{align}
We then use the definition of the $\CC_p[SU(N=pr+l)]$ theory, so that the r.h.s.~of \eqref{fig:KPp=2_proof1} becomes\footnote{Here we discuss the case $l \neq 0$. The case $l=0$ is similar.}
\begin{align}\label{fig:KPproof2}
\includegraphics[scale=1]{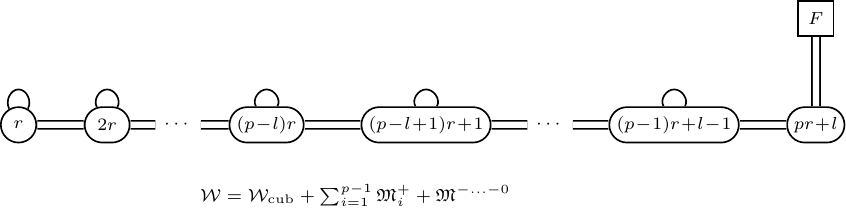}
\end{align}
We recall that $\M_i^+=\M^{0^{i-1}+0^{p-i}}$ is the monopole with charge $+1$ under the topological symmetry of the $i$-th node, counting the nodes from the left.

Starting from the rightmost gauge node, which does not have the adjoint and does not have the monopole superpotential, we dualize sequentially the quiver using the Aharony duality. After the first dualization we get
\begin{align}\label{fig:KPproof3}
\includegraphics[scale=1]{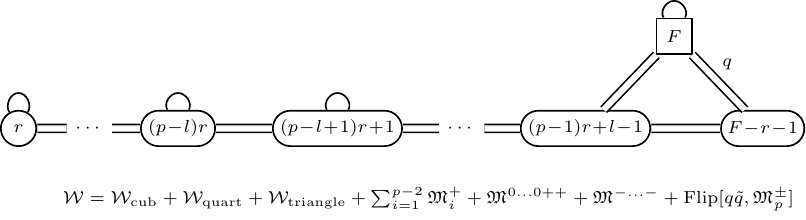}
\end{align}
where $\CW_{\text{triangle}}$ encodes the cubic couplings associated to the triangle. Also, the term $\CW_{\text{quart}}$ encodes the quartic superpotential between bifundamentals on the sides of the gauge nodes without an adjoint chiral. The dualization extended the rightmost $\M^{+}_i$ monopole and the $\M^{-\ldots-0}$ monopole, according to the rules of \cite{Benvenuti:2020wpc} (see Appendix \ref{app:monopoles}), and produced $F^2+2$ flippers. 

The adjoint for the next to last node became massive, so now we can use the Aharony duality on this node. Defining  $s=F-r-1$ we get
\begin{align}\label{fig:KPproof4}
\includegraphics[scale=1]{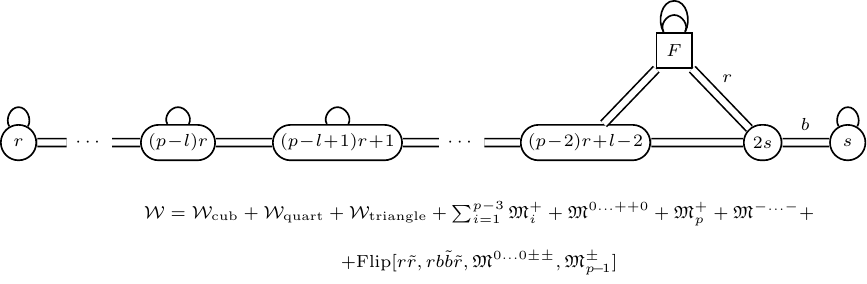}
\end{align}
We keep dualizing sequentially $l$ nodes in the quiver, from right to left. This builds a tail on the right with nodes $U(j s)$, $j=l, \ldots, 2, 1$. At each step the flavor node moves left, and we produce a new set of $F^2+2$ flippers. Let us also comment about the monopole superpotential. Each $\M^+_i$ term in the superpotential will be extended at one step and shortened at the next step. 
The $\M^{-\ldots-}$ stays the same until the last step,  when it is shortened to $\M^{0-\ldots-}$. The flipped monopoles are extended at each step after the step where they have been created. After $l$ iterations we reach the following quiver:
\begin{align}\label{fig:KPproof5}
\includegraphics[scale=1]{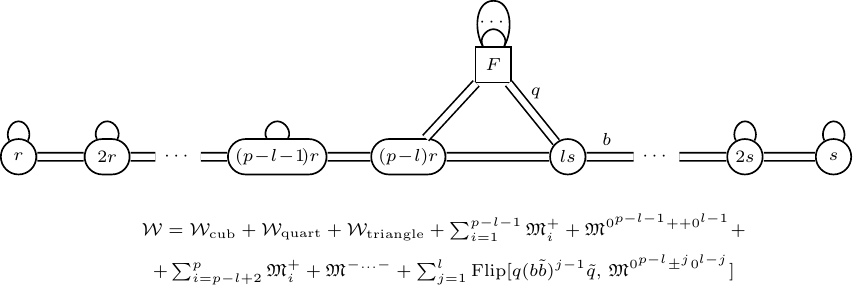}
\end{align}

Now we have to dualize the node which was overbalanced in the original definition of the $\CC^p[SU(N)]$ theory.
After $p$ more iterations of the Aharony duality, we reach the following quiver:
\begin{align}\label{fig:KPproof6}
\includegraphics[scale=0.95]{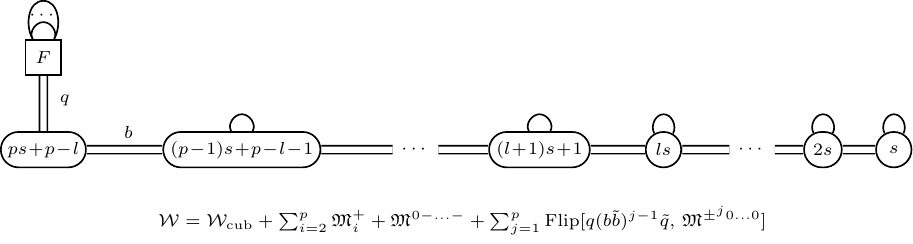}
\end{align}
The leftmost node has rank $ps+p-l=pF-pr-l = p F - N$. Moreover, we can recognize the quiver and superpotential of $\CC_p[SU(p F - N)]$ with its global symmetry gauged as $U(p F - N)$ with $F$ flavors, plus $p(F^2+2)$ flippers. When can then reconfine the $\CC_p[SU(p F - N)]$ to an $SU(p F - N)$ adjoint using \eqref{fig:conf_genk}
\begin{align}\label{fig:KPproof7}
    \includegraphics[scale=1]{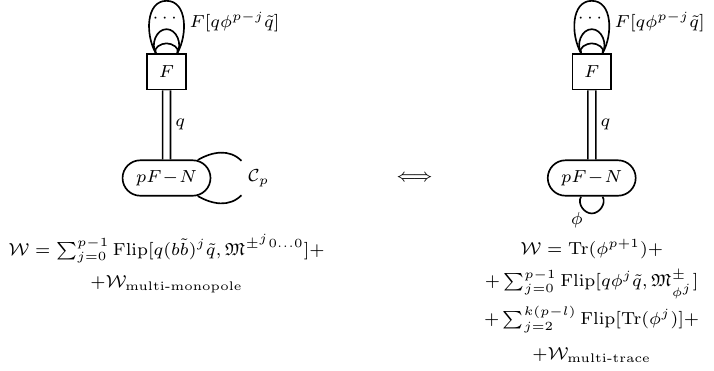}
\end{align}
In the last step we have also reintroduced explicitly the exactly marginal deformations that were present in the starting point \eqref{fig:KPproof1}.

Some comments regarding the result are in order. First of all let us remark that KP duality relates a $U(N)$ gauge group with a $U(pF-N)$. If we take $N=pr+l$ then $pF-N = pr' + l'$ with $r' = F-r-1$ and $l'=p-l$.
The formula \eqref{eq:flipperkl}, which determines the set of flippers, is invariant under $l \to p-l$, therefore the flippers that we have on the l.h.s.~in the first step \eqref{fig:KPproof1} are the same as those on the r.h.s.~of the last step \eqref{fig:KPproof7}.
Moreover, one can check that the list of all the possible exactly marginal deformations generated by multi-monopole operators is the same for a $\CC_p[SU(N=pr+l)]$ and $\CC_p[SU(N'=pr'+l')]$ if $l' = p-l$. We therefore claim that starting from the $\CC_p[SU(N)]$ theory on the r.h.s.~of \eqref{fig:KPproof1} at the most generic point of the conformal manifold, after performing the steps described above, we get to the $\CC_p[SU(pF-N)]$ theory on the l.h.s.~of \eqref{fig:KPproof7} which, again, has all the exactly marginal deformations turned on.

To conclude, let us stress one more time that our derivation of the KP duality assumes that we are at a generic point of the conformal manifold. It would be interesting to understand if our technique could be used to probe how the conformal manifolds of the two dual theories are related. However, we leave this problem to future investigation.

\section{Relation to the 4d $\mathcal{N}=2$ $D_p[SU(N)]$ SCFTs}
\label{sec:DpSUN}

In this section we discuss the relation between the family of 3d $\mathcal{N}=2$ $\mathcal{C}_p[SU(N)]$ theories and the dimensional reduction of a class of 4d $\mathcal{N}=2$ SCFTs, called $D_p[SU(N)]$ theories \cite{Cecotti:2012jx,Cecotti:2013lda,Wang:2015mra}. In particular, we will discuss how the confining duality \eqref{fig:conf_genk} enjoyed by the $\mathcal{C}^p[SU(N)]$ theories that we presented in Section \ref{sec3dandDp} can be understood as the three-dimensional version of the result of \cite{Maruyoshi:2023mnv}, that a 4d $\mathcal{N}=1$ chiral in the adjoint $\Phi$ of an $SU(N)$ symmetry with superpotential $\mathcal{W}\sim\Tr(\Phi^{p+1})$ can be deconfined into a $D_p[SU(N)]$ theory with a suitable CB deformation when $\gcd(p,N)=1$.


\subsection{Review of $D_p[SU(N)]$ theories and their dimensional reduction}

In this section we review some properties of the $D_p[SU(N)]$ theories and of the Lagrangians for their circle reduction to 3d. The $D_p[SU(N)]$ theories are 4d $\mathcal{N}=2$ SCFTs that admit various different realizations \cite{Cecotti:2012jx,Cecotti:2013lda,Wang:2015mra}. They can for example be obtained via geometric engineering of Type IIB string theory on certain Calabi--Yau 3-fold isolated hypersurface singularities, or by compactifying the 6d $(2,0)$ $A_N$-type SCFT on a sphere with one irregular and one regular puncture, along the lines of the class $\mathcal{S}$ construction \cite{Gaiotto:2009we}. Each realization has its utility for understanding various properties of these theories.

All of these theories possess at least an $SU(N)$ flavor symmetry, which in the construction via the 6d to 4d compactification is manifest in the regular puncture. However, generically there can also be additional abelian flavor symmetries, as well as sporadic symmetry enhancements. In general, the flavor symmetry is at least \cite{Giacomelli:2017ckh}
\begin{align}\label{eq:symmDpSUN}
G_F=SU(N)\times U(1)^{m-1}\,,\qquad m=\gcd(p,N)\,.
\end{align}
The flavor central charge of the $SU(N)$ factor, which is related to the anomaly with the $U(1)_r$ factor of the 4d $\mathcal{N}=2$ R-symmetry, is given by \cite{Xie:2016evu}
\begin{align}\label{eq:flavorcentralcharge}
k_{SU(N)}=-2\,\mathrm{Tr}\,r\,SU(N)^2=2N-\frac{2N}{p}\,.
\end{align}
One can also consider the SCFTs obtained via a HB RG flow triggered by a nilpotent vev, labelled by a partition $[Y]$ of $N$, for the $SU(N)$ moment map operator $\mu$. This can be understood as partially closing the regular puncture in the class $\mathcal{S}$ construction. Following the notation of \cite{Beem:2023ofp}, we denote these theories by $D_p(SU(N),[Y])$ and simplify this notation to $D_p(SU(N),[Y=1^N])=D_p[SU(N)]$ when the puncture is maximal.

The CB spectrum of the theory is given by all the operators $u_{i,j}$ with $0\leq i\leq p$ and $0\leq j\leq N-2$ whose scaling dimensions
\begin{align}
\Delta(u_{i,j})=N-j-\frac{N}{p}i
\end{align}
are strictly greater than two. When $m=\gcd(p,N)=1$ this is simplified to \cite{Maruyoshi:2023mnv}
\begin{align}
\Delta(u_i)=\frac{p+1+i}{p}\,,\qquad 0\leq i\leq p-2\,.
\end{align}
In particular, the operator $u_{\text{min}}$ with the minimal scaling dimension has
\begin{align}
\Delta(u_{\text{min}})=\frac{p+1}{p}\,.
\end{align}

Another important piece of data that characterize these theories are the conformal central charges. The $c$ central charge is given by \cite{Cecotti:2013lda,Giacomelli:2017ckh}
\begin{align}\label{eq:ccentralcharge}
c_{\mathcal{N}=2}=\frac{1}{2}\mathrm{Tr}\,r\,I_3^2-\frac{1}{24}\Tr\,r^3=\frac{p-1}{12}(N^2-1)\,.
\end{align}
where $I_3$ is the Cartan generator of the $SU(2)_R$ factor of the 4d $\mathcal{N}=2$ R-symmetry. Together with the knowledge of the CB spectrum, this allows us to determine also the $a$ central charge via the Shapere--Tachikawa relation \cite{Shapere:2008zf}
\begin{align}
2c_{\mathcal{N}=2}-a_{\mathcal{N}=2}=\frac{1}{4}\sum_{i=1}^r(2\Delta_i-1)\,,
\end{align}
where $r$ denotes the rank of the theory, i.e.~the dimension of the CB, and $\Delta_i$ is the scaling dimension of the $i$-th CB operator. For $m=\gcd(p,N)=1$ one finds more explicitly
\begin{align}\label{eq:acentralcharge}
a_{\mathcal{N}=2}=\frac{1}{2}\mathrm{Tr}\,r\,I_3^2-\frac{1}{48}\Tr\,r^3=\frac{(4p-1)(p-1)}{48p}(N^2-1)\,.
\end{align}

Finally, let us consider the dimensional reduction of the $D_p[SU(N)]$ theory on a circle, which yields a 3d $\mathcal{N}=4$ SCFT. Even though the 4d theory is generically non-Lagrangian, it turns out that the 3d SCFT obtained via the $S^1$ compactification admits a Lagrangian description in terms of a linear quiver gauge theory. In particular, in \cite{Closset:2020afy,Giacomelli:2020ryy} it has been shown that when $m=\gcd(p,N)=1$ and focusing on the case $p\leq N$ that is relevant to us, such quiver corresponds with the one in \eqref{fig:conf_genk} for the $\mathcal{C}_p[SU(N)]$ theory, but with the superpotential given only by the cubic interaction $\mathcal{W}_{\text{cub}}$ without the monopole superpotential, so to have $\mathcal{N}=4$ supersymmetry. As we have already pointed out previously, these coincide with the $T^\sigma_\rho[SU(N)]$ theories \cite{Gaiotto:2008ak} for
\begin{align}
\sigma=[1^N]\,,\qquad \rho=[(r+1)^{l},r^{p-l}]\,,
\end{align}
where we recall the notation $N=pr+l$ with $0\leq l\leq p-1$. The quiver for the theory $D_p(SU(N),[Y])$ with a partially closed regular puncture, again for $m=\gcd(p,N)=1$ and $p<N$, is simply obtained by taking $\sigma=[Y]$ in the definition of $T^\sigma_\rho[SU(N)]$.

When $m=\gcd(p,N)\neq 1$ the situation is instead more complicated for the following reason. In such case, the $D_p[SU(N)]$ theory generically has (except for some special cases) $m-1$ exactly marginal deformations, and one can find a point of the conformal manifold where the theory can be described as a partially weakly coupled gauge theory, consisting of a linear sequence of conformal gaugings via special unitary groups of theories of the same family $D_{p_i}(SU(N_i),[Y_i])$ with $m_i=\gcd(p_i,N_i)=1$ and thus no exactly marginal deformations. This is captured by the following expression (which actually holds for any $p$ and $N$) \cite{Closset:2020afy,Giacomelli:2020ryy,Beem:2023ofp}:
\begin{align}\label{eq:DpSUNgauging}
D_p[SU(N)]=D_{(m-1)q}(SU(N-n)) \longleftarrow SU(N-n) \longrightarrow D_q(SU(qN-n),[(q-1)^N,1^{N-n}])\,,
\end{align}
where we recall that $m=\gcd(p,N)$ and we defined
\begin{align}
q=\frac{p}{m}\,,\qquad n=\frac{N}{m}\,.
\end{align}
The above expression means that we are gauging a diagonal $SU(N-n)$ subgroup of the symmetries of the component theories. This gauging then leaves a $SU(N)\times U(1)$ symmetry from the right theory, which also has $\gcd(q,qN-n)=\gcd(q,n)=1$. The same is not true for the left theory, which instead has $\gcd((m-1)q,N-n)=\gcd((m-1)q,(m-1)n)=m-1$. However, iterating this relation for a total of $m-1$ times, we obtain that the $D_p[SU(N)]$ theory can be expressed as a linear sequence of $m-1$ gaugings of $D_{p_i}(SU(N_i),[Y_i])$ theories with $m_i=\gcd(p_i,N_i)=1$. Moreover, the flavor symmetry preserved by these gaugings is exactly the one predicted in \eqref{eq:symmDpSUN}. We have just reviewed that the 3d quiver for the component theories are linear unitary quivers. The observation of \cite{Closset:2020afy,Giacomelli:2020ryy} is that one can use \eqref{eq:DpSUNgauging} to express the 3d quivers for the $D_p[SU(N)]$ theories with $m=\gcd(p,N)\neq 1$ as a linear quiver with mixed unitary and special unitary gauge groups.

Let us consider for example the $D_4(SU(6))$ theory. In this case $m=2$ and we have one exactly marginal deformation. Indeed, following \eqref{eq:DpSUNgauging} we can express the theory as a gauging of two theories with no exactly marginal deformations
\begin{align}
D_4(SU(6))=D_2(SU(3)) \longleftarrow SU(3) \longrightarrow D_2(SU(9)) \,.
\end{align}
The 3d reduction of $D_2(SU(3))$ is $U(1)$ SQED with 3 flavors, while that of $D_2(SU(9))$ is $U(4)$ SQCD with 9 flavors. Gauging a diagonal combination of the $SU(3)$ symmetry of the former theory and of the one acting on 3 out of the 9 flavors of the latter theory, we obtain that the 3d reduction of $D_4(SU(6))$ can be described by the quiver
\begin{equation}\label{eq:3dquivD4SU6}
\includegraphics[scale=1.1]{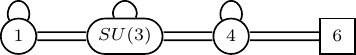}
\end{equation}

\subsection{Comparison between $\mathcal{C}_p[SU(N)]$ and 3d quivers for $D_p[SU(N)]$}

We will now discuss more in details the relation between the family of 3d $\mathcal{N}=2$ $\mathcal{C}_p[SU(N)]$ quivers and those obtained from the circle reduction of the $D_p[SU(N)]$ theories.  For the entirety of this section, as we did in the rest of the paper, we focus on $p\leq N$.

\paragraph{\boldmath$\gcd(p,N)=1$.}
Let us begin from the case $\gcd(p,N)=1$. As we have just reviewed, the quiver for the 3d reduction of $D_p[SU(N)]$ is given by the $T^\sigma_\rho[SU(N)]$ theory with $\sigma=[1^N]$ and $\rho=[(r+1)^{l},r^{p-l}]$, whose matter and gauge content turns out to coincide with that of the $\mathcal{C}_p[SU(N)]$ theory. The only difference is in the superpotential: the 3d reduction of $D_p[SU(N)]$ has $\mathcal{N}=4$ supersymmetry and so it only has the cubic interaction $\mathcal{W}_{\text{cub}}$ between the bifundamentals and the adjoint chirals, while the $\mathcal{C}_p[SU(N)]$ theory as we defined it in Section \ref{sec3dandDp} has also a monopole superpotential which breaks supersymmetry down to $\mathcal{N}=2$. Moreover, in the same section we have shown that such monopole deformation of the $\mathcal{N}=4$ theory makes it flow to a WZ theory, consisting of a chiral in the adjoint of the $SU(N)$ flavor symmetry with interaction $\mathcal{W}\sim\Tr(\Phi^{p+1})$. It is then natural to wonder whether there exists a deformation of the $D_p[SU(N)]$ theory directly in 4d that breaks supersymmetry from $\mathcal{N}=2$ to $\mathcal{N}=1$ and that makes it flow to the same WZ theory. Interestingly, such a deformation has been recently studied in \cite{Maruyoshi:2023mnv} (see also \cite{Xie:2021omd,Kang:2023dsa,Bajeot:2023gyl} for analogous previous observations), so let us briefly review their result.

As we reviewed above, all $D_p[SU(N)]$ theories with $\gcd(p,N)=1$ have a relevant chiral CB operator with the minimal scaling dimension $\Delta(u_{\text{min}})=(p+1)/p$. Turning this on in the superpotential breaks supersymmetry from $\mathcal{N}=2$ to $\mathcal{N}=1$, since it breaks the $\mathcal{N}=2$ R-symmetry $SU(2)_R\times U(1)_r$ to an $\mathcal{N}=1$ R-symmetry $U(1)_R$ that is a combination of the Cartan $U(1)_{I_3}\subset SU(2)_R$ and of $U(1)_r$. The exact combination can be determined from the knowledge of the scaling dimension of the operator $u_{\text{min}}$ and recalling that for a 4d $\mathcal{N}=2$ CB operator $r=2\Delta$ and $I_3=0$
\begin{align}\label{eq:embRsymm}
U(1)_R=\frac{1}{p+1}\left(2\,U(1)_{I_3}+p\,U(1)_r\right)\,.
\end{align}
Since a moment map operator in a 4d $\mathcal{N}=2$ SCFT has $r=0$ and $I_3=1$, one has that under such a symmetry the moment map operator $\mu$ of $D_p[SU(N)]$ for the $SU(N)$ flavor symmetry has R-charge
\begin{align}\label{eq:Rmomentmap}
R[\mu]=\frac{2}{p+1}\,.
\end{align}
This is mapped to the chiral operator $\Phi$ of the $\mathcal{N}=1$ theory after the deformation that is in the adjoint of $SU(N)$ and couples as $\mathcal{W}\sim\Tr(\Phi^{p+1})$. Computing the $\mathcal{N}=1$ $a$ and $c$ central charges \cite{Anselmi:1997am} by exploiting the anomalies \eqref{eq:flavorcentralcharge}-\eqref{eq:ccentralcharge}-\eqref{eq:acentralcharge} of $D_p[SU(N)]$ and the embedding \eqref{eq:embRsymm}\footnote{We also need that $\mathrm{Tr}\,r=\mathrm{Tr}\,r^3=48(a_{\mathcal{N}=2}-c_{\mathcal{N}=2})$.}
\begin{align}
a_{\mathcal{N}=1}&=\frac{3}{32}\left(3\mathrm{Tr}\,R^3-\mathrm{Tr}\,R\right)=-\frac{3  (p-1) (p^2-4 p+1)}{16 (p+1)^3}(N^2-1) \,,\nn\\
c_{\mathcal{N}=1}&=\frac{1}{32}\left(9\mathrm{Tr}\,R^3-5\mathrm{Tr}\,R\right)=- \frac{(p-1) (p^2-7 p+1)}{8 (p+1)^3}(N^2-1) \,,
\end{align}
one also finds that these coincide with those of a chiral in the adjoint of $SU(N)$ and with R-charge $2/(p+1)$. An additional check of this claim that was done in \cite{Maruyoshi:2023mnv} was at the level of a limit of the index, which is related to the Schur limit \cite{Gadde:2011uv} of the index of the $D_p[SU(N)]$ theory \cite{Xie:2016evu,Song:2017oew}.

The confining dualities for the 3d $\mathcal{C}_p[SU(N)]$ theories that we discussed in Section \ref{sec3dandDp} thus represent an additional check for the result of \cite{Maruyoshi:2023mnv} about the 4d $D_p[SU(N)]$ theories. When compactifying a 4d $\mathcal{N}=2$ theory on a circle of radius $\beta$, we obtain a 3d effective theory with KK modes whose CB is obtained from the 4d CB with $2r$ additional compact directions, where we recall that $r$ is the rank of the 4d theory. Flowing to energies smaller than $\frac{1}{\beta}$ the KK tower is integrated out and these directions decompactify, so that we obtain a 3d CB whose complex dimension is twice that of the 4d CB. Since the CB of the 3d theory is parametrized by monopole operators, it is then natural to expect that the 4d CB operator $u_{\text{min}}$ descends to a monopole operator in the 3d $\mathcal{N}=4$ quiver theory for the dimensional reduction of $D_p[SU(N)]$, although determining the precise mapping between the 4d and the 3d CB operators is difficult due to the fact that the 4d theory is non-Lagrangian. Moreover, the monopole superpotential of $\mathcal{C}_p[SU(N)]$ consists of many terms, while the deformation of $D_p[SU(N)]$ that makes it flow to a WZ model is just given by one operator $u_{\text{min}}$. The generation of monopole superpotentials in the dimensional reduction of 4d $\mathcal{N}=1$ theories to 3d $\mathcal{N}=2$ theories is actually quite common, see e.g.~\cite{Aharony:2013dha,Aharony:2013kma}, but repeating a similar analysis here is difficult again due to the non-Lagrangian nature of the $D_p[SU(N)]$ theory. However, we have observed that the role of the monopole superpotential of $\mathcal{C}_p[SU(N)]$ is to set the R-charge of the $SU(N)$ moment map, consisting of the meson matrix constructed with the flavors at the end of the tail of the quiver \eqref{fig:conf_genk}, to $2/(p+1)$, which is exactly what happens also in 4d \eqref{eq:Rmomentmap}. Moreover, the monopole superpotential also breaks the topological symmetries except for a residual $\mathbb{Z}_{p+1}$, which is also the only other symmetry present in 4d on top of $SU(N)$ and $U(1)_R$. Hence, we expect the full monopole superpotential of $\mathcal{C}_p[SU(N)]$ to be present in the dimensional reduction of $D_p[SU(N)]$ deformed by $u_{\text{min}}$ in order to match the symmetries and the charges between the 4d and the 3d theories.

\paragraph{\boldmath$\gcd(p,N)\neq 1$ and $N\neq pr$.}
Let us move now to the case where $\gcd(p,N)\neq 1$, but excluding the possibility that $N$ is an integer multiple of $p$, that is $l\neq 0$. The 3d $\mathcal{C}_p[SU(N)]$ theory is still given by the $T^\sigma_\rho[SU(N)]$ theory with $\sigma=[1^N]$ and $\rho=[(r+1)^{l},r^{p-l}]$ deformed by the monopole superpotential, but now the 3d reduction of $D_p[SU(N)]$ is a linear quiver with mixed unitary and special unitary nodes. However, we can find a deformation of the latter quiver that makes it flow to the $T^\sigma_\rho[SU(N)]$ theory.

Consider for example the case $p=4$ and $N=6$, or equivalently $r=1$ and $l=2$. As we have seen above, the quiver for the 3d reduction of $D_4(SU(6))$ is given by the quiver in \eqref{eq:3dquivD4SU6}. On the other hand, the 3d $\mathcal{N}=4$ theory obtained from $\mathcal{C}_4[SU(6)]$ by removing the monopole superpotential is the $T^{[1^6]}_{[2^2,1^2]}[SU(6)]$ theory, whose quiver is
\begin{equation}\label{eq:3dquivC4SU6}
\includegraphics[scale=1.1]{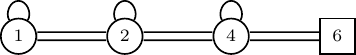}
\end{equation}
Nevertheless, we observe that the quiver \eqref{eq:3dquivC4SU6} can be obtained from \eqref{eq:3dquivD4SU6} by replacing the $SU(3)$ gauge node, which is connected to 5 flavors, with a $U(2)$ gauge node with the same number of flavors. 

This is a general feature that persists for other values of $p$ and $N$: the 3d quiver of $D_p[SU(N)]$ generically has $m-1$ special unitary gauge nodes and the $i$-th node is $SU(N-in)$ with $2(N-in)-1$ flavors (recall that $m=\gcd(p,N)$ and $n=N/m$), and the 3d $\mathcal{N}=4$ quiver associated to $\mathcal{C}_p[SU(N)]$ by removing the monopole superpotential has such nodes replaced by $U(N-in-1)$ nodes again with $2(N-in)-1$ flavors. As shown in \cite{Closset:2020afy}, there exists in general a deformation that makes us flow from a $SU(M)$ theory with $2M-1$ flavors to a $U(M-1)$ theory with $2M-1$ flavors. This consists of a vev for the $SU(M)$ adjoint chiral of the form
\begin{align}\label{eq:adjvev}
\langle\Phi\rangle=\mathrm{diag}(\mathfrak{m},\cdots,\mathfrak{m},-(M-1)\mathfrak{m})\,,
\end{align}
which also induces an equal mass for the flavors $Q$, $\tilde{Q}$
\begin{align}
\delta\mathcal{W}=\mathfrak{m}\,\mathrm{Tr}_{U(2M-1)}\mathrm{Tr}_{SU(M)}Q\tilde{Q}\,,
\end{align}
where $\mathrm{Tr}_{SU(M)}$ is the trace over the color indices while $\mathrm{Tr}_{U(2M-1)}$ is over the flavor indices. Notice indeed that such a vev breaks the $SU(M)$ gauge group down to the expected $U(M-1)$ subgroup. This deformation can be studied in the mirror dual of the $SU(M)$ SQCD with $2M-1$ flavors, where it corresponds to a vev for a meson which can be induced by an FI deformation that can be easily analyzed. The result is precisely the mirror dual of the $U(M-1)$ SQCD with $2M-1$ flavors \cite{Closset:2020afy}.

After such a deformation of the 3d quiver of $D_p[SU(N)]$, the unitary nodes that were not directly affected by the deformation might have become ugly, so they need to be dualized with the effect that their rank is reduced by one and a free twisted hyper is produced \cite{Gaiotto:2008ak}. It is only after such dualizations that we recover the 3d $\mathcal{N}=4$ unitary linear quiver corresponding to $\mathcal{C}_p[SU(N)]$ but with the monopole deformation removed. This did not happen for the example of the $D_4(SU(6))$ theory that we have seen before, but it is usually the case when $m>2$ for the unitary nodes that are in between two special unitary nodes in the 3d quiver of $D_p[SU(N)]$. 

Consider for example the case $p=6$ and $N=9$, for which $m=\gcd(6,9)=3$. The quiver for the 3d reduction of $D_6(SU(9))$ is
\begin{equation}\label{eq:3dquivD6SU9}
\includegraphics[scale=1.1]{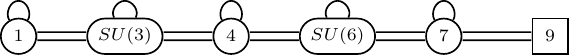}
\end{equation}
which reflects the fact that, according to \eqref{eq:DpSUNgauging}, $D_6(SU(9))$ can be understood as the $SU(6)$ gauging of $D_4(SU(6))$ and $D_2(SU(15))$. Turning on the deformation \eqref{eq:adjvev} at the two special unitary nodes we obtain the quiver
\begin{equation}\label{eq:3dquivD6SU9def}
\includegraphics[scale=1.1]{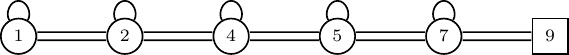}
\end{equation}
As anticipated, the $U(4)$ node that used to be in between the special unitary nodes that we deformed has 7 flavors and is thus ugly. If we dualize it, we produce a free twisted hyper and obtain the quiver
\begin{equation}\label{eq:3dquivC6SU9}
\includegraphics[scale=1.1]{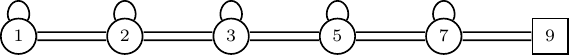}
\end{equation}
This is precisely the $T^{[1^9]}_{[2^3,1^3]}[SU(9)]$ theory that is related to $\mathcal{C}_6[SU(9)]$ upon turning on the monopole superpotential deformation

Summarizing, the 3d $\mathcal{N}=2$ $\mathcal{C}_p[SU(N)]$ theory for $\gcd(p,N)\neq 1$ and $N\neq pr$ can be obtained from the 4d $\mathcal{N}=2$ $D_p[SU(N)]$ SCFT as follows:
\begin{enumerate}
\item compactify the $D_p[SU(N)]$ theory on a circle so to get a 3d $\mathcal{N}=4$ SCFT that is the IR limit of a 3d $\mathcal{N}=4$ linear quiver with mixed unitary and special unitary gauge nodes;
\item turn on deformations of the form \eqref{eq:adjvev} which effectively replace each special unitary gauge node $SU(M)$ of this quiver with a unitary gauge node $U(M-1)$ of the same rank;
\item dualize each ugly node $U(M)$ with $2M-1$ flavors into the good node $U(M-1)$ with $2M-1$ flavors plus a free twisted hyper;
\item turn on a suitable monopole superpotential.
\end{enumerate}
In particular, the last deformation breaks supersymmetry from $\mathcal{N}=4$ to $\mathcal{N}=2$ and makes the theory flow to a WZ model consisting of an adjoint chiral $\Phi$ for the $SU(N)$ flavor symmetry with superpotential $\mathcal{W}\sim\Tr(\Phi^{p+1})$. 

\paragraph{Some open questions.} We conclude this section by commenting on some open questions about the relationship between the 3d $\mathcal{N}=2$ $\CC_p[SU(N)]$ theories and the 4d $\mathcal{N}=2$ $D_p[SU(N)]$ SCFTs, which would be interesting to further investigate in future works.

We have seen that for $\gcd(p,N)=1$ the 3d reduction of $D_p[SU(N)]$ flows to $\CC_p[SU(N)]$ upon turning on the monopole superpotential deformation, which is consistent with the fact that both $\CC_p[SU(N)]$ and $D_p[SU(N)]$ deformed by the CB operator $u_{\text{min}}$ flow to a WZ theory. However, for $\gcd(p,N)\neq1$ and $N\neq pr$ an additional deformation, corresponding to the steps 2.~and 3.~listed above, is needed. It would be interesting to understand whether these operations can be uplifted to 4d, so to find a deformation of $D_p[SU(N)]$ that makes it flow to a WZ model which generalizes the result of \cite{Maruyoshi:2023mnv} to the case of $\gcd(p,N)\neq 1$. 

Another open question is the relationship between $\mathcal{C}_p[SU(N)]$ and $D_p[SU(N)]$ when $N=pr$. In this case, $\mathcal{C}_p[SU(N)]$ is given by a sequence of balanced gauge nodes $U(jr)$ for $j=1,\ldots,p-1$, while $D_p[SU(N)]$ is fully Lagrangian already in 4d and given by a similar quiver but where the gauge nodes are special unitary $SU(jr)$ rather than unitary (see e.g.~\cite{Cecotti:2013lda,Giacomelli:2017ckh}). Hence, the two quivers cannot be related by the deformation \eqref{eq:adjvev} as for $N\neq pr$.

Finally, in our discussion we focused on the quivers of these theories and we neglected possible gauge singlet fields. In fact, the confining duality for the $\mathcal{C}_p[SU(N)]$ theory also predicts the presence of singlet chirals flipping traces of powers of the adjoint field $\Phi$, which are not present in the analogous duality for $D_p[SU(N)]$. On the other hand, the 3d reduction of $D_p[SU(N)]$ generically also has some decoupled twisted hypers, which are in number \cite{Giacomelli:2020ryy}
\begin{equation}
    H_{\text{free}}=\frac{(N-m)(p-N-m)}{2m}\,,
\end{equation}
where we recall that $m=\gcd(p,N)$. These are in particular crucial in order to match the CB dimension between 4d and 3d. Moreover, they encode the fact that generically on the most generic point of the HB of $D_p[SU(N)]$ we do not have a collection of free hypers but also some non-Higgsable (i.e.~with trivial HB) SCFT, such as the $(A_1,A_2)$ AD theory, which in 3d reduces just to twisted hypers. It would be interesting to understand how to reconcile the presence of the singlet chirals in the confining duality of $\CC_p[SU(N)]$ and of the twisted hypers in the 3d reduction of $D_p[SU(N)]$.

\section{The 4d $\mathcal{N}=1$ $\mathsf{C}_p[USp(2N)]$ family}\label{sec:4dCp}

In this section we introduce the 4d $\mathcal{N}=1$ $\mathsf{C}_p[USp(2N)]$ theories, where $N$ and $p$ are positive integers such that $N>p$. We show that the $\mathsf{C}_p[USp(2N)]$ theories confine to a WZ model of a chiral $\Phi$ in the traceless antisymmetric representation of $USp(2N)$ with superpotential $\CW \sim \Tr(\Phi^{p+1})$. Upon a circle reduction and suitable deformations, they reduce to the $\CC_p[SU(N)]$ theories we introduced in Section \ref{sec3dandDp} (see e.g.~\cite{Benini:2017dud,Amariti:2018wht,Pasquetti:2019hxf,Bajeot:2022lah,Bottini:2021vms,Comi:2022aqo,Bajeot:2023gyl} for similar dimensional reduction limits), and their respective confining dualities are similarly related.


\subsection{The  $\mathsf{C}_p[USp(2N)]$ family}
The $\mathsf{C}_p[USp(2N)]$ theories are 4d $\CN=1$ quiver theories with $p-1$ $USp(2n_j)$ gauge nodes and a $USp(2N)$ flavor symmetry.\footnote{Unlike what we did in Section \ref{sec3dandDp}, here for simplicity we ignore possible finite symmetries and only focus on the Lie algebras of the global symmetries.} As in Section \ref{sec3dandDp}, we distinguish two situations depending on whether $N$ is a multiple of $p$ or not. It is then useful to define
\begin{align}
    N = pr + l \,, \qquad l=0,\ldots,p-1 \,.
\end{align}
The ranks $n_j$ of the gauge groups are determined as in \eqref{eq:ranks_Cp}.

\paragraph{\boldmath{$l\neq 0 $}.} In this case the $\mathsf{C}_p[USp(2(pr+l))]$ theory can be depicted as\footnote{In this section and in the following, where we work solely in four dimensions, we adopt a different notation for quivers than what we did in 3d. Each node, round or square, denotes a gauge or flavor symplectic group. Lines are chirals in the fundamental representation of the nodes to whom they are linked. Square nodes labelled with $1$ do not represent flavor symmetries and are used only to encode the number of chiral fields. Unless stated otherwise, arcs are traceful antisymmetric tensors and crosses denotes flipping fields.}
\begin{align}\label{fig:4dCp_quiver_ln0}
    \includegraphics[scale=1]{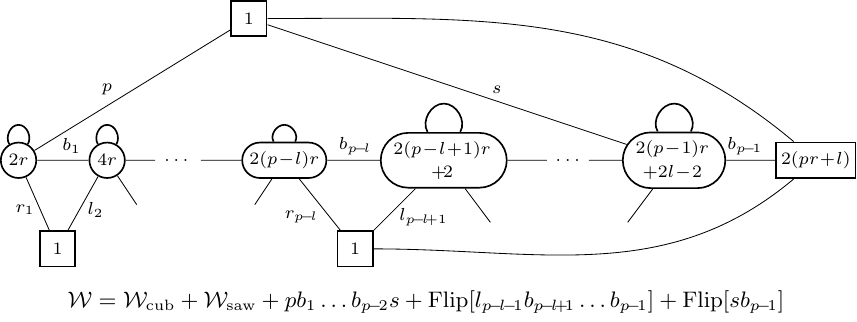}
\end{align}
In analogy with Section \ref{sec3dandDp}, we call a $USp(2N_c)$ gauge node \emph{balanced} if it sees exactly $2F=4N_c$ fundamental chirals, counting only the horizontal bifundamentals and not chirals in the saw. If instead it sees $2F>4N$ it is called \emph{overbalanced}. In figure \eqref{fig:4dCp_quiver_ln0} all the gauge nodes are balanced except the $(p-l)$-th one which is overbalanced with excess number 1, similarly to 3d. We will also refer to the first $p-l-1$ nodes as the \emph{left tail} and to the last $l-1$ as the \emph{right tail}. 

The $\mathsf{C}_p[USp(2N)]$ theory in \eqref{fig:4dCp_quiver_ln0} has also a superpotential consisting of several terms, which we now illustrate. $\CW_{\text{cub}}$ encodes cubic couplings between each antisymmetric field and the bifundamentals on its sides. $\CW_{\text{saw}}$ contains all the cubic superpotential terms for the triangles in the saw of the form $r_j b_j l_{j+1}$. Then we have also the $p b_1 \ldots b_{p-2} s$ term and two flippers that in the figure appear as the two singlet arcs. 

The vanishing of the NSVZ $\b$-functions and the superpotential constrain the R-charges as follows:
\begin{align}\label{eq:adjbif_rch}
    & R[\text{antisymmetrics}] = 2 - \frac{2}{p+1} \,, \nn \\
    & R[b_j] = \frac{1}{p+1} \,,
\end{align}
as well as\footnote{To obtain the following relations we can  fix the charges of antisymmetrics and bifundamentals as in \eqref{eq:adjbif_rch}. One then solves the NSVZ $\b$-function condition to obtain that $R[r_j] + R[l_j]$ equals $\tfrac{p-1}{p+1}$ if the $j$-th node is balanced or $\tfrac{2p-1}{p+1}$ if the $j$-th node is the unbalanced one. Then one also takes into account the constraint due to $\CW_{\text{saw}}$ imposing $R[r_j]+R[l_{j+1}] = 2 - \tfrac{1}{p+1}$. Analogously $R[s]$ and $R[p]$ follow.}
\begin{align}\label{eq:chir_rch}
    & R[r_j] = \begin{cases}
        R[r_{j-1}] - \frac{3}{p+1} \qquad & \text{if} \quad j \neq p-l \\
        R[r_{p-l}] + \frac{2p-3}{p+1} \qquad & \text{if} \quad j = p-l
    \end{cases}  \,, \nn \\
    & R[l_j] = 2 - \frac{1}{p+1} - R[r_{j-1}] \,, \nn \\
    & R[p] = \begin{cases}
        2\frac{2p-1}{p+1} - R[r_1] \qquad & \text{if} \quad l = 1 \\
        2\frac{p-1}{p+1} - R[r_1] \qquad & \text{otherwise}
    \end{cases} \,, \nn \\
    & R[s] = \begin{cases}
        2\frac{2p-1}{p+1} - R[l_{p\!-\!1}] \qquad & \text{if} \quad l = p-1 \\
        2\frac{p-1}{p+1} - R[l_{p\!-\!1}] \qquad & \text{otherwise}
    \end{cases} \,.
\end{align}
All the R-charges are fixed from the choice of $R[r_1]$. This implies that there is a non-anomalous $U(1)_C$ global symmetry that can mix with the R-symmetry. We will comment on this symmetry later. 

\paragraph{\boldmath{$l=0$}.} In this case the $\mathsf{C}_p[USp(2pr)]$ theory is given by the following quiver theory:
\begin{align}\label{fig:4dCp_quiver_l0}
	\includegraphics[scale=1]{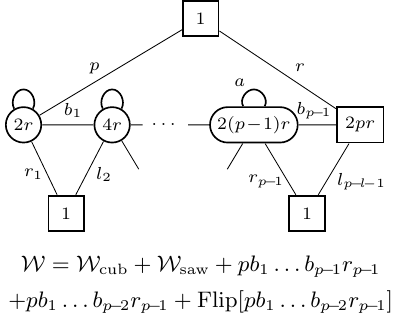}
\end{align}
Notice that in this case all nodes are balanced, similarly to 3d.

The superpotential contains, other than the cubic and saw terms, a term linear in the meson $p b_1 \ldots b_{p\!-\!2} ar_{p\!-\!1}$\footnote{The antisymmetric $a$ used to dress the meson can be taken to be any of the antisymmetrics. All the possible choices are identified in the chiral ring of the theory. For simplicity we assume that the antisymmetric $a$ is the one charged under the $USp(2(p-1)r)$ gauge group.} and also a flipping term for the similar meson $p b_1 \ldots b_{p\!-\!2} r_{p\!-\!1}$. These are the analogues of the dressed monopole superpotential and the flip of the undressed monopole that we had in 3d. 

The R-charges of the bifundamental and antisymmetric chirals are as in \eqref{eq:adjbif_rch}, while the R-charge of the rest of the fields can again be obtained imposing the vanishing of the NSVZ $\b$-function and the superpotential constraints, finding relations similar to those in \eqref{eq:chir_rch}.

\paragraph{Global symmetries and gauge invariant operators.}
The $\mathsf{C}_p[USp(2N)]$ theory is characterized by a $USp(2N)$ flavor symmetry. There is also an extra $U(1)_C$ symmetry that is compatible with the superpotential in figure \eqref{fig:4dCp_quiver_ln0} and \eqref{fig:4dCp_quiver_l0}. The charge assignment for this symmetry is:
\begin{align}
    & C[b_j] = C[\text{antisymmetrics}] = 0 \,, \nn \\
    & C[r_j] = C[s] = +1 \,, \nn \\
    & C[l_j] = C[r] = -1 \,.
\end{align}
From which it follows that also the flippers $F[s b_{p\!-\!1}]$ and $F[l_{p\!-\!l\!-\!1} b_{p\!-\!l\!+\!1} \ldots b_{p\!-\!1} ]$ have $C$-charge $+1$ and $-1$ respectively. We claim that the $U(1)_C$ symmetry is actually unfaithful, i.e.~it trivializes in the low energy theory. Therefore all the gauge invariant operators that can be constructed with a non-zero charge under $U(1)_C$ are expected to vanish in the chiral ring. 
This claim is supported by the fact that all the anomalies involving $U(1)_C$ vanish and by the validity of the confining duality, since on the dual side this symmetry is not present.

Below we discuss some interesting gauge invariant operators of the $\mathsf{C}_p[USp(2N)]$ theory which we will later map to the singlets in the confined dual frame \eqref{fig:4dCpconfdual}.
There is a traceless antisymmetric operator  $\mu = b_{p\!-\!1}^2$, with R-charge
\begin{align}
    R[\mu] = \frac{2}{p+1} \,.
\end{align}
Sometimes we will, by abuse of notation, refer to this operator as ``moment map", in analogy with the 3d story in Section \ref{sec3dandDp}.

In addition we have also other gauge invariant operators that are singlets of the $USp(2N)$ global symmetry. In the theories \eqref{fig:4dCp_quiver_ln0} and \eqref{fig:4dCp_quiver_l0} we can construct mesons using the fields in the saw of the quiver
\begin{align}
    \mathsf{M}^{(k)} = \begin{cases}
        l_j b_j \ldots b_{j-1+k} r_{j+k} \\
        p b_1 \ldots b_{k-1} r_{k} \\
        l_{p-k} b_{p-k} \ldots b_{p-2} s 
    \end{cases} \qquad \text{for} \quad k \geq 1 \,, 
\end{align}
such that the nodes $j$ and $j-1+k$ nodes belong to either the left or right balanced tails. These operators have R-charge
\begin{align}
    R[\mathsf{M}^{(k)}] = 2 - 2\frac{k+1}{p+1} \qquad \text{for} \quad k \geq 1 \,.
\end{align}
In fact, for any fixed value $k$ there are many such operators, however we claim that the mesons of the same length $k$ constructed from the same balanced tail are degenerate. We are then left with two sets of operators, those constructed from the left tail and those from the right one, which we name $\mathsf{M}_{L/R}^{(k)}$. By looking at the quivers in \eqref{fig:4dCp_quiver_ln0} and \eqref{fig:4dCp_quiver_l0} one finds that the operators are
\begin{align}\label{eq:4dcpmesons}
    & l \neq 0 \quad : \quad \begin{cases}
        \mathsf{M}^{(k)}_L & \quad k = 1, \ldots, p-l-2 \qquad \text{from the left tail}\,, \\
        \mathsf{M}^{(k)}_R & \quad k = 1, \ldots, l-2 \qquad \text{from the right tail}\,, \\
    \end{cases} \nn \\
    & l = 0 \quad : \quad \mathsf{M}^{(k)} \quad k = 1,\ldots, p-3 \qquad \text{from the (only) tail}\,.
\end{align}
These operators are not charged under the $USp(2N)$ global symmetry, therefore we can construct exactly marginal deformations for the $\mathsf{C}_p[USp(2N)]$ theory given by the product of $\mathsf{M}^{(k)}_{L/R}$ operators that have total R-charge 2. For example, considering only the product of two $\mathsf{M}^{(k)}_{L/R}$ operators we have that $\mathsf{M}^{(k)}_{L/R} \mathsf{M}^{(p-j-1)}_{L/R}$ always have R-charge 2, provided that the operators exists in the given $\mathsf{C}^p[USp(2N)]$ theory.

By simply looking at the quivers in figures \eqref{fig:4dCp_quiver_ln0} and \eqref{fig:4dCp_quiver_l0}, one may notice that there are many more gauge invariant operators that we have not considered. However, we claim that all the gauge invariant operators carrying a non-zero charge under the $U(1)_C$ unfaithful symmetry are set to zero in the chiral ring, as we already mentioned previously.

\subsubsection{Statement of the confining duality}
We claim that the $\mathsf{C}_p[USp(2N)]$ theories confine to a traceless antisymmetric field with $\CW \sim \Tr(\Phi^{p+1})$ superpotential plus flippers
\begin{align}\label{fig:4dCpconfdual}
\begin{tikzpicture}[scale=1.2,every node/.style={scale=1.2},font=\scriptsize]
\begin{scope}[shift={(-1,0)}]
    \node[flavor] (f1) at (-1,0) {$2N$};
    \node (Cp) at (-1,0.75) {$\mathsf{C}_p$};
    \draw[thick] (f1) to[out=120,in=-135] (Cp) to[out=-45,in=60] (f1);
    \node (W1) at (-1,-0.75) {$\CW = \CW_{\text{multi-meson}}$};
\end{scope}
    \node (=) at (1,0) {$\Longleftrightarrow$};
\begin{scope}[shift={(1,0)}]
    \node[flavor] (f2) at (2,0) {$2N$};
    \node[left] at (1.9,0.6) {$\Phi$};
    \draw[thick] (f2) to[out=120,in=180] ([yshift=7mm] f2) to[out=0,in=60] (f2);
    \node (W2) at (2,-0.75) {$\CW = \Tr(\Phi^{p+1}) +$};
    \node (W22) at (2,-1.25) {$ + \sum_{j=2}^{k(l)} \text{Flip}[\Tr(\Phi^j)]+$};
    \node (W23) at (2,-1.75) {$ + \CW_{\text{multi-trace}}$};
\end{scope}
\end{tikzpicture}
\end{align}
where in the figure we depict the $\mathsf{C}_p[USp(2N)]$ theory in short as the label $\mathsf{C}_p$ embedded in an arc. \\
On the r.h.s.~of figure \eqref{fig:4dCpconfdual} we have flipping fields for the traces of $\Phi$ from power 2 up to $k(l)$, with
\begin{align}
    k(l) = \min(l,p-l) \,.
\end{align}
The above formula enjoys the property $k(l) = k(p-l)$. This will turn out to be important when we will use this duality to prove the IKS duality, which maps the gauge ranks from $N=pr+l$ to $N'=pr'+(p-l)$. \\
The superpotential on the l.h.s.~of the duality in \eqref{fig:4dCpconfdual} labelled as $\CW_{\text{multi-meson}}$ contains all the possible exactly marginal deformations given as the product of two or more $\mathsf{M}^{(k)}_{L/R}$ operators. Correspondingly on the r.h.s.~we have the superpotential $\CW_{\text{multi-trace}}$ which contains all the exactly marginal double trace operators. The reason why this duality is considered at a generic point of the conformal manifold is the same as in the 3d confining duality for the $\CC^p[SU(N)]$ theory, discussed below figure \eqref{fig:conf_genk}. 

As a first simple check of the confining duality in \eqref{fig:4dCpconfdual} we checked that the anomalies $\Tr( U(1)_R )$, $\Tr( U(1)_R^3 ) $ and $\Tr( U(1)_R USp(2N)^2 )$ of the $\mathsf{C}_p[USp(2N)]$ theory check with those of the WZ model.
In the following section we will provide a derivation of this duality.


\subsubsection{The mirror theories $\check{\mathsf{C}}_p[USp(2N)]$}
It is possible to construct a \emph{mirror-like} dual of the 4d $\CN=1$ theories $\check{\mathsf{C}}_p[USp(2N)]$ following the strategies initiated in \cite{Hwang:2020wpd} (see also \cite{Bottini:2021vms,Hwang:2021ulb,Comi:2022aqo}). The mirror duals are different for $l \neq 0$ and $l = 0$
\begin{align}\label{fig:4dCpmir}
	&\includegraphics[scale=0.75]{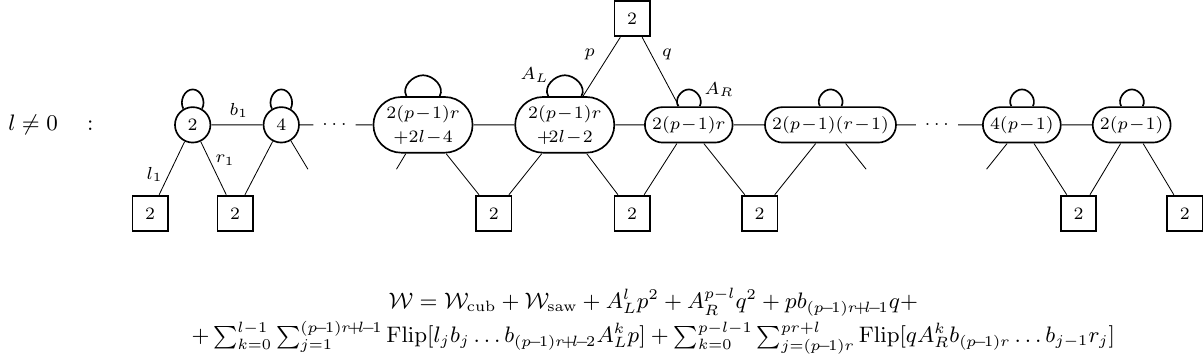}\nn\\
	&\includegraphics[scale=0.75]{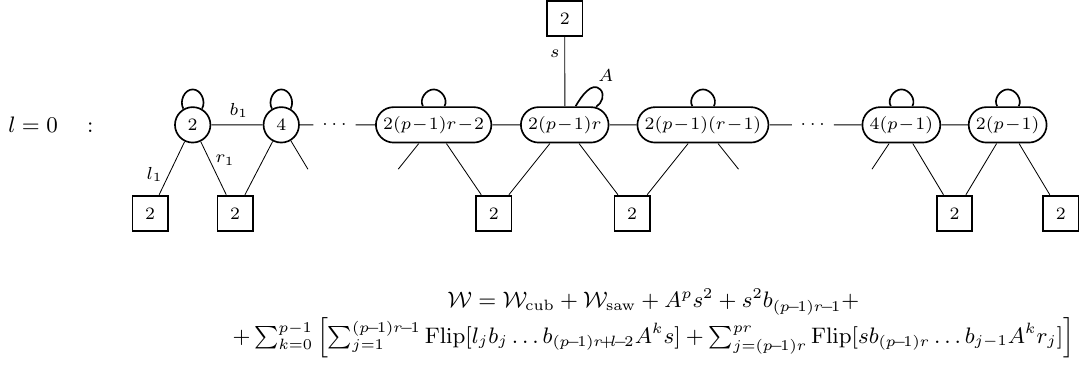}
\end{align}
In the theories in \eqref{fig:4dCpmir} the superpotential contains $\CW_{\text{cub}}$, encoding all the cubic couplings between horizontal bifundamentals and the antisymmetric fields on their sides, and $\CW_{\text{saw}}$, encoding the cubic couplings for each tooth composing the saw. On top of that we turn on dressed mesonic operators constructed from the fundamental flavors in the midlle of the quiver and flipping fields.

The R-charges of all the fields can be fixed from the superpotential and vanishing $\b$-function constraints. We get that all the horizontal bifundamentals and the antisymmetric fields have R-charges (for both $l \neq 0$ and $l = 0$)
\begin{align}
    R[\text{bif}] = \frac{1}{2} - \frac{1}{2(p+1)} \,, \nn \\
    R[\text{antisymmetrics}] = \frac{2}{p+1} \,.
\end{align}
The R-charges of the $p,q$ fields for $l \neq 0$ and of the $s$ field for $l=0$ are fixed to be
\begin{align}
    & l \neq 0 \quad : \quad \begin{cases}
        & R[p] = 1 - \frac{l}{p+1} \\
        & R[q] = 1 - \frac{p-l}{p+1} 
    \end{cases} \,, \nn \\
    & l = 0 \quad : \quad R[s] = 1 - \frac{p}{p+1} \,.
\end{align}
The R-charge of the fields in the saw are
\begin{align}
    & R[l_j] = 1 - \frac{1}{p+1} + R[l_{j-1}] \,, \nn \\
    & R[r_j] = 1 + \frac{1}{p+1} - R[l_{j+1}] \,.
\end{align}
Notice that all the R-charges are fixed from the assignment of $R[l_1]$, indicating again the existance of a non-anomalous abelian $U(1)_{C'}$ symmetry. The above formula holds both for $l \neq 0$ and for $l = 0$.

\paragraph{Global symmetries and gauge invariant operators.}
The UV global symmetry of the quiver theories in \eqref{fig:4dCpmir} is:
\begin{align}
    SU(2)^N \times SU(2)_F \times U(1)_{C'} \,,
\end{align}
where $SU(2)^N$ are the $N$ symmetries rotating the chirals in the saw, $SU(2)_F$ is the symmetry acting on the fundamental flavors in the middle of the quiver, and $U(1)_{C'}$ is a non-anomalous abelian global symmetry consistent with the superpotential as explained at the end of the previous section. In the IR the $SU(2)^N$ symmetries enhance to form a $USp(2N)$ group, which is mapped to the $USp(2N)$ global symmetry of the $\mathsf{C}_p[USp(2N)]$ theory. 
The remaining $SU(2)_F \times U(1)_{C'}$ symmetries are unfaithful and thus we do not map them to some symmetry of the $\mathsf{C}_p[USp(2N)]$ theory.
All the gauge invariant operators that can be constructed and that are charged under these symmetries thus vanish in the chiral ring. Moreover, we checked  that all the anomalies involving $SU(2)_F$ and $U(1)_{C'}$ are exactly zero. 

We can construct a gauge invariant operator which is the traceless antisymmetric representation of the emergent $USp(2N)$ symmetry. This operator is built by casting together mesonic operators in a $N\times N$ matrix. The $N-1$ elements on the diagonal are given as the traces of the $N-1$ antisymmetric fields. The off-diagonal entries are filled with mesons built from the saw as $l_j b_j \ldots b_{k-1} r_k$. The resulting operator has R-charge $\tfrac{2}{p+1}$ and is naturally mapped to the moment map $\mu$ of the $\mathsf{C}_p[USp(2N)]$ theory.

There are also gauge invariant operators that are singlets under the $USp(2N)$ global symmetry. These are constructed by taking the square meson built from central flavors and dressing them with powers of the antisymmetric. Their R-charges are
\begin{align}
    & l \neq 0 \quad : \quad \begin{cases}
        R[p^2 A_L^j] = 2 - 2\frac{l-j}{p+1} \qquad & j = 0,\ldots, l-2 \\
        R[q^2 A_R^j] = 2 - 2\frac{p-l-j}{p+1} \qquad & j=0,\ldots,p-l-2
    \end{cases} \,, \nn \\
    & l = 0 \quad : \quad R[s^2 A^j] = 2\frac{j+1}{p+1} \qquad j = 0,\ldots,p-2 \,.
\end{align}
These operators for $l\neq0$ are mapped as
\begin{align}
    p^2 A_L^j \quad & \leftrightarrow \quad \mathsf{M}_L^{(l-1-j)} \qquad j = 0,\ldots,l-2 \,, \nn \\
    q^2 A_R^j \quad & \leftrightarrow \quad \mathsf{M}_R^{(p-l-1-j)} \qquad j=0,\ldots,p-l-2 \,.
\end{align}
For $l=0$ they are mapped as
\begin{align}
    s^2 A^j \quad \leftrightarrow \quad \mathsf{M}^{(p-1-j)} \qquad j=0,\ldots,p-l-1 \,.
\end{align}

\subsubsection*{Deriving the mirror theories}

The strategy to derive these dualities is very similar to the 3d case in Section \ref{sec3dandDp}. We start from a pair of 4d $\CN=1$ theories that enjoy a mirror-like duality. These theories have been introduced in \cite{Hwang:2020wpd} (see also \cite{Hwang:2021ulb,Comi:2022aqo}) and are called $E_\r^\s[USp(2N)]$. These theories are inspired by and, under a suitable dimensional reduction, they reduce to the 3d $\CN=4$ $T_\r^\s[SU(N)]$ theories., where $\r$ and $\s$ labels are partitions of $N$. $\s$ encodes the subgroup of $USp(2N)$ which is the manifest global symmetry of the theory (analogous to the 3d Higgs branch symmetry), while $\r$ defines a subgroup of $USp(2N)$ which is global symmetry that emerges in the IR from the enhancement of a collection of $SU(2)$ UV symmetries (analogous to the 3d Coulomb branch symmetry). 
For the case of interest, the $\mathsf{C}_p[USp(2N=2(pr+l))]$ theories are obtained starting from a $E_\r^\s[USp(2(pr+l)]$ theory with partitions
\begin{align}
    \s = [ 1^N ] \,, \qquad\r = [(r+1)^l, r^{p-l}] \,. \nn \\
\end{align}
Notice that these are the same partitions in \eqref{eq:Cppartitions} that defined the $\CC_p[SU(N)]$ theories. In particular for $l=0$ we have $\r = [r^p]$. From the partition $\s$ we see that the theory has a $USp(2N)$ manifest global symmetry while from the partition $\r$ we observe a $USp(2l) \times USp(2(p-l))$ enhanced global symmetry. Similarly to 3d, the mirrors are obtained by swapping $\r \leftrightarrow \s$, meaning that the manifest and emergent global symmetries are swapped. 

Starting from this mirror-like pair we want to deform the theories in order to break completely the $USp(2l) \times USp(2(p-l))$ symmetry, leaving a theory with only a $USp(2N)$ global symmetry group which is that of the $\mathsf{C}_p[USp(2N)]$ theory. The precise step followed in order to deform the mirror-like pair are different for $l \neq 0$ and $l = 0$. Let us then discuss separately the two cases in the following paragraphs.

\paragraph{\boldmath $l \neq 0$. }
In this case the strategy to derive the $\check{\mathsf{C}}_p[USp(2N)]$ mirror theory is schematically depicted below.
On the l.h.s.~of the figure we depict the steps to obtain $\mathsf{C}_p[USp(2N)]$ from the relevant $E_\r^\s[USp(2N)]$ theory, while on the r.h.s.~we depict the mirror dual procedure. Notice that we do not write all the singlets and all the superpotential terms at each step to make the discussion simpler. Indeed, one can keep track of all the details to find precisely the stated duality between the $\mathsf{C}_p[USp(2N)]$ theories and their mirror $\check{\mathsf{C}}_p[USp(2N)]$.
\begin{align}\label{fig:4dCpmir_ln0}
    \includegraphics[scale=0.55]{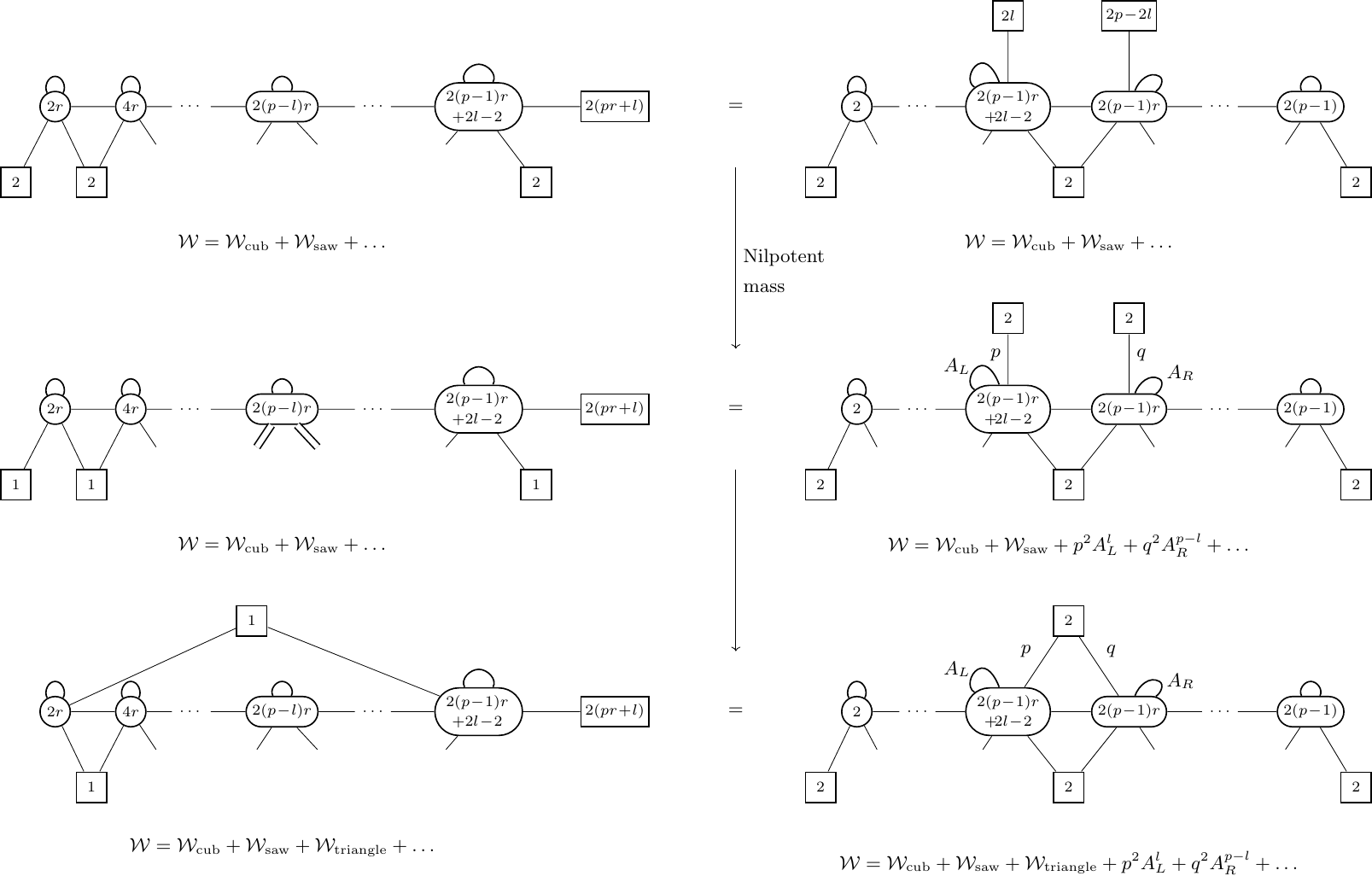}
\end{align}
We start from the 4d $\CN=1$ mirror pair in the first line of the figure above. The two starting theories have a $USp(2(pr+l)) \times USp(2l) \times USp(2p-2l)$ global symmetry. On the electric side the first group is manifest while the remaining two are emergent in the IR and come from the enhancement of the $SU(2)^p = SU(2)^l \times SU(2)^{p-l}$ groups. Similarly, on the magnetic side the $USp(2(pr+l))$ group is emergent while the remaining two groups are manifest.\footnote{To actually see the claimed enhancement it is very important to consider the correct singlets in the theory. However, these are not very important in the following discussion and therefore we do not consider them. For more details one can see \cite{Hwang:2020wpd,Comi:2022aqo}} These theories can be deformed by an analogues of the nilpotent deformations that are usually performed in 3d, as discussed extensively in \cite{Hwang:2020wpd}. Even though these are not nilpotent in 4d, we will still refer to them as ``nilpotent" by abuse of notation. In the first step, leading to the second line in figure \eqref{fig:4dCpmir_ln0}, we turn on a next-to-maximal nilpotent masses for the antisymmetric operators charged under $USp(2l)$ and $USp(2p-2l)$. On the electric side this deformation has the effect of integrating out part of the fields in the saw, while on the magnetic side the mass has the effect of integrating out many of the fundamental flavors and breaking $USp(2p) \to SU(2)_{V_1}$ and $USp(2p-2l) \to SU(2)_{V_2}$. In the second step, leading to the third line of figure \eqref{fig:4dCpmir_ln0}, on the magnetic side we turn on the superpotential corresponding to $bpq$, with the effect of breaking the $SU(2)_{V_1} \times SU(2)_{V_2}$ symmetry to a $SU(2)_F$ diagonal subgroup. On the electric side, the superpotential dual to $pbq$, has the effect of integrating two out of the four chirals seen by the $USp(2(p-l)r)$ gauge node and also turns on the $\CW_{\text{triangle}}$ superpotential.
All in all, we land on the duality in the third line of figure \eqref{fig:4dCpmir_ln0}, which is precisely the stated duality between the theory $\mathsf{C}_p[USp(2N)]$ theory in \eqref{fig:4dCp_quiver_ln0} and its mirror in the first line of \eqref{fig:4dCpmir}. Again we point out that in the presentation we did not keep track of the singlet fields for simplicity, but this can be done explicitly.

\paragraph{\boldmath $l = 0$.}
In this case the strategy is schematically depicted in the figure below. Notice that we depict only some singlets and we do not write all the superpotential terms at each step, to make the discussion simpler. Indeed, one can keep track of all the details to find precisely the stated duality between the $\mathsf{C}_p[USp(2N)]$ theories and their mirror $\check{\mathsf{C}}_p[USp(2N)]$.
\begin{align}\label{fig:4dCpmir_l0}
    \includegraphics[scale=0.6]{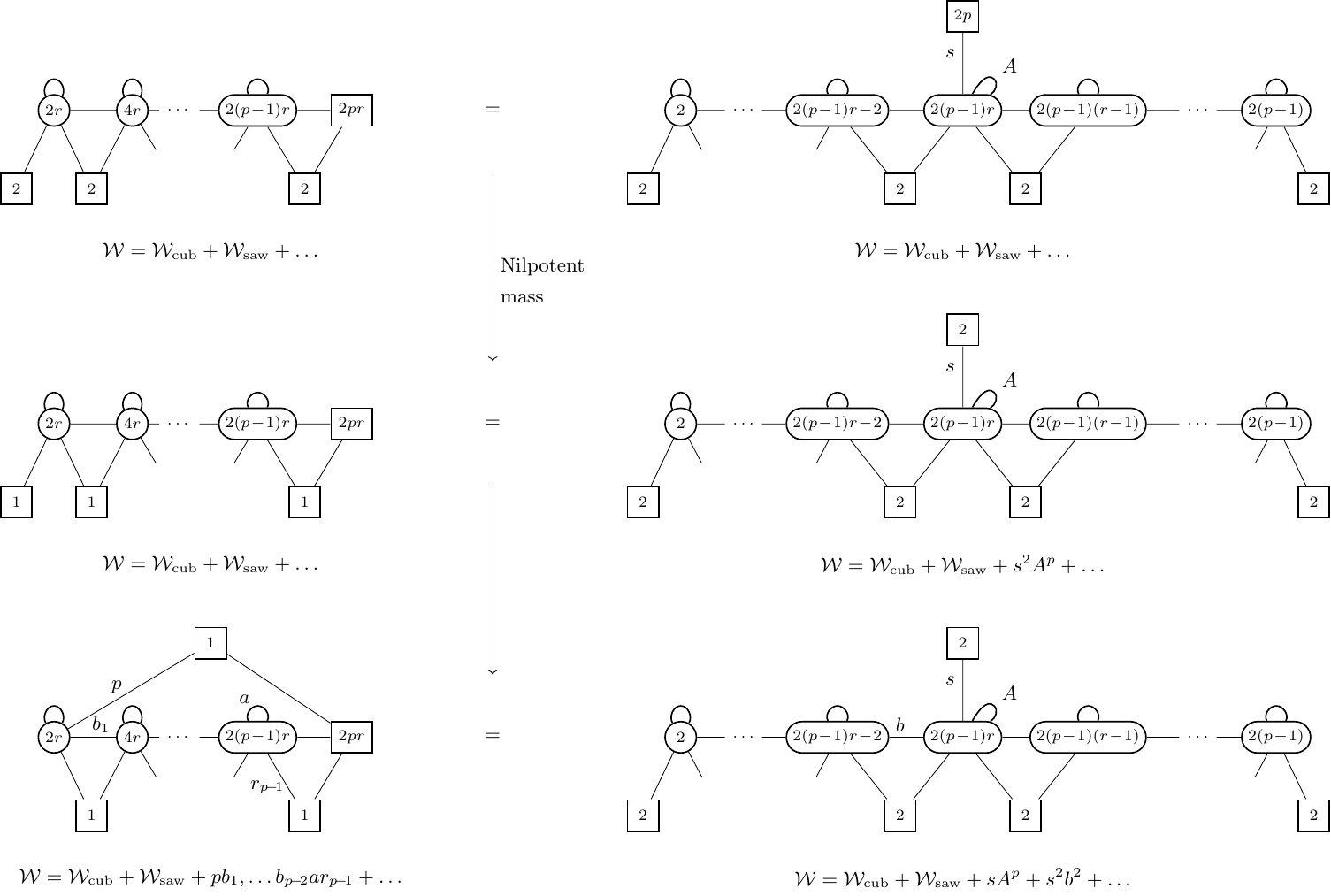}
\end{align}
We start from the 4d $\CN=1$ duality in the first line of the figure above. The global symmetry is $USp(2pr) \times USp(2p)$, where the first group is emergent on the r.h.s.~while the second is emergent on the l.h.s. In the first step, leading to the second line of figure \eqref{fig:4dCpmir_l0}, we perform a nilpotent mass for the antisymmetric operator of $USp(2p)$ with the effect of breaking $USp(2p) \to SU(2)$. On the electric side the mass has the effect of integrating out half of the saw, so that we are left with a single $U(1)$ symmetry acting on the fields of the saw which is the Cartan of the $SU(2)$. On the magnetic side the effect is that we are left with a single flavor in a $s^2 A^p$ superpotential. In the second step, leading to the third line of figure \eqref{fig:4dCpmir_l0}, we turn on $q^2b^2$ on the magnetic side. The dual deformation on the electric side is the long triangle superpotential. After this deformation the $SU(2)$ symmetry becomes unfaithful. All in all we are left with the duality in the third line of figure \eqref{fig:4dCpmir_l0}, which is precisely the duality between the $\mathsf{C}_p[USp(2N)]$ theory is \eqref{fig:4dCp_quiver_l0} and its mirror in the second line of \eqref{fig:4dCpmir}.

\subsection{Derivation of the confining duality}
The confining duality in \eqref{fig:4dCpconfdual} can be derived iteratively with the same strategy as in the 3d case decribed in Section \ref{deriv3dconf}. We will not describe the iterations explicitly since these are identical to those in 3d, but we just present the inductive step and the inductive basis that are needed. 

\paragraph{Inductive step.}
If $N=pr+l > p+1$, the $\mathsf{C}_p[USp(2N)]$ theory ejoys a duality with a $\mathsf{C}_p[USp(2(N-1))]$ theory plus singlets
\begin{equation}\label{fig:4dlemma1}
    \includegraphics[scale=1]{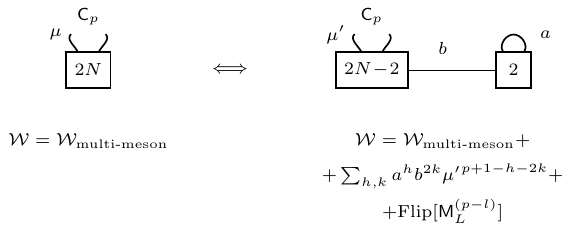}
\end{equation}
where the operators of the $\mathsf{C}_p[USp(2(N-1))]$ theory are primed in order to distinguish them from those of the $\mathsf{C}_p[USp(2N)]$. On the r.h.s.~the superpotential includes all the possible terms between $\mu',b,a$ of degree $p+1$, such that $b$ appears only with even powers. There is also a flipping term for the mesons $\mathsf{M}_L^{(p-l)}$. Notice that for $l=0,1$ this operator does not exist, therefore we do not have any flipping terms and thus there is no extra singlet. 

For $l \neq 0$ the operator map of the duality \eqref{fig:4dlemma1} is
For $l\neq 0$ the map of the operators in the duality \eqref{lemma1} is
\begin{align}\label{eq:lemma1_opmap}
    \mu &\quad \longleftrightarrow \quad \{ \mu',b,a \}  \, \,, \nn \\
    \mathsf{M}_L^{(j)} &\quad \longleftrightarrow \quad {\mathsf{M}'}_L^{(j)} \,, \nn \\
    \mathsf{M}_R^{(j)} &\quad \longleftrightarrow \quad \{ {\mathsf{M}'}_R^{(j)} , F[{\mathsf{M}'}_L^{(p-l)}] \} \,.
\end{align}
For $l= 0$ the situation is simpler, since on the electric side we have no $\mathsf{M}_R^j$ operators and in the dual we have no ${\mathsf{M}'}_L^j$. The operator map is then simply
\begin{align}\label{eq:lemma1_opmap}
    \mu &\quad \longleftrightarrow \quad \{ \mu',b,a \}  \, \,, \nn \\
    \mathsf{M}^{(j)} &\quad \longleftrightarrow \quad  {\mathsf{M}'}_R^{(j)} \,.
\end{align}
Notice that on the r.h.s.~we have $l=p-1$ and looking at \eqref{eq:4dcpmesons} this implies that the set ${\mathsf{M}'}_L^{(j)}$ is empty.

\paragraph{Inductive basis.} The basis of the inductive derivation consists in the confining duality \eqref{fig:4dCpconfdual} for the $\mathsf{C}_p[USp(2(p+1))]$ theory, that is $N=p+1$ or equivalently $r=1$, $l=1$
\begin{align}\label{fig:4dlemma2}
\includegraphics[scale=1.1]{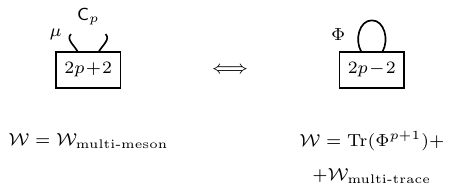}
\end{align}

We do not provide a complete derivation of the two above dualities \eqref{fig:4dlemma1} and \eqref{fig:4dlemma2}. The derivation can be done by repeating the same strategy used in 3d in Section \ref{deriv3dconf}, with the only difference being that we now employ the new 4d mirror theories $\check{\mathsf{C}}_p[USp(2N)]$ and the 4d Intriligator--Pouliot duality \cite{Intriligator:1995ne} instead of the Aharony duality.

\section{Derivation of the 4d $\CN=1$ $USp(2N)$ Intriligator duality}\label{sec:KSI}
In this section we use the confining duality \eqref{fig:4dCpconfdual} for the 4d $\CN=1$ $\mathsf{C}_p[USp(2N)]$ theories to provide a derivation of the version with symplectic gauge group of the Kutasov--Schwimmer duality (IKS), which was proposed by Intriligator in \cite{Intriligator:1995ff}. The derivation only assumes the validity of the confinement property of the $\mathsf{C}_p[USp(2N)]$ theories which, as we discussed in the previous section, is a consequence of the Intriligator--Pouliot (IP) duality. This implies that the IKS duality can be derived only assuming the validity of IP duality. 

The IKS duality relates two gauge theories with symplectic gauge group, $2F$ fundamental chirals and an antisymmetric field as
\begin{align}\label{fig:KSIduality}
    \includegraphics[scale=1]{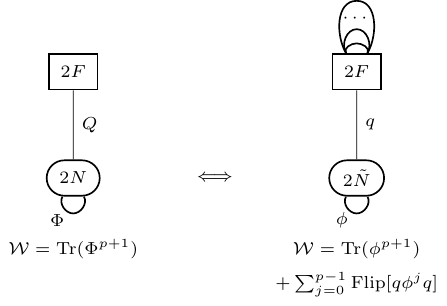}
\end{align}
with $\tilde{N} = p(F-2)-N$. The mapping of the chiral ring generators is
\begin{equation}
\left\{
\begin{tabular}{c}
 $Q \Phi^j Q$  \\
 $\Tr( \Phi^j)$  
\end{tabular}
\right\}
\qquad \longleftrightarrow \qquad
\left\{
\begin{tabular}{c}
 $F[q \phi^{p-1-j} q]$  \\
 $\Tr( \phi^j )$
\end{tabular}
\right\}
\qquad
\begin{tabular}{c}
$ j = 0, 1, \ldots, p-1 $ \,,\\
$ j = 2, 3, \ldots, p-1 $ \,.
\end{tabular}
\end{equation}
Notice that for $p=1$ the antisymmetric chiral is massive and we recover the Intriligator--Pouliot (IP) duality \cite{Intriligator:1995ne}.

The derivation of the IKS duality is very similar to that of the 3d KP duality described in Section \ref{sec:KPproof}. In short, we first deconfine the antisymmetric chiral with a suitable $\mathsf{C}_p[USp(2N)]$ theory, we then apply iteratively the basic IP duality along the quiver and finally we reconfine the resulting $\mathsf{C}_p[USp(2\tilde{N})]$ theory to land on the expected IKS dual frame. Let us next describe this procedure a bit more in detail, focusing on the case $l\neq0$ for concreteness. The derivation for $l=0$ works similarly with just some minor differences.

 We first consider a different version of the confining duality \eqref{fig:4dCpconfdual} in which we flip the flipper $F[l_{p\!-\!l\!-\!1}b_{p\!-\!l\!+\!1}\ldots b_{p\!-\!1}]$ of the $\mathsf{C}_p[USp(2N)]$ theory in figure \eqref{fig:4dCp_quiver_ln0},\footnote{Similarly, for $l=0$ one flips the singlet $r$ in figure \eqref{fig:4dCp_quiver_l0}.} which is in the fundamental representation of the global $USp(2N)$ symmetry. The resulting theory will then confine to a $USp(2N)$ antisymmetric plus a singlet. We name the flipped theory as $\mathsf{C}^{\text{F}}_p[USp(2N)]$ and the new confining duality can be depicted as:
\begin{align}\label{fig:4dCpconfdual_mod}
\begin{tikzpicture}[scale=1.2,every node/.style={scale=1.2},font=\scriptsize]
\begin{scope}[shift={(-1,0)}]
    \node[flavor] (f1) at (-1,0) {$2N$};
    \node (Cp) at (-1,0.8) {$\!\mathsf{C}^{\text{F}}_p\!$};
    \draw[thick] (f1) to[out=120,in=-135] (Cp) to[out=-45,in=60] (f1);
    \node (W1) at (-1,-0.75) {$\CW = \CW_{\text{multi-meson}}$};
\end{scope}
    \node (=) at (0.7,0) {$\Longleftrightarrow$};
\begin{scope}[shift={(1,0)}]
    \node[flavor] (f2) at (2,0) {$2N$};
    \node[flavor] (f3) at (3.5,0) {$1$};
    \draw[-] (f2) -- (f3);
    \node[left] at (1.9,0.6) {$\Phi$};
    \node at (2.75,0.3) {$q$};
    \draw[thick] (f2) to[out=120,in=180] ([yshift=7mm] f2) to[out=0,in=60] (f2);
    \node (W2) at (2.75,-0.75) {$\CW = \Tr(\Phi^{p+1}) +$};
    \node (W22) at (2.75,-1.25) {$ + \sum_{j=2}^{k(l)} \text{Flip}[\Tr(\Phi^j)]+$};
    \node (W23) at (2.75,-1.75) {$ + \CW_{\text{multi-trace}}$};
\end{scope}
\end{tikzpicture}
\end{align}
Notice that the extra fundamental chiral $q$ on the r.h.s.~does not enter in the superpotential. 

To derive the IKS duality we start from the l.h.s.~of figure \eqref{fig:KSIduality} and use the modified confining duality in \eqref{fig:4dCpconfdual_mod} to deconfine the antisymmetric with superpotential $\CW \sim \Tr(\Phi^j)$ and a fundamental chiral. Writing explicitly the quiver of the $\mathsf{C}^{\text{F}}_p[USp(2N)]$ theory, we have
\begin{align}\label{fig:KSIproof1}
    \includegraphics[scale=0.75]{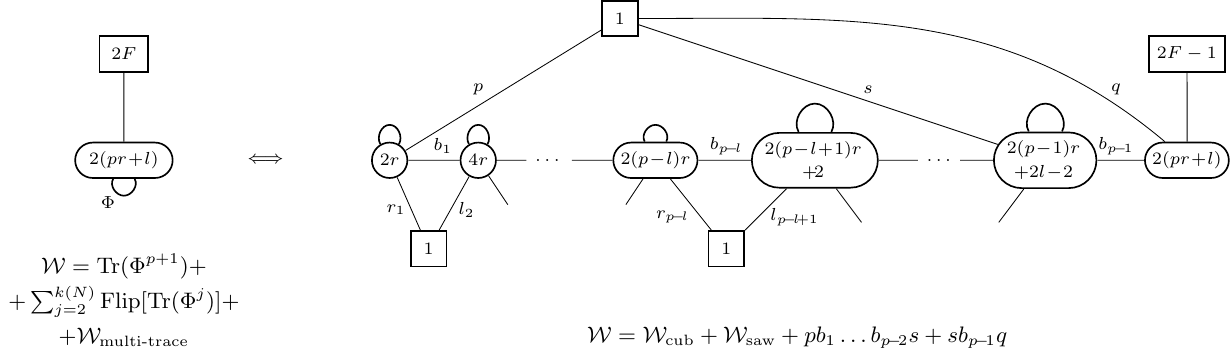}
\end{align}
where on the l.h.s.~we added for convenience a set of flippers for the traces of $\Phi$ from $2$ up to $k(l)$, where we recall that
\begin{align}
    k(l) = \min(l,p-l) \,.
\end{align}
We also added $\CW_{\text{multi-trace}}$ in order to use the deconfining duality in \eqref{fig:4dCpconfdual_mod}. On the r.h.s.~of \eqref{fig:KSIproof1} we are not keeping track of all the exactly marginal deformations corresponding to $\CW_{\text{multi-meson}}$ to make the discussion simpler. We will comment about them at the end of the subsection. 

At this point, we notice that on the r.h.s.~\eqref{fig:KSIproof1} the rightmost gauge node does not have an antisymmetric field, therefore we can apply the IP duality. Also, some new fields and superpotential terms are created. We do not give all the details, which can be recovered by studying in detail the precise effect of IP duality. The result is the following theory:
\begin{align}
    \includegraphics[scale=0.8]{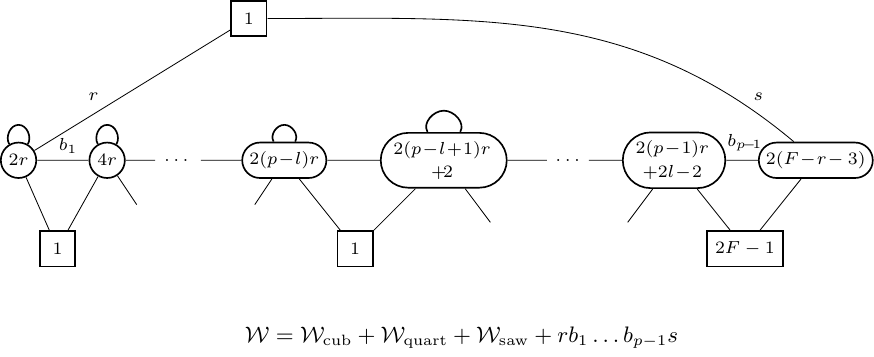}
\end{align}
The main effect of the dualization is that the antisymmetric and one fundamental of the adjacent node are removed. We then apply the IP duality on the second-to-last gauge node to obtain
\begin{align}
    \includegraphics[scale=0.8]{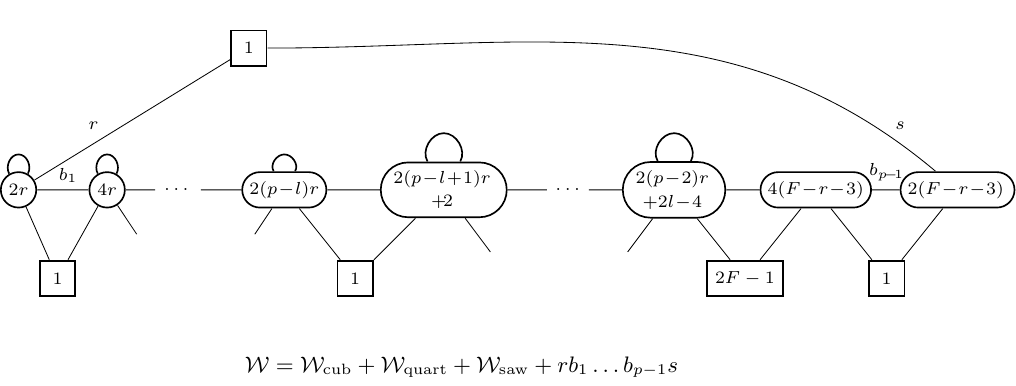}
\end{align}

We keep iterating the IP duality to reach finally
\begin{align}\label{fig:KSIproof4}
    \includegraphics[scale=0.8]{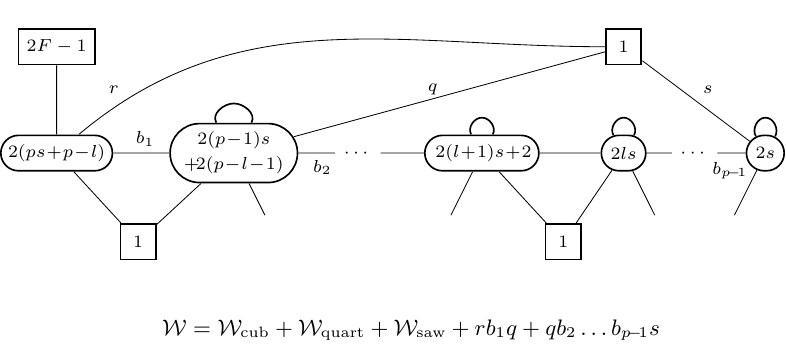}
\end{align}
Where we have defined $s=F-r-3$. We can now recognize the tail to be the modified $\mathsf{C}^{\text{F}}_p[USp(2\tilde{N})]$ theory with $N=p(F-r-3)+(p-l)$. Confining it using the duality in \eqref{fig:4dCpconfdual_mod} we obtain 
\begin{align}\label{fig:KSIproof5}
    \includegraphics[]{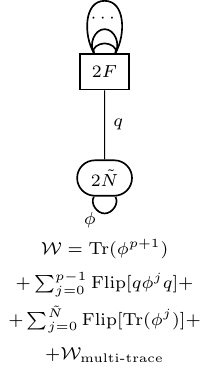}
\end{align}
By keeping track of all the superpotential terms involving the $\mathsf{M}_{L/R}^{(j)}$ operators, we obtain the theory in \eqref{fig:KSIproof4} with all the exactly marginal deformations turned on. Therefore, the confinement generates precisely the theory in \eqref{fig:KSIproof5} including the exactly marginal interactions $\mathcal{W}_{\text{multi-trace}}$. Moreover, the flippers produced at this last step map directly to those we introduced in \eqref{fig:KSIproof1}, so we can just remove them from both sides to recover the IKS duality \eqref{fig:KSIduality} on a generic point of the conformal manifold.

\section{Conclusions and outlook}
\label{conclusions}

In this paper we provided a derivation of some Kutasov--Schwimmer-like (KS-like) dualities which involve a rank-2 field and a degree-$(p+1)$ superpotential, specifically the 3d $\mathcal{N}=2$ Kim--Park duality for unitary gauge group and adjoint matter, and the 4d $\mathcal{N}=1$ Intriligator duality for symplectic gauge group and antisymmetric matter. The derivation uses the technique of the deconfinement, which consists of trading the rank-2 field for an auxiliary quiver gauge theory that can then be further dualized to the desired dual frame.

We proceeded in two steps. We first introduced two new classes of linear quivers, the 3d $\mathcal{N}=2$ $\CC_p[SU(N)]$ and the 4d $\mathcal{N}=1$ $\mathsf{C}_p[USp(2N)]$ theories, and showed that they respectively confine into an $SU(N)$ adjoint and a $USp(2N)$ antisymmetric chiral with degree-$(p+1)$ superpotential. We then used these confining dualities to deconfine the rank-$2$ fields in the KS-like dualities. This allowed us to obtain quiver gauge theories that we could further dualize with known Seiberg-like dualities, which only involve fields in the fundamental representation, so to obtain the expected dual gauge group connected to an additional linear quiver. Finally, such dual quiver could be traded back for the rank-2 field using again our new confining dualities, so to find the desired KS-like dual with the correct superpotential and flipping fields.

All the steps in our derivation assume the validity of only some fundamental Seiberg-like duality, which are the Intriligator--Pouliot duality in 4d and the Aharony duality in 3d (since the variants with monopole superpotential can be obtained as deformations of it). As a byproduct, we are thus able to provide a proof, which was previously missing, of the integral identities of partition functions and supersymmetric indices of the KS-like dualities by only assuming the validity of those for the Seiberg-like dualities, which instead have been already proven in the math literature.

Our $\CC_p[SU(N)]$ theories and their confining duality, which played a crucial role in the derivation of the KS-like dualities, are deeply connected to  the 4d $\mathcal{N}=2$ $D_p[SU(N)]$ SCFTs. Indeed, for $\gcd(p,N)=1$ we showed that the quiver of the former can be obtained as a monopole superpotential deformation of the 3d $\mathcal{N}=4$ quiver for the circle compactification of the latter. Such monopole deformation is expected to be related to the CB deformation of $D_p[SU(N)]$ that makes it flow to an $SU(N)$ adjoint chiral with degree-$(p+1)$ superpotential. Moreover, we showed that a similar relationship holds for $\gcd(p,N)\neq1$ and $N\neq pr$ as well, but it involves a further deformation in 3d that Higgses the special unitary nodes in the quiver for the 3d reduction of $D_p[SU(N)]$ to the unitary subgroups of the same rank.

There are some details about the relationship between $\CC_p[SU(N)]$ and $D_p[SU(N)]$ that deserve further investigation, as we have already discussed at the end of Section \ref{sec:DpSUN}. First of all, it would be interesting to understand whether the additional deformations needed in the case $\gcd(p,N)\neq1$ and $N\neq pr$ can be uplifted to 4d. This would give a deformation of $D_p[SU(N)]$ that makes it flow to the $SU(N)$ adjoint also in such cases. Moreover, we are still missing a way to relate $\CC_p[SU(N)]$ and $D_p[SU(N)]$ for $N=pr$. Finally, in our analysis we only focused on the quiver content of the theories and neglected possible gauge singlet fields. It is still not clear how the singlets in the confining duality of $\CC_p[SU(N)]$ that flip the traces of powers of the adjoint field and the twisted hypers that can show up in the 3d reduction of $D_p[SU(N)]$ are related. 

We would also like to mention that in \cite{Elitzur:1997fh,Elitzur:1997hc}, building on the earlier work \cite{Hanany:1996ie}, it was argued that a 4d $\CN=1$ $SU(N)$ gauge theory with $F$ flavors and an adjoint $\Phi$ with $\CW = \Tr(\Phi^{p+1})$ admits a brane engineering in Type IIA as $N$ D4-branes suspended between a stack of $p$ coincident NS5-branes and a single NS5'-brane. The branes are taken so that the D4's extend in the $(01236)$ directions of the 10d spacetime, while the NS5's fill $(012345)$ and the NS5' fills $(012389)$. As usual, flavors are insertion of D6-branes filling $(0123789)$. With this choice the brane setup always preserves 4 supercharges. 
One may also consider a similar setup in Type IIB engineering a 3d $\CN=2$ theory with $U(N)$ gauge group, $F$ flavors and an adjoint chiral $\Phi$ with $\CW = \Tr(\Phi^{p+1})$, where the D4-branes are replaced by D3's and the D6's by D5's. From this perspective the $\CC_p[SU(N)]$ theories, that we recall are always linear quivers with $p-1$ gauge nodes, seem a candidate to describe the brane system with a specific choice on how many D3 terminate on each NS5 when the latter are pulled apart. It would be interesting to investigate more this field theory interpretation of the brane setup, in particular how the monopole superpotential that is present in the $\CC_p[SU(N)]$ theory should arise. 

A different possible line of future investigation would be to search for more confining dualities for matter in tensor representation with a polynomial superpotential. For example, one can still focus on $SU(N)$ or $USp(2N)$ but consider more general rank-2 representations, or take other gauge groups like $SO(2N)$. In fact in \cite{Bajeot:2023gyl} a confining duality for orthogonal gauge group with matter in the symmetric representation and a cubic superpotential has already been found, and it would be interesting to extended it to a superpotential of higher degree. 
These new results could shed light on the vast landscape of Kutasov--Schwimmer-like dualities that exist for $SU/USp/SO$ gauge theories with matter in the adjoint/symmetric/antisymmetric representation \cite{Intriligator:1995ax,Spiridonov:2009za}. 

Alternatively, one can also consider theories with multiple copies of the same tensor field. For example, the 4d $\CN=1$ $SU(N)$ SQCD with two adjoint chirals has been studied in \cite{Intriligator:2003mi} (see also \cite{Intriligator:2016sgx}), where it has been shown that this admits an ADE classification depending on the form of the superpotential. For some of these theories dual frames are known, specifically for $A_n$ and $D_n$ type superpotential they have been discussed in \cite{Intriligator:2003mi}, while for $E_7$ in \cite{Kutasov:2014yqa}. It would be interesting to understand whether the deconfinement technique can also be applied to the two adjoint SQCD, to not only derive the known dualities but also find possible unknown ones.

Another fruitful approach for investigating the dynamics of 4d $\mathcal{N}=1$ theories consists of realizing them as compactifications of 6d SCFTs on Riemann surfaces with fluxes for global symmetries (see e.g.~the review \cite{Razamat:2022gpm} and references therein). Various infrared dualities have been derived from this perspective. Examples include the confining Seiberg duality for $SU(2)$ with six flavors and its generalization to $USp(2N)$ with one antisymmetric and six flavors from compactifications of the 6d E-string theory \cite{Kim:2017toz,Pasquetti:2019hxf,Razamat:2020bix}, and the Intriligator--Pouliot duality from the compactification of the 6d $(D,D)$ conformal matter \cite{Kim:2018bpg,Kim:2018lfo,Razamat:2020bix}. Similarly one can derive 3d $\mathcal{N}=2$ dualities such as the Aharony duality from compactification of 5d SCFTs on Riemann surfaces with fluxes \cite{Sacchi:2021afk,Sacchi:2021wvg,Sacchi:2023rtp,Sacchi:2023omn}. It would be interesting to find a similar higher-dimensional origin also for the KS-like dualities such as those that we studied in this paper.

\acknowledgments
We would like to thank Antonio Amariti, Simone Giacomelli,  Craig Lawrie and  Simone Rota for discussions and  comments on the draft. 
SB is partially supported by the INFN “Iniziativa Specifica GAST”.
SB and SP are partially supported by the MUR-PRIN grant No. 2022NY2MXY. MS is partially supported by the ERC Consolidator Grant \#864828 “Algebraic Foundations of Supersymmetric Quantum Field Theory (SCFTAlg)” and by the Simons Collaboration for the Nonperturbative Bootstrap under grant \#494786 from the Simons Foundation. 

\appendix

\section{Review of Aharony duality and variants with monopole superpotentials}\label{appendix}

\subsection{Statement of the dualities}
\label{app:aharony}

The Aharony duality \cite{Aharony:1997gp} relates a 3d $\CN=2$ $U(N)$ gauge theory with $F$ flavors to a $U(F-N)$ theory with $F$ flavors and  $F^2+2$ singlets flipping the meson matrix and the two fundamental monopoles of the dual theory.
We can depict it in the following way:
\begin{align}\label{fig:aharony}
\includegraphics[scale=1.1]{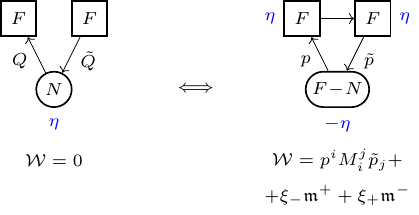}
\end{align}
The global symmetry of the dual theories is
\begin{equation}
    SU(F)_U\times SU(F)_V\times U(1)_m\times U(1)_\eta\,,
\end{equation}
where $U(1)_m$ is the axial symmetry under which $Q,\tilde{Q}$ have charge $+1$, while $U(1)_\eta$ is the topological symmetry. 
The map between the gauge invariant operators is
\begin{align}
    Q^i\tilde{Q}_j \quad &\leftrightarrow \quad M_j^i\,, \nn \\
    \M^+ \quad &\leftrightarrow \quad \xi_+\,, \nn \\
    \M^- \quad &\leftrightarrow \quad \xi_- \,.
\end{align}

We can consider the global symmetries in the duality as $U(F)$ rather than $SU(F)$. This is useful when the duality is used inside a bigger quiver, where the $U(F)$, or a subgroup thereof, is gauged. We encountered many examples of this throughout Section \ref{sec3dandDp}. In \eqref{fig:aharony} we are writing in blue the FI parameter associated to each gauge node. The blue parameters close to the flavor nodes then indicate that there is a BF coupling between the $U(1)_\eta$ topological symmetry and the $U(1) \subset U(F)$ flavor symmetries.

The partition function identity associated to the duality is
\begin{align}
   & \frac{1}{N!} \int \prod_{j=1}^N dZ_j e^{2\pi i \eta \sum_{j=1}^N Z_j}
   \frac{ \prod_{j=1}^N \prod_{a=1}^F s_b(\tfrac{iQ}{2} - \u_a \pm ( Z_j + M_a ) ) }
   { \prod_{i<j}^N s_b( \tfrac{iQ}{2} \pm (Z_j - Z_k)) } = \nn \\
   & = e^{- 2\pi i \eta \sum_{j=1}^F M_a}
   s_b(\tfrac{iQ}{2} - \tfrac{ iQ(F-N+1) - 2\sum_{a=1}^{F}\u_a }{2} \pm \eta ) \nn \\
   & \prod_{j,k=1}^F s_b( \tfrac{iQ}{2} - (\u_a + \u_b - M_a + M_b)) \nn \\
   & \frac{1}{(F-N)!} \int \prod_{j=1}^{F-N} dZ_j e^{-2\pi i \eta \sum_{j=1}^N Z_j}
   \frac{ \prod_{j=1}^{F-N} \prod_{a=1}^F s_b( \u_a \pm (Z_j - M_a)) }
   {\prod_{j<k}^{F-N} s_b( \tfrac{iQ}{2} \pm (Z_j - Z_k) )} \,,
\end{align}
where $\Vec{M}$ and $\Vec{\u}$ parameterize the $SU(F)_U \times SU(F)_V \times U(1)_m$ symmetries as
\begin{align}
    & U_a = \u_a - M_a \,, \nn \\
    & V_a = - \u_a - M_a \,, \nn \\
    & m = \sum_{a=1}^F (U_a - V_a) \,.
\end{align}
We can see the aforementioned BF coupling as encoded in the exponential prefactor on the r.h.s.~of the equality. This trivializes if we impose that $\sum_{a=1}^F (U_a + V_a) = 0$, which can be achieved using the freedom of shifting the gauge parameters $Z_j$ since the actual faithful flavor symmetry is $SU(F)_U\times SU(F)_V$. However, this cannot be done when the duality is applied inside a quiver where the diagonal $U(F)_M$, or a subgroup thereof, is gauged. In such a case, the BF coupling produced by the duality becomes a shift of the FI of the gauge nodes adjacent to the one that was dualized, as it has been discussed in detail in \cite{Pasquetti:2019uop}. 

One can also deform the duality by introducing one or two monopoles in the superpotential. As shown in \cite{Benini:2017dud}, the duality for the $U(N)$ SQCD with a single monopole superpotential is
\begin{align}\label{fig:1monop}
    \includegraphics[scale=1.1]{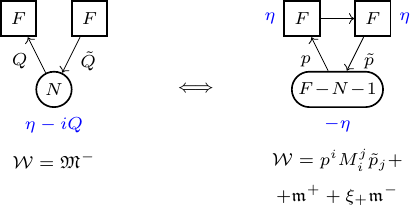}
\end{align}
The effect of the linear monopole superpotential is to break the $U(1)_m \times U(1)_{\eta} \to U(1)_a$, so that now the global symmetry group is
\begin{align}
    SU(F)_U \times SU(F)_V \times U(1)_a \,.
\end{align}
The map between the gauge invariant operators reads
\begin{align}
    Q^i\tilde{Q}_j \quad &\leftrightarrow \quad M_j^i \nn \\
    \M^+ \quad &\leftrightarrow \quad \xi_+ 
\end{align}
The corresponding partition function identity is
\begin{align}
   & \frac{1}{N!} \int \prod_{j=1}^N dZ_j e^{2\pi i (\eta - i\frac{Q}{2}) \sum_{j=1}^N Z_j}
   \frac{ \prod_{j=1}^N \prod_{a=1}^F s_b(\tfrac{iQ}{2} - \u_a \pm (Z_j + M_a) ) }
   { \prod_{i<j}^N s_b( \tfrac{iQ}{2} \pm (Z_j - Z_k)) } = \nn \\
   & = e^{ -2i\pi (\sum_{a=1}^F M_a \u_a + (\eta - i\frac{Q}{2}) \sum_{a=1}^F M_a ) }
   s_b(\tfrac{iQ}{2} - 2\eta ) \nn \\
   & \prod_{j,k=1}^F s_b( \tfrac{iQ}{2} - ( \u_a + \u_b - M_a + M_b) ) \nn \\
   & \frac{1}{(F-N-1)!} \int \prod_{j=1}^{F-N-1} dZ_j e^{-2\pi i \eta \sum_{j=1}^N Z_j}
   \frac{ \prod_{j=1}^{F-N-1} \prod_{a=1}^F s_b( \u_a \pm (Z_j - M_a) ) }
   {\prod_{j<k}^{F-N-1} s_b( \tfrac{iQ}{2} \pm (Z_j - Z_k) )} \,,
\end{align}
with the constraint imposed by the superpotential
\begin{align}
    \eta + \sum_{a=1}^F \u_a = i\frac{Q}{2}(F-N) \,.
\end{align}

We can further deform the duality by introducing a second monopole superpotential term. We obtain the following duality:
\begin{align}\label{fig:2monop}
    \includegraphics[scale=1.1]{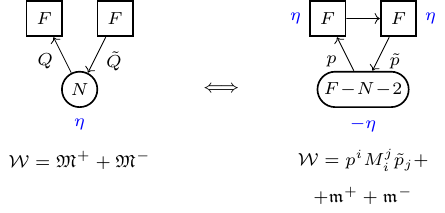}
\end{align}
The superpotential now breaks also $U(1)_a$ so that the global symmetry group is
\begin{align}
    SU(F)_U \times SU(F)_V \,.
\end{align}
There is only one gauge invariant operator, the meson matrix $Q^i \tilde{Q}_j$, which is again mapped to $M^i_j$. 
The corresponding partition function identity is
\begin{align}
   & \frac{1}{N!} \int \prod_{j=1}^N dZ_j
   \frac{ \prod_{j=1}^N \prod_{a=1}^F s_b(\tfrac{iQ}{2} - \u_a \pm (Z_j + M_a) ) }
   { \prod_{i<j}^N s_b( \tfrac{iQ}{2} \pm (Z_j - Z_k)) } = \nn \\
   & = \prod_{j,k=1}^F s_b( \tfrac{iQ}{2} - ( \u_a + \u_b - M_a + M_b) ) \nn \\
   & \frac{1}{(F-N-2)!} \int \prod_{j=1}^{F-N-2} dZ_j e^{-2\pi i \eta \sum_{j=1}^N Z_j}
   \frac{ \prod_{j=1}^{F-N-2} \prod_{a=1}^F s_b( \u_a \pm (Z_j - M_a) ) }
   {\prod_{j<k}^{F-N-2} s_b( \tfrac{iQ}{2} \pm (Z_j - Z_k) )} \,,
\end{align}
with the constraint imposed by the superpotential:
\begin{align}
    \sum_{a=1}^F \u_a = i\frac{Q}{2}(F-N-1) \,.
\end{align}

As a final comment we discuss below the details regarding the confining case of the three dualities.
\begin{itemize}
    \item For the Aharony duality in \eqref{fig:aharony}, when $F=N$ the magnetic theory is a WZ model of $N^2$ chirals $M^i_j$ and two chirals $\xi_+$ and $\xi_-$ with $\CW = \xi_+\xi_- \det(M)$. 
    
    \item For the one-monopole duality in \eqref{fig:1monop}, when $F=N+1$ the magnetic theory is a WZ model of $(N+1)^2$ chirals $M^i_j$ and a chiral $\xi_+$ with $\CW = \xi_+ \det(M)$.
    
    \item For the two-monopole duality in \eqref{fig:2monop}, when $F=N+2$ the magnetic theory is a WZ model of $(N+2)^2$ chirals $M^i_j$ with $\CW = \det(M)$.
\end{itemize}

\subsection{Mapping of extended monopoles under Aharony duality}\label{app:monopoles}

We now briefly review the results of \cite{Benvenuti:2020wpc}, regarding how the extended monopoles of a linear quiver theory map under Aharony duality, which we extensively used in sections \ref{sec3dandDp} and \ref{sec:KPproof}. 

The setup is a generic linear quiver with a sequence of three unitary gauge nodes with generic ranks $n_1,n_2,n_3$, connected by bifundamental chirals
\begin{align}
    \includegraphics[]{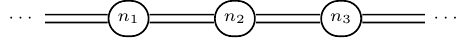}
\end{align}
We want to study how the monopoles of the theory are mapped after we apply the Aharony duality on the $U(n_2)$ central node. We consider monopoles with minimal flux $\M^{\ldots \bullet \bullet \bullet \ldots}$, where each blank entry ``$\bullet$" can either be $\pm1$ or $0$. The monopoles of the theory after the dualization will be primed in order to distinguish them with those of the theory before the dualization. The map works as follows:
\begin{itemize}
    \item The monopole charged only under the $U(n_2)$ maps to singlets flipping the monopoles of the same node after the duality
    \begin{align}
        \M^{\ldots 0 \pm 0 \ldots} \quad \leftrightarrow \quad F[\M'^{\ldots 0 \pm 0 \ldots}] \,.
    \end{align}
    
    \item The monopoles charged under only the $U(n_1)$ or $U(n_3)$ nodes \emph{extend} to the central node, which means that they map as
    \begin{align}
        \M^{\ldots \pm 0 0 \ldots} &\quad \leftrightarrow \quad \M'^{\ldots \pm \pm 0 \ldots} \,, \nn \\
        \M^{\ldots 0 0 \pm \ldots} &\quad \leftrightarrow \quad \M'^{\ldots 0 \pm \pm \ldots} \,.
    \end{align}
    
    \item The monopoles charged under both $U(n_1)$ and $U(n_2)$, or similarly under $U(n_3)$ and $U(n_2)$, \emph{shorten}, which means that they map as
    \begin{align}
        \M^{\ldots \pm \pm 0 \ldots} &\quad \leftrightarrow \quad \M'^{\ldots \pm 0 0 \ldots} \,, \nn \\
        \M^{\ldots 0 \pm \pm \ldots} &\quad \leftrightarrow \quad \M'^{\ldots 0 0 \pm \ldots} \,.
    \end{align}

    \item The monopoles charged under all the three nodes remain the same
    \begin{align}
        \M^{\ldots \pm \pm \pm \ldots} \quad \leftrightarrow \quad \M'^{\ldots \pm \pm \pm \ldots} \,.
    \end{align}
\end{itemize}

\bibliographystyle{JHEP}
\bibliography{ref}

\providecommand{\href}[2]{#2}\begingroup\raggedright\begin{thebibliography}{10}

\bibitem{Seiberg:1994rs}
N.~Seiberg and E.~Witten, \emph{{Electric - magnetic duality, monopole condensation, and confinement in N=2 supersymmetric Yang-Mills theory}}, \href{https://doi.org/10.1016/0550-3213(94)90124-4}{\emph{Nucl. Phys. B} {\bfseries 426} (1994) 19} [\href{https://arxiv.org/abs/hep-th/9407087}{{\ttfamily hep-th/9407087}}].

\bibitem{1Kutasov_1995}
D.~Kutasov and A.~Schwimmer, \emph{On duality in supersymmetric yang-mills theory}, \href{https://doi.org/10.1016/0370-2693(95)00676-c}{\emph{Physics Letters B} {\bfseries 354} (1995) 315–321}.

\bibitem{Kutasov_1995}
D.~Kutasov, \emph{A comment on duality in n = 1 supersymmetric non-abelian gauge theories}, \href{https://doi.org/10.1016/0370-2693(95)00392-x}{\emph{Physics Letters B} {\bfseries 351} (1995) 230–234}.

\bibitem{Intriligator:1995ff}
K.A.~Intriligator, \emph{{New RG fixed points and duality in supersymmetric SP(N(c)) and SO(N(c)) gauge theories}}, \href{https://doi.org/10.1016/0550-3213(95)00296-5}{\emph{Nucl. Phys. B} {\bfseries 448} (1995) 187} [\href{https://arxiv.org/abs/hep-th/9505051}{{\ttfamily hep-th/9505051}}].

\bibitem{Kim_2013}
H.~Kim and J.~Park, \emph{Aharony dualities for 3d theories with adjoint matter}, \href{https://doi.org/10.1007/jhep06(2013)106}{\emph{Journal of High Energy Physics} {\bfseries 2013} (2013) }.

\bibitem{Park_2013}
J.~Park and K.-J.~Park, \emph{Seiberg-like dualities for 3d $ \mathcal{N}=2 $ theories with su(n) gauge group}, \href{https://doi.org/10.1007/jhep10(2013)198}{\emph{Journal of High Energy Physics} {\bfseries 2013} (2013) }.

\bibitem{Kinney:2005ej}
J.~Kinney, J.M.~Maldacena, S.~Minwalla and S.~Raju, \emph{{An Index for 4 dimensional super conformal theories}}, \href{https://doi.org/10.1007/s00220-007-0258-7}{\emph{Commun. Math. Phys.} {\bfseries 275} (2007) 209} [\href{https://arxiv.org/abs/hep-th/0510251}{{\ttfamily hep-th/0510251}}].

\bibitem{Romelsberger:2005eg}
C.~Romelsberger, \emph{{Counting chiral primaries in N = 1, d=4 superconformal field theories}}, \href{https://doi.org/10.1016/j.nuclphysb.2006.03.037}{\emph{Nucl. Phys. B} {\bfseries 747} (2006) 329} [\href{https://arxiv.org/abs/hep-th/0510060}{{\ttfamily hep-th/0510060}}].

\bibitem{Dolan:2008qi}
F.A.~Dolan and H.~Osborn, \emph{{Applications of the Superconformal Index for Protected Operators and q-Hypergeometric Identities to N=1 Dual Theories}}, \href{https://doi.org/10.1016/j.nuclphysb.2009.01.028}{\emph{Nucl. Phys. B} {\bfseries 818} (2009) 137} [\href{https://arxiv.org/abs/0801.4947}{{\ttfamily 0801.4947}}].

\bibitem{Bhattacharya:2008zy}
J.~Bhattacharya, S.~Bhattacharyya, S.~Minwalla and S.~Raju, \emph{{Indices for Superconformal Field Theories in 3,5 and 6 Dimensions}}, \href{https://doi.org/10.1088/1126-6708/2008/02/064}{\emph{JHEP} {\bfseries 02} (2008) 064} [\href{https://arxiv.org/abs/0801.1435}{{\ttfamily 0801.1435}}].

\bibitem{Bhattacharya:2008bja}
J.~Bhattacharya and S.~Minwalla, \emph{{Superconformal Indices for N = 6 Chern Simons Theories}}, \href{https://doi.org/10.1088/1126-6708/2009/01/014}{\emph{JHEP} {\bfseries 01} (2009) 014} [\href{https://arxiv.org/abs/0806.3251}{{\ttfamily 0806.3251}}].

\bibitem{Kim:2009wb}
S.~Kim, \emph{{The Complete superconformal index for N=6 Chern-Simons theory}}, \href{https://doi.org/10.1016/j.nuclphysb.2009.06.025}{\emph{Nucl. Phys. B} {\bfseries 821} (2009) 241} [\href{https://arxiv.org/abs/0903.4172}{{\ttfamily 0903.4172}}].

\bibitem{Imamura:2011su}
Y.~Imamura and S.~Yokoyama, \emph{{Index for three dimensional superconformal field theories with general R-charge assignments}}, \href{https://doi.org/10.1007/JHEP04(2011)007}{\emph{JHEP} {\bfseries 04} (2011) 007} [\href{https://arxiv.org/abs/1101.0557}{{\ttfamily 1101.0557}}].

\bibitem{Kapustin:2011jm}
A.~Kapustin and B.~Willett, \emph{{Generalized Superconformal Index for Three Dimensional Field Theories}},  \href{https://arxiv.org/abs/1106.2484}{{\ttfamily 1106.2484}}.

\bibitem{Dimofte:2011py}
T.~Dimofte, D.~Gaiotto and S.~Gukov, \emph{{3-Manifolds and 3d Indices}}, \href{https://doi.org/10.4310/ATMP.2013.v17.n5.a3}{\emph{Adv. Theor. Math. Phys.} {\bfseries 17} (2013) 975} [\href{https://arxiv.org/abs/1112.5179}{{\ttfamily 1112.5179}}].

\bibitem{Csaki:1996zb}
C.~Csaki, M.~Schmaltz and W.~Skiba, \emph{{Confinement in N=1 SUSY gauge theories and model building tools}}, \href{https://doi.org/10.1103/PhysRevD.55.7840}{\emph{Phys. Rev. D} {\bfseries 55} (1997) 7840} [\href{https://arxiv.org/abs/hep-th/9612207}{{\ttfamily hep-th/9612207}}].

\bibitem{Berkooz:1995km}
M.~Berkooz, \emph{{The Dual of supersymmetric SU(2k) with an antisymmetric tensor and composite dualities}}, \href{https://doi.org/10.1016/0550-3213(95)00400-M}{\emph{Nucl. Phys. B} {\bfseries 452} (1995) 513} [\href{https://arxiv.org/abs/hep-th/9505067}{{\ttfamily hep-th/9505067}}].

\bibitem{Pouliot:1995me}
P.~Pouliot, \emph{{Duality in SUSY SU(N) with an antisymmetric tensor}}, \href{https://doi.org/10.1016/0370-2693(95)01427-6}{\emph{Phys. Lett. B} {\bfseries 367} (1996) 151} [\href{https://arxiv.org/abs/hep-th/9510148}{{\ttfamily hep-th/9510148}}].

\bibitem{Luty:1996cg}
M.A.~Luty, M.~Schmaltz and J.~Terning, \emph{{A Sequence of duals for Sp(2N) supersymmetric gauge theories with adjoint matter}}, \href{https://doi.org/10.1103/PhysRevD.54.7815}{\emph{Phys. Rev. D} {\bfseries 54} (1996) 7815} [\href{https://arxiv.org/abs/hep-th/9603034}{{\ttfamily hep-th/9603034}}].

\bibitem{Pasquetti:2019uop}
S.~Pasquetti and M.~Sacchi, \emph{{From 3$d$ dualities to 2$d$ free field correlators and back}}, \href{https://doi.org/10.1007/JHEP11(2019)081}{\emph{JHEP} {\bfseries 11} (2019) 081} [\href{https://arxiv.org/abs/1903.10817}{{\ttfamily 1903.10817}}].

\bibitem{Sacchi:2020pet}
M.~Sacchi, \emph{{New 2d $ \mathcal{N} $ = (0, 2) dualities from four dimensions}}, \href{https://doi.org/10.1007/JHEP12(2020)009}{\emph{JHEP} {\bfseries 12} (2020) 009} [\href{https://arxiv.org/abs/2004.13672}{{\ttfamily 2004.13672}}].

\bibitem{Benvenuti:2020gvy}
S.~Benvenuti, I.~Garozzo and G.~Lo~Monaco, \emph{{Sequential deconfinement in 3d $ \mathcal{N} $ = 2 gauge theories}}, \href{https://doi.org/10.1007/JHEP07(2021)191}{\emph{JHEP} {\bfseries 07} (2021) 191} [\href{https://arxiv.org/abs/2012.09773}{{\ttfamily 2012.09773}}].

\bibitem{Etxebarria:2021lmq}
I.n.G.~Etxebarria, B.~Heidenreich, M.~Lotito and A.K.~Sorout, \emph{{Deconfining $ \mathcal{N} $ = 2 SCFTs or the art of brane bending}}, \href{https://doi.org/10.1007/JHEP03(2022)140}{\emph{JHEP} {\bfseries 03} (2022) 140} [\href{https://arxiv.org/abs/2111.08022}{{\ttfamily 2111.08022}}].

\bibitem{Benvenuti:2021nwt}
S.~Benvenuti and G.~Lo~Monaco, \emph{{A toolkit for ortho-symplectic dualities}}, \href{https://doi.org/10.1007/JHEP09(2023)002}{\emph{JHEP} {\bfseries 09} (2023) 002} [\href{https://arxiv.org/abs/2112.12154}{{\ttfamily 2112.12154}}].

\bibitem{Bajeot:2022kwt}
S.~Bajeot and S.~Benvenuti, \emph{{S-confinements from deconfinements}}, \href{https://doi.org/10.1007/JHEP11(2022)071}{\emph{JHEP} {\bfseries 11} (2022) 071} [\href{https://arxiv.org/abs/2201.11049}{{\ttfamily 2201.11049}}].

\bibitem{Bottini:2022vpy}
L.E.~Bottini, C.~Hwang, S.~Pasquetti and M.~Sacchi, \emph{{Dualities from dualities: the sequential deconfinement technique}}, \href{https://doi.org/10.1007/JHEP05(2022)069}{\emph{JHEP} {\bfseries 05} (2022) 069} [\href{https://arxiv.org/abs/2201.11090}{{\ttfamily 2201.11090}}].

\bibitem{Bajeot:2022lah}
S.~Bajeot and S.~Benvenuti, \emph{{Sequential deconfinement and self-dualities in 4d$ \mathcal{N} $ = 1 gauge theories}}, \href{https://doi.org/10.1007/JHEP10(2022)007}{\emph{JHEP} {\bfseries 10} (2022) 007} [\href{https://arxiv.org/abs/2206.11364}{{\ttfamily 2206.11364}}].

\bibitem{Bajeot:2022wmu}
S.~Bajeot and S.~Benvenuti, \emph{{4d N=1 dualities from 5d dualities}},  \href{https://arxiv.org/abs/2212.11217}{{\ttfamily 2212.11217}}.

\bibitem{Bajeot:2023gyl}
S.~Bajeot, S.~Benvenuti and M.~Sacchi, \emph{{S-confining gauge theories and supersymmetry enhancements}}, \href{https://doi.org/10.1007/JHEP08(2023)042}{\emph{JHEP} {\bfseries 08} (2023) 042} [\href{https://arxiv.org/abs/2305.10274}{{\ttfamily 2305.10274}}].

\bibitem{Amariti:2023wts}
A.~Amariti, F.~Mantegazza and D.~Morgante, \emph{{Sporadic dualities from tensor deconfinement}}, \href{https://doi.org/10.1007/JHEP05(2024)188}{\emph{JHEP} {\bfseries 05} (2024) 188} [\href{https://arxiv.org/abs/2307.14146}{{\ttfamily 2307.14146}}].

\bibitem{Amariti:2024sde}
A.~Amariti and F.~Mantegazza, \emph{{A new 4d $ \mathcal{N} $ = 1 duality from the superconformal index}}, \href{https://doi.org/10.1007/JHEP06(2024)206}{\emph{JHEP} {\bfseries 06} (2024) 206} [\href{https://arxiv.org/abs/2402.00609}{{\ttfamily 2402.00609}}].

\bibitem{Amariti:2024gco}
A.~Amariti and F.~Mantegazza, \emph{{Confinement for 3d $\mathcal{N}=2$$SU(N)$ with a Symmetric tensor}},  \href{https://arxiv.org/abs/2405.11972}{{\ttfamily 2405.11972}}.

\bibitem{Hwang:2021ulb}
C.~Hwang, S.~Pasquetti and M.~Sacchi, \emph{{Rethinking mirror symmetry as a local duality on fields}}, \href{https://doi.org/10.1103/PhysRevD.106.105014}{\emph{Phys. Rev. D} {\bfseries 106} (2022) 105014} [\href{https://arxiv.org/abs/2110.11362}{{\ttfamily 2110.11362}}].

\bibitem{Comi:2022aqo}
R.~Comi, C.~Hwang, F.~Marino, S.~Pasquetti and M.~Sacchi, \emph{{The SL(2, \ensuremath{\mathbb{Z}}) dualization algorithm at work}}, \href{https://doi.org/10.1007/JHEP06(2023)119}{\emph{JHEP} {\bfseries 06} (2023) 119} [\href{https://arxiv.org/abs/2212.10571}{{\ttfamily 2212.10571}}].

\bibitem{Giacomelli:2023zkk}
S.~Giacomelli, C.~Hwang, F.~Marino, S.~Pasquetti and M.~Sacchi, \emph{{Probing bad theories with the dualization algorithm. Part I}}, \href{https://doi.org/10.1007/JHEP04(2024)008}{\emph{JHEP} {\bfseries 04} (2024) 008} [\href{https://arxiv.org/abs/2309.05326}{{\ttfamily 2309.05326}}].

\bibitem{Giacomelli:2024laq}
S.~Giacomelli, C.~Hwang, F.~Marino, S.~Pasquetti and M.~Sacchi, \emph{{Probing bad theories with the dualization algorithm II}},  \href{https://arxiv.org/abs/2401.14456}{{\ttfamily 2401.14456}}.

\bibitem{Benvenuti:2023qtv}
S.~Benvenuti, R.~Comi and S.~Pasquetti, \emph{{Mirror dualities with four supercharges}},  \href{https://arxiv.org/abs/2312.07667}{{\ttfamily 2312.07667}}.

\bibitem{Intriligator:1995ne}
K.A.~Intriligator and P.~Pouliot, \emph{{Exact superpotentials, quantum vacua and duality in supersymmetric SP(N(c)) gauge theories}}, \href{https://doi.org/10.1016/0370-2693(95)00618-U}{\emph{Phys. Lett. B} {\bfseries 353} (1995) 471} [\href{https://arxiv.org/abs/hep-th/9505006}{{\ttfamily hep-th/9505006}}].

\bibitem{Aharony:1997gp}
O.~Aharony, \emph{{IR duality in d = 3 N=2 supersymmetric USp(2N(c)) and U(N(c)) gauge theories}}, \href{https://doi.org/10.1016/S0370-2693(97)00530-3}{\emph{Phys. Lett. B} {\bfseries 404} (1997) 71} [\href{https://arxiv.org/abs/hep-th/9703215}{{\ttfamily hep-th/9703215}}].

\bibitem{Benini:2017dud}
F.~Benini, S.~Benvenuti and S.~Pasquetti, \emph{{SUSY monopole potentials in 2+1 dimensions}}, \href{https://doi.org/10.1007/JHEP08(2017)086}{\emph{JHEP} {\bfseries 08} (2017) 086} [\href{https://arxiv.org/abs/1703.08460}{{\ttfamily 1703.08460}}].

\bibitem{Amariti:2014iza}
A.~Amariti and C.~Klare, \emph{{A journey to 3d: exact relations for adjoint SQCD from dimensional reduction}}, \href{https://doi.org/10.1007/JHEP05(2015)148}{\emph{JHEP} {\bfseries 05} (2015) 148} [\href{https://arxiv.org/abs/1409.8623}{{\ttfamily 1409.8623}}].

\bibitem{Nii_2015}
K.~Nii, \emph{3d duality with adjoint matter from 4d duality}, \href{https://doi.org/10.1007/jhep02(2015)024}{\emph{Journal of High Energy Physics} {\bfseries 2015} (2015) }.

\bibitem{Amariti:2018wht}
A.~Amariti and L.~Cassia, \emph{{USp(2N$_{c}$) SQCD$_{3}$ with antisymmetric: dualities and symmetry enhancements}}, \href{https://doi.org/10.1007/JHEP02(2019)013}{\emph{JHEP} {\bfseries 02} (2019) 013} [\href{https://arxiv.org/abs/1809.03796}{{\ttfamily 1809.03796}}].

\bibitem{Pasquetti:2019hxf}
S.~Pasquetti, S.S.~Razamat, M.~Sacchi and G.~Zafrir, \emph{{Rank $Q$ E-string on a torus with flux}}, \href{https://doi.org/10.21468/SciPostPhys.8.1.014}{\emph{SciPost Phys.} {\bfseries 8} (2020) 014} [\href{https://arxiv.org/abs/1908.03278}{{\ttfamily 1908.03278}}].

\bibitem{Bottini:2021vms}
L.E.~Bottini, C.~Hwang, S.~Pasquetti and M.~Sacchi, \emph{{4d S-duality wall and SL(2, \ensuremath{\mathbb{Z}}) relations}}, \href{https://doi.org/10.1007/JHEP03(2022)035}{\emph{JHEP} {\bfseries 03} (2022) 035} [\href{https://arxiv.org/abs/2110.08001}{{\ttfamily 2110.08001}}].

\bibitem{2003math......9252R}
E.M.~{Rains}, \emph{{Transformations of elliptic hypergometric integrals}}, \href{https://doi.org/10.48550/arXiv.math/0309252}{\emph{arXiv Mathematics e-prints} (2003) math/0309252} [\href{https://arxiv.org/abs/math/0309252}{{\ttfamily math/0309252}}].

\bibitem{Fokko}
F.J.~{van de Bult}, \emph{{Hyperbolic Hypergeometric Functions}}, Ph.D. thesis, Thomas Stieltjes Institute For Mathematics, 2007.

\bibitem{Spiridonov:2009za}
V.P.~Spiridonov and G.S.~Vartanov, \emph{{Elliptic Hypergeometry of Supersymmetric Dualities}}, \href{https://doi.org/10.1007/s00220-011-1218-9}{\emph{Commun. Math. Phys.} {\bfseries 304} (2011) 797} [\href{https://arxiv.org/abs/0910.5944}{{\ttfamily 0910.5944}}].

\bibitem{Spiridonov:2011hf}
V.P.~Spiridonov and G.S.~Vartanov, \emph{{Elliptic hypergeometry of supersymmetric dualities II. Orthogonal groups, knots, and vortices}}, \href{https://doi.org/10.1007/s00220-013-1861-4}{\emph{Commun. Math. Phys.} {\bfseries 325} (2014) 421} [\href{https://arxiv.org/abs/1107.5788}{{\ttfamily 1107.5788}}].

\bibitem{Cecotti:2012jx}
S.~Cecotti and M.~Del~Zotto, \emph{{Infinitely many N=2 SCFT with ADE flavor symmetry}}, \href{https://doi.org/10.1007/JHEP01(2013)191}{\emph{JHEP} {\bfseries 01} (2013) 191} [\href{https://arxiv.org/abs/1210.2886}{{\ttfamily 1210.2886}}].

\bibitem{Cecotti:2013lda}
S.~Cecotti, M.~Del~Zotto and S.~Giacomelli, \emph{{More on the N=2 superconformal systems of type $D_p(G)$}}, \href{https://doi.org/10.1007/JHEP04(2013)153}{\emph{JHEP} {\bfseries 04} (2013) 153} [\href{https://arxiv.org/abs/1303.3149}{{\ttfamily 1303.3149}}].

\bibitem{Wang:2015mra}
Y.~Wang and D.~Xie, \emph{{Classification of Argyres-Douglas theories from M5 branes}}, \href{https://doi.org/10.1103/PhysRevD.94.065012}{\emph{Phys. Rev. D} {\bfseries 94} (2016) 065012} [\href{https://arxiv.org/abs/1509.00847}{{\ttfamily 1509.00847}}].

\bibitem{Maruyoshi:2023mnv}
K.~Maruyoshi, E.~Nardoni and J.~Song, \emph{{Dualities of adjoint SQCD and supersymmetry enhancement}}, \href{https://doi.org/10.1007/JHEP09(2023)082}{\emph{JHEP} {\bfseries 09} (2023) 082} [\href{https://arxiv.org/abs/2306.08867}{{\ttfamily 2306.08867}}].

\bibitem{Xie:2021omd}
D.~Xie and W.~Yan, \emph{{A study of $ \mathcal{N} $ = 1 SCFT derived from $ \mathcal{N} $ = 2 SCFT: index and chiral ring}}, \href{https://doi.org/10.1007/JHEP03(2023)201}{\emph{JHEP} {\bfseries 03} (2023) 201} [\href{https://arxiv.org/abs/2109.04090}{{\ttfamily 2109.04090}}].

\bibitem{Kang:2023dsa}
M.J.~Kang, C.~Lawrie, K.-H.~Lee and J.~Song, \emph{{Emergent N=4 Supersymmetry from N=1}}, \href{https://doi.org/10.1103/PhysRevLett.130.231601}{\emph{Phys. Rev. Lett.} {\bfseries 130} (2023) 231601} [\href{https://arxiv.org/abs/2302.06622}{{\ttfamily 2302.06622}}].

\bibitem{Closset:2020afy}
C.~Closset, S.~Giacomelli, S.~Schafer-Nameki and Y.-N.~Wang, \emph{{5d and 4d SCFTs: Canonical Singularities, Trinions and S-Dualities}}, \href{https://doi.org/10.1007/JHEP05(2021)274}{\emph{JHEP} {\bfseries 05} (2021) 274} [\href{https://arxiv.org/abs/2012.12827}{{\ttfamily 2012.12827}}].

\bibitem{Giacomelli:2020ryy}
S.~Giacomelli, N.~Mekareeya and M.~Sacchi, \emph{{New aspects of Argyres--Douglas theories and their dimensional reduction}}, \href{https://doi.org/10.1007/JHEP03(2021)242}{\emph{JHEP} {\bfseries 03} (2021) 242} [\href{https://arxiv.org/abs/2012.12852}{{\ttfamily 2012.12852}}].

\bibitem{Hwang}
C.~Hwang and S.~Kim, \emph{{ S-confinement of 3d Argyres–Douglas theories and the Seiberg-like duality with an adjoint matter}}, .

\bibitem{Borokhov:2002ib}
V.~Borokhov, A.~Kapustin and X.-k.~Wu, \emph{{Topological disorder operators in three-dimensional conformal field theory}}, \href{https://doi.org/10.1088/1126-6708/2002/11/049}{\emph{JHEP} {\bfseries 11} (2002) 049} [\href{https://arxiv.org/abs/hep-th/0206054}{{\ttfamily hep-th/0206054}}].

\bibitem{Borokhov:2002cg}
V.~Borokhov, A.~Kapustin and X.-k.~Wu, \emph{{Monopole operators and mirror symmetry in three-dimensions}}, \href{https://doi.org/10.1088/1126-6708/2002/12/044}{\emph{JHEP} {\bfseries 12} (2002) 044} [\href{https://arxiv.org/abs/hep-th/0207074}{{\ttfamily hep-th/0207074}}].

\bibitem{Gaiotto:2008ak}
D.~Gaiotto and E.~Witten, \emph{{S-Duality of Boundary Conditions In N=4 Super Yang-Mills Theory}}, \href{https://doi.org/10.4310/ATMP.2009.v13.n3.a5}{\emph{Adv. Theor. Math. Phys.} {\bfseries 13} (2009) 721} [\href{https://arxiv.org/abs/0807.3720}{{\ttfamily 0807.3720}}].

\bibitem{Benna:2009xd}
M.K.~Benna, I.R.~Klebanov and T.~Klose, \emph{{Charges of Monopole Operators in Chern-Simons Yang-Mills Theory}}, \href{https://doi.org/10.1007/JHEP01(2010)110}{\emph{JHEP} {\bfseries 01} (2010) 110} [\href{https://arxiv.org/abs/0906.3008}{{\ttfamily 0906.3008}}].

\bibitem{Bashkirov:2010kz}
D.~Bashkirov and A.~Kapustin, \emph{{Supersymmetry enhancement by monopole operators}}, \href{https://doi.org/10.1007/JHEP05(2011)015}{\emph{JHEP} {\bfseries 05} (2011) 015} [\href{https://arxiv.org/abs/1007.4861}{{\ttfamily 1007.4861}}].

\bibitem{Amariti:2021lhk}
A.~Amariti, M.~Fazzi, S.~Rota and A.~Segati, \emph{{Conformal S-dualities from O-planes}}, \href{https://doi.org/10.1007/JHEP01(2022)116}{\emph{JHEP} {\bfseries 01} (2022) 116} [\href{https://arxiv.org/abs/2108.05397}{{\ttfamily 2108.05397}}].

\bibitem{Amariti:2022dyi}
A.~Amariti, M.~Bianchi, M.~Fazzi, S.~Mancani, F.~Riccioni and S.~Rota, \emph{{$ \mathcal{N} $ = 1 conformal dualities from unoriented chiral quivers}}, \href{https://doi.org/10.1007/JHEP09(2022)235}{\emph{JHEP} {\bfseries 09} (2022) 235} [\href{https://arxiv.org/abs/2207.10100}{{\ttfamily 2207.10100}}].

\bibitem{Benvenuti:2020wpc}
S.~Benvenuti, I.~Garozzo and G.~Lo~Monaco, \emph{{Monopoles and dualities in 3d$ \mathcal{N} $ = 2 quivers}}, \href{https://doi.org/10.1007/JHEP10(2021)191}{\emph{JHEP} {\bfseries 10} (2021) 191} [\href{https://arxiv.org/abs/2012.08556}{{\ttfamily 2012.08556}}].

\bibitem{Aprile:2018oau}
F.~Aprile, S.~Pasquetti and Y.~Zenkevich, \emph{{Flipping the head of $T[SU(N)]$: mirror symmetry, spectral duality and monopoles}}, \href{https://doi.org/10.1007/JHEP04(2019)138}{\emph{JHEP} {\bfseries 04} (2019) 138} [\href{https://arxiv.org/abs/1812.08142}{{\ttfamily 1812.08142}}].

\bibitem{Hwang:2020wpd}
C.~Hwang, S.~Pasquetti and M.~Sacchi, \emph{{4d mirror-like dualities}}, \href{https://doi.org/10.1007/JHEP09(2020)047}{\emph{JHEP} {\bfseries 09} (2020) 047} [\href{https://arxiv.org/abs/2002.12897}{{\ttfamily 2002.12897}}].

\bibitem{Cremonesi:2013lqa}
S.~Cremonesi, A.~Hanany and A.~Zaffaroni, \emph{{Monopole operators and Hilbert series of Coulomb branches of $3d$ $\mathcal{N} = 4$ gauge theories}}, \href{https://doi.org/10.1007/JHEP01(2014)005}{\emph{JHEP} {\bfseries 01} (2014) 005} [\href{https://arxiv.org/abs/1309.2657}{{\ttfamily 1309.2657}}].

\bibitem{Gaiotto:2009we}
D.~Gaiotto, \emph{{N=2 dualities}}, \href{https://doi.org/10.1007/JHEP08(2012)034}{\emph{JHEP} {\bfseries 08} (2012) 034} [\href{https://arxiv.org/abs/0904.2715}{{\ttfamily 0904.2715}}].

\bibitem{Giacomelli:2017ckh}
S.~Giacomelli, \emph{{RG flows with supersymmetry enhancement and geometric engineering}}, \href{https://doi.org/10.1007/JHEP06(2018)156}{\emph{JHEP} {\bfseries 06} (2018) 156} [\href{https://arxiv.org/abs/1710.06469}{{\ttfamily 1710.06469}}].

\bibitem{Xie:2016evu}
D.~Xie, W.~Yan and S.-T.~Yau, \emph{{Chiral algebra of the Argyres-Douglas theory from M5 branes}}, \href{https://doi.org/10.1103/PhysRevD.103.065003}{\emph{Phys. Rev. D} {\bfseries 103} (2021) 065003} [\href{https://arxiv.org/abs/1604.02155}{{\ttfamily 1604.02155}}].

\bibitem{Beem:2023ofp}
C.~Beem, M.~Martone, M.~Sacchi, P.~Singh and J.~Stedman, \emph{{Simplifying the Type $A$ Argyres-Douglas Landscape}},  \href{https://arxiv.org/abs/2311.12123}{{\ttfamily 2311.12123}}.

\bibitem{Shapere:2008zf}
A.D.~Shapere and Y.~Tachikawa, \emph{{Central charges of N=2 superconformal field theories in four dimensions}}, \href{https://doi.org/10.1088/1126-6708/2008/09/109}{\emph{JHEP} {\bfseries 09} (2008) 109} [\href{https://arxiv.org/abs/0804.1957}{{\ttfamily 0804.1957}}].

\bibitem{Anselmi:1997am}
D.~Anselmi, D.Z.~Freedman, M.T.~Grisaru and A.A.~Johansen, \emph{{Nonperturbative formulas for central functions of supersymmetric gauge theories}}, \href{https://doi.org/10.1016/S0550-3213(98)00278-8}{\emph{Nucl. Phys. B} {\bfseries 526} (1998) 543} [\href{https://arxiv.org/abs/hep-th/9708042}{{\ttfamily hep-th/9708042}}].

\bibitem{Gadde:2011uv}
A.~Gadde, L.~Rastelli, S.S.~Razamat and W.~Yan, \emph{{Gauge Theories and Macdonald Polynomials}}, \href{https://doi.org/10.1007/s00220-012-1607-8}{\emph{Commun. Math. Phys.} {\bfseries 319} (2013) 147} [\href{https://arxiv.org/abs/1110.3740}{{\ttfamily 1110.3740}}].

\bibitem{Song:2017oew}
J.~Song, D.~Xie and W.~Yan, \emph{{Vertex operator algebras of Argyres-Douglas theories from M5-branes}}, \href{https://doi.org/10.1007/JHEP12(2017)123}{\emph{JHEP} {\bfseries 12} (2017) 123} [\href{https://arxiv.org/abs/1706.01607}{{\ttfamily 1706.01607}}].

\bibitem{Aharony:2013dha}
O.~Aharony, S.S.~Razamat, N.~Seiberg and B.~Willett, \emph{{3d dualities from 4d dualities}}, \href{https://doi.org/10.1007/JHEP07(2013)149}{\emph{JHEP} {\bfseries 07} (2013) 149} [\href{https://arxiv.org/abs/1305.3924}{{\ttfamily 1305.3924}}].

\bibitem{Aharony:2013kma}
O.~Aharony, S.S.~Razamat, N.~Seiberg and B.~Willett, \emph{{3$d$ dualities from 4$d$ dualities for orthogonal groups}}, \href{https://doi.org/10.1007/JHEP08(2013)099}{\emph{JHEP} {\bfseries 08} (2013) 099} [\href{https://arxiv.org/abs/1307.0511}{{\ttfamily 1307.0511}}].

\bibitem{Elitzur:1997fh}
S.~Elitzur, A.~Giveon and D.~Kutasov, \emph{{Branes and N=1 duality in string theory}}, \href{https://doi.org/10.1016/S0370-2693(97)00375-4}{\emph{Phys. Lett. B} {\bfseries 400} (1997) 269} [\href{https://arxiv.org/abs/hep-th/9702014}{{\ttfamily hep-th/9702014}}].

\bibitem{Elitzur:1997hc}
S.~Elitzur, A.~Giveon, D.~Kutasov, E.~Rabinovici and A.~Schwimmer, \emph{{Brane dynamics and N=1 supersymmetric gauge theory}}, \href{https://doi.org/10.1016/S0550-3213(97)00446-X}{\emph{Nucl. Phys. B} {\bfseries 505} (1997) 202} [\href{https://arxiv.org/abs/hep-th/9704104}{{\ttfamily hep-th/9704104}}].

\bibitem{Hanany:1996ie}
A.~Hanany and E.~Witten, \emph{{Type IIB superstrings, BPS monopoles, and three-dimensional gauge dynamics}}, \href{https://doi.org/10.1016/S0550-3213(97)00157-0}{\emph{Nucl. Phys. B} {\bfseries 492} (1997) 152} [\href{https://arxiv.org/abs/hep-th/9611230}{{\ttfamily hep-th/9611230}}].

\bibitem{Intriligator:1995ax}
K.A.~Intriligator, R.G.~Leigh and M.J.~Strassler, \emph{{New examples of duality in chiral and nonchiral supersymmetric gauge theories}}, \href{https://doi.org/10.1016/0550-3213(95)00473-1}{\emph{Nucl. Phys. B} {\bfseries 456} (1995) 567} [\href{https://arxiv.org/abs/hep-th/9506148}{{\ttfamily hep-th/9506148}}].

\bibitem{Intriligator:2003mi}
K.A.~Intriligator and B.~Wecht, \emph{{RG fixed points and flows in SQCD with adjoints}}, \href{https://doi.org/10.1016/j.nuclphysb.2003.10.033}{\emph{Nucl. Phys. B} {\bfseries 677} (2004) 223} [\href{https://arxiv.org/abs/hep-th/0309201}{{\ttfamily hep-th/0309201}}].

\bibitem{Intriligator:2016sgx}
K.~Intriligator and E.~Nardoni, \emph{{Deformations of $W_{A,D,E}$ SCFTs}}, \href{https://doi.org/10.1007/JHEP09(2016)043}{\emph{JHEP} {\bfseries 09} (2016) 043} [\href{https://arxiv.org/abs/1604.04294}{{\ttfamily 1604.04294}}].

\bibitem{Kutasov:2014yqa}
D.~Kutasov and J.~Lin, \emph{{Exceptional N=1 Duality}},  \href{https://arxiv.org/abs/1401.4168}{{\ttfamily 1401.4168}}.

\bibitem{Razamat:2022gpm}
S.S.~Razamat, E.~Sabag, O.~Sela and G.~Zafrir, \emph{{Aspects of 4d supersymmetric dynamics and geometry}}, \href{https://doi.org/10.21468/SciPostPhysLectNotes.78}{\emph{SciPost Phys. Lect. Notes} {\bfseries 78} (2024) 1} [\href{https://arxiv.org/abs/2203.06880}{{\ttfamily 2203.06880}}].

\bibitem{Kim:2017toz}
H.-C.~Kim, S.S.~Razamat, C.~Vafa and G.~Zafrir, \emph{{E-String Theory on Riemann Surfaces}}, \href{https://doi.org/10.1002/prop.201700074}{\emph{Fortsch. Phys.} {\bfseries 66} (2018) 1700074} [\href{https://arxiv.org/abs/1709.02496}{{\ttfamily 1709.02496}}].

\bibitem{Razamat:2020bix}
S.S.~Razamat and E.~Sabag, \emph{{SQCD and pairs of pants}}, \href{https://doi.org/10.1007/JHEP09(2020)028}{\emph{JHEP} {\bfseries 09} (2020) 028} [\href{https://arxiv.org/abs/2006.03480}{{\ttfamily 2006.03480}}].

\bibitem{Kim:2018bpg}
H.-C.~Kim, S.S.~Razamat, C.~Vafa and G.~Zafrir, \emph{{D-type Conformal Matter and SU/USp Quivers}}, \href{https://doi.org/10.1007/JHEP06(2018)058}{\emph{JHEP} {\bfseries 06} (2018) 058} [\href{https://arxiv.org/abs/1802.00620}{{\ttfamily 1802.00620}}].

\bibitem{Kim:2018lfo}
H.-C.~Kim, S.S.~Razamat, C.~Vafa and G.~Zafrir, \emph{{Compactifications of ADE conformal matter on a torus}}, \href{https://doi.org/10.1007/JHEP09(2018)110}{\emph{JHEP} {\bfseries 09} (2018) 110} [\href{https://arxiv.org/abs/1806.07620}{{\ttfamily 1806.07620}}].

\bibitem{Sacchi:2021afk}
M.~Sacchi, O.~Sela and G.~Zafrir, \emph{{Compactifying 5d superconformal field theories to 3d}}, \href{https://doi.org/10.1007/JHEP09(2021)149}{\emph{JHEP} {\bfseries 09} (2021) 149} [\href{https://arxiv.org/abs/2105.01497}{{\ttfamily 2105.01497}}].

\bibitem{Sacchi:2021wvg}
M.~Sacchi, O.~Sela and G.~Zafrir, \emph{{On the 3d compactifications of 5d SCFTs associated with SU(N + 1) gauge theories}}, \href{https://doi.org/10.1007/JHEP05(2022)053}{\emph{JHEP} {\bfseries 05} (2022) 053} [\href{https://arxiv.org/abs/2111.12745}{{\ttfamily 2111.12745}}].

\bibitem{Sacchi:2023rtp}
M.~Sacchi, O.~Sela and G.~Zafrir, \emph{{Trinions for the 3d compactification of the 5d rank 1 $ {E}_{N_{f+1}} $ SCFTs}}, \href{https://doi.org/10.1007/JHEP06(2023)085}{\emph{JHEP} {\bfseries 06} (2023) 085} [\href{https://arxiv.org/abs/2301.06561}{{\ttfamily 2301.06561}}].

\bibitem{Sacchi:2023omn}
M.~Sacchi, O.~Sela and G.~Zafrir, \emph{{5d to 3d compactifications and discrete anomalies}}, \href{https://doi.org/10.1007/JHEP10(2023)185}{\emph{JHEP} {\bfseries 10} (2023) 185} [\href{https://arxiv.org/abs/2305.08185}{{\ttfamily 2305.08185}}].

\end{thebibliography}\endgroup

\end{document}